\pdfoutput=1
\documentclass[useAMS,usenatbib]{mn2e}
\usepackage[caption=false]{subfig}
\usepackage{fancyhdr}
\usepackage{graphicx}
\usepackage{ifpdf}
\ifpdf
  \DeclareGraphicsExtensions{.pdf}
\else
  \DeclareGraphicsExtensions{.eps}
\fi
\usepackage{longtable}
\usepackage{tabularx}
\usepackage{threeparttablex}
\usepackage{amsmath}
\usepackage{lscape}
\usepackage{subfig}
\usepackage{amssymb}

\newcommand{\Msun}{\mbox{\,$\rm M_{\odot}$}}
\newcommand{\Lsun}{\mbox{\,$\rm L_{\odot}$}}
 
\newcommand{\teff}{$T_{\rm eff}$}

\newcommand{\logg}{$\log g$}

\title[Post-AGBs, post-RGBs and YSOs in the LMC]
{Optically Visible Post-AGB Stars, Post-RGB Stars and Young Stellar Objects in the Large Magellanic 
Cloud}
\author[D. Kamath et al.]{D. Kamath$^{1}$\thanks{E-mail:
Devika.Kamath@ster.kuleuven.be (DK)}, P. R. Wood$^{2}$ and H. Van Winckel$^{1}$\\
$^{1}$Instituut voor Sterrenkunde, K.U.Leuven, Celestijnenlaan 200D bus 
2401, B-3001 Leuven, Belgium\\
$^{2}$Research School of Astronomy and Astrophysics, Mount Stromlo 
Observatory, Weston Creek, ACT 2611, Australia\\}

\begin{document}

\date{Accepted ... Received ...; in original form 2012 February 1}

\pagerange{\pageref{firstpage}--\pageref{lastpage}} \pubyear{2012}

\maketitle

\label{firstpage}

\begin{abstract}
We have carried out a search for optically visible post-Asymptotic
Giant Branch (post-AGB) stars in the Large Magellanic Cloud
(LMC). First, we selected candidates with a mid-IR excess and then
obtained their optical spectra. We disentangled contaminants with
unique spectra such as M-stars, C-stars, planetary nebulae,
quasi-stellar objects and background galaxies. Subsequently, we
performed a detailed spectroscopic analysis of the remaining
candidates to estimate their stellar parameters such as effective
temperature, surface gravity ($\log\,g$), metallicity ([Fe/H]),
reddening and their luminosities. This resulted in a sample of 35
likely post-AGB candidates with late-G to late-A spectral types, low
$\log\,g$, and [Fe/H]\,$<$\,$-$0.5. Furthermore, our study confirmed
the existence of the dusty post-Red Giant Branch (post-RGB) stars,
discovered previously in our SMC survey, by revealing 119 such objects
in the LMC. These objects have mid-IR excesses and stellar
parameters ($T_{\rm eff}$, $\log\,g$, [Fe/H]) similar to those of post-AGB stars except
that their luminosities ($<$\,2500\,\Lsun), and hence masses and radii,
are lower. These post-RGB stars are likely to be products
of binary interaction on the RGB. The post-AGB and post-RGB objects
show SED properties similar to the Galactic post-AGB stars, where some
have a surrounding circumstellar shell, while some others have a
surrounding stable disc similar to the Galactic post-AGB
binaries. This study also resulted in a new sample of 162 young
stellar objects, identified based on a robust $\log\,g$
criterion. Other interesting outcomes include objects with an UV
continuum and an emission line spectrum; luminous supergiants; hot
main-sequence stars; and 15 B[e] star candidates, 12 of which are
newly discovered in this study.
\end{abstract}

\section{Introduction}
\label{intro}

Post asymptotic giant branch (post-AGB) stars are low- to
intermediate-mass stars that have evolved off the 
asymptotic giant branch (AGB) because a strong dusty mass loss has removed 
almost the whole stellar envelope. For single stars, this mass loss occurs during a 
phase of very high mass-loss rates called the ’superwind phase’
\citep[e.g.,][]{vw93}. During the post-AGB phase, which lasts for
around $\sim$\,10$^2$\,$-$\,10$^5$ years (depending on the core mass),
the radius of the star decreases and the star evolves to higher temperatures (from 3\,$\times$\,10$^3$\,K to 
$>$\,2\,$\times$\,10$^4$\,K) with a constant luminosity
\citep[e.g.,][]{vw93,vw94}. When the hot central star ionizes the surrounding mass-loss shell, the 
object is referred to as a planetary nebula
\citep[PN,][]{balick02}. During the post-AGB phase, the warm stellar
photosphere makes it possible to 
quantify the chemical abundances in post-AGB stars for a very wide range of 
elements from CNO up to the most heavy $s$-process elements, well beyond
the Ba peak \citep{reyniers03}, 
that are brought to the stellar surface during the AGB phase \citep{karakas07b,karakas14a}. 

For low- to intermediate-mass AGB stars in binary systems, mass loss can be 
induced via binary interaction \citep[e.g.,][]{waters97,chen11,nie12,abate13,soker14}. 
A significant fraction of the ejected matter may end 
up in a circumbinary disc of dust and gas, and inside the disc is a binary system 
containing the post-AGB star \citep[e.g.,][]{deruyter06}. Binary 
interaction alters the intrinsic properties (such as: chemical composition, 
pulsation, mass-loss, dust-formation, circumstellar envelope morphology etc.) 
of the star and plays a dominant role in determining its ultimate fate. 
A variety of peculiar objects such as barium stars and bipolar
planetary nebulae result from such binary interactions. 
Therefore, an in-depth study using post-AGB stars is essential to 
constrain both single and binary stellar evolution and to study the evolutionary 
connection to possible precursors and progeny. 

Owing to their dusty circumstellar environments, a large mid-infrared (mid-IR) excess is a characteristic 
feature of post-AGB stars and a detection of cold circumstellar material using mid-IR 
photometry can be used to identify these objects. The first extensive 
search for these objects was initiated in the mid-80's using results from the 
Infrared Astronomical Satellite \citep{neugebauer84} which enabled the identification 
of post-AGB stars in our Galaxy \citep{kwok93}. The Toru$\acute{\rm n}$ catalogue 
\citep{szczerba07} for Galactic post-AGB stars lists around 391 very 
likely post-AGB objects. The Galactic sample of post-AGB stars have been found to be a 
very diverse group of objects \citep{vanwinckel03}. 
Studies showed that the majority of the optically visible Galactic post-AGB stars could be classified 
based on their spectral energy distributions (SEDs) into two groups:
shell-sources and disc-sources \citep{vanwinckel03}. The shell-sources show
a double-peaked SED with the hot central star peaking at shorter
wavelengths while the cold, detached, expanding dust shell peaks at
longer wavelengths. This type of SED is considered to be characteristic of objects
that follow the single star evolution
scenario mentioned above. The disc-sources do not show two distinct flux peaks in the mid-IR but they do display a clear near-infrared (near-IR) excess indicating that circumstellar dust must be close 
to the central star, near sublimation temperature.  It is now well established that 
this feature in the SED indicates the presence of a stable compact circumbinary disc, 
and therefore these sources are referred to as 
disc-sources \citep{deruyter06,deroo07,gielen11-pro,hillen13}. 
The rotation of the disc was resolved with the ALMA array \citep{bujarrabal13a} in 
one object and using single dish observations \citet{bujarrabal13b} confirmed that disc rotation is indeed 
widespread. Moreover, these disc-sources are
confirmed to be binaries 
and show orbital periods between 100 and 2000 days \citep{vanwinckel09,gorlova14}. In 
contrast, for the Galactic shell-sources 
long-term radial velocity monitoring efforts have not yet resulted in any 
clear detected binary orbit \citep{hrivnak11}, which either confirms the
single-star nature of these objects or introduces a possibility that
these systems can have companions on very wide orbits. 

The Galactic post-AGB stars show a large diversity in their chemical composition. 
Some objects are the most $s$-process
enriched objects known to date \citep{reyniers04} while others are not
enriched at all. These $s$-process rich
objects are considered to be progeny of carbon stars 
that have had $s$-process enrichment while on the AGB. Studies also show
that many of the binary Galactic post-AGB stars are depleted in refractory elements 
\citep[e.g.,][]{giridhar05,maas05,rao12}.

Though the Galactic post-AGB sample is observationally well studied, the poorly constrained 
distances pose a severe limitation on our ability to 
fully exploit this assorted group of objects.  For
instance, unknown distances (and hence luminosities and initial masses)
hampers the interpretation of the diversity in the chemical abundances of these objects as a function of
luminosity and mass. Also, for the binary post-AGB stars, it is not
possible to associate the binary orbital parameters to the luminosity or mass. 

To overcome this limitation and to better understand the post-AGB evolution, we
need to exploit populations of post-AGB stars with known distances
like those in the Magellanic
Clouds [MCs]. Until recently, in the Small Magellanic Cloud (SMC), only 5
post-AGB candidates had been
identified
\citep[][]{whitelock89,kuzinskas00,kraemer06,volk11,desmedt12}. Similarly,
in the Large Magellanic Cloud (LMC), \citet{wood01-pr} had identified
around 25 post-AGB stars based on a near-IR photometric and optical spectroscopic
study of point sources from the Midcourse Space Experiment (MSX) IR
survey \citep{egan01}. Subsequently, based on mid-IR and optical
photometry, \citet{vanaarle11} constructed a catalogue of 1337
optically visible post-AGB candidates in the LMC. They carried out a 
classification based on visual analysis of optical spectra of 105 of these candidates which 
resulted in 70 post-AGB candidates in the LMC. 
Recently, \citet{matsuura14} confirmed the evolutionary nature of
seven post-AGB stars that were previously identified by
\citet{wood01-pr} and \citet{vanaarle11}. Based on polycyclic aromatic hydrocarbon
(PAH) features that were
found in Spitzer Space Telescope (SST) IRS spectra of these objects, they
characterised the circumstellar dust which resulted
in five objects being classified as C-rich post-AGB stars and two
being classified as O-rich post-AGB stars.

To fully understand the post-AGB star population in the Magellanic
Clouds and to provide constraints for stellar evolutionary and
nucleosynthesis models, a more complete sample of these objects 
is definitely required. Therefore, in our recent studies,
we exploited the release of the mid-IR SST surveys SAGE for the LMC 
\citep{meixner06,blum06} and S$^{3}$MC \citep{bolatto07} and 
SAGE-SMC \citep{gordon11} for the SMC to identify post-AGB stars in
the MCs with 
mid-IR excesses indicative of a past history of heavy dusty mass loss. We also 
performed an extensive low-resolution optical spectroscopic survey to 
systematically characterise the optically bright post-AGB stars in the 
SMC and LMC. 

The results of our SMC study are presented in 
\citet{kamath14} (hereafter referred to as Paper I). That study
provided the first extensive spectroscopically 
verified catalogue of optically visible post-AGB candidates in the
SMC. We were able to
find a sample of 21 likely post-AGB 
candidates. These objects showed SED properties similar to the
Galactic post-AGB stars mentioned above. 
The SMC study also resulted in the discovery of a sample of 42 
new, low-luminosity, dusty objects that are likely to be
post-RGB candidates. These objects 
have mid-IR excesses, stellar parameters and SEDs similar to post-AGB stars 
(late-G to late-A spectral types, low log g values, and 
low metallicities with [Fe/H]\,$\lesssim$\,$-$1.0). However, their 
luminosities ($\approx$\,100\,$-$\,2500\Lsun), and
hence masses and radii, are much lower than that expected
for post-AGB stars. Since RGB stars are known to have luminosities
$\gtrsim$\,2500\Lsun\footnote{Based on observational studies in the
  Magellanic Clouds \citep[e.g.,][]{frogel83,wood99-pr,cioni99} and 
evolutionary tracks of \citet{bertelli08} corresponding to LMC and SMC 
metallicities}, it is likely that these objects are evolved, dusty
post-RGB stars. Such objects have, so far, not been identified in the
Galaxy, because of the unknown distance and hence luminosities to the
Galactic objects. Furthermore, our study also resulted in the identification of other interesting contaminating
objects with similar IR colours to post-AGB stars, such as: 
M-stars, C-stars, red-shifted galaxies, planetary nebulae (PNe), 
quasi-stellar objects (QSOs), objects with strong emission lines,
and luminous young stellar objects (YSOs). All objects but the YSOs
have unique spectra that allow them to be easily identified. However, 
disentangling the YSOs from the post-AGB/post-RGB stars is 
not very easy as YSOs are also surrounded by large 
amounts of circumstellar material and have SEDs, luminosities and
spectral types similar to 
post-AGB/post-RGB stars. Therefore, we devised a rather robust
separation criterion 
based on the \logg\, of the central star which resulted in a sample of 
40 likely YSO candidates. Our SMC survey revealed a rich
sample of post-AGB objects similar to the Galactic post-AGB stars, a  
new class of dusty post-RGB objects, a significant YSO population and 
other interesting contaminants with similar IR colours. 

To complement our above mentioned SMC study and to extend the photometric
classification of optically visible post-AGB stars in the LMC by
\citet{vanaarle11}, we carried out an extensive low-resolution optical 
spectroscopic survey to identify and characterise optically visible
post-AGB stars. In this paper, we present the 
results of our LMC survey. 

The structure of this paper is as follows: In Section~\ref{sampleselection} 
we present an overview of the selection criteria used to obtain an initial sample of 
post-AGB/post-RGB candidates in the 
LMC. The sample of objects considered for the spectroscopic analysis,
along with the specifications of the optical spectra, the data
reduction procedure and a preliminary spectral classification 
are presented in Section~\ref{specobv} and Section~\ref{preclass}. In Section~\ref{rv} 
we present the results of the radial velocity study used to establish LMC membership of the 
objects. In Section~\ref{SP} we present our detailed spectroscopic analysis 
to estimate the stellar parameters (\teff, \logg, [Fe/H]) from the 
spectra, and the reddening (E[$B-V$]) from SED fitting. 
Subsequently, in Section~\ref{pagbprgbyso} we present the final catalogues of the 
spectroscopically verified post-AGB, post-RGB and YSO candidates. 
In Sections~\ref{sedanalysis}\,$-$~\ref{hrtext} we analyse different 
characteristics of these objects by examining their SEDs, their
optical spectra and the evolutionary phase of the 
individual candidates. The completeness of 
the survey is presented in Section~\ref{completeness}. In
Section~\ref{evolrate}, we empirically estimate the evolutionary
rates of the post-AGB and post-RGB phase. Finally, we
conclude with an evaluation and summary of our survey.

\section{Initial sample selection}
\label{sampleselection}

To obtain an initial sample selection for our spectroscopic survey, we adopted the
sample selected by \citet{vanaarle11}, who identified 
optically visible post-AGB star candidates in the LMC based on
photometry. Full details of
their selection of objects can be found in \citet{vanaarle11}. In
brief, the post-AGB candidates were selected  from the $\sim4
\times 10^{6}$ sources in the SST SAGE LMC survey \citep{meixner06,blum06}, such that
they possess a mid-IR excess compared to the mid-IR flux estimated
from their optical fluxes. The selection 
criteria required the sources to have both 
24$\mu$m and 8$\mu$m magnitudes which satisfy $F_{24\mu\rm{m}} > 0.4
\times F_{8\mu\rm{m}}$, (where, $F_{24\mu\rm{m}}$ and $F_{24\mu\rm{m}}$ are
the fluxes at 24$\mu$m and 8$\mu$m, respectively) or
[8]\,$-$\,[24]\,$>$\,1.384, (where [24] and [8] are the 24$\mu$m and
8$\mu$m magnitudes, respectively). Furthermore, the objects were also required to have a catalogued $U, B, V, R,$ or $I$ magnitude from 
either the Massey $U,B,V,R$ CCD survey of the Magellanic Clouds
\citep{massey02}, the LMC stellar catalogue \citep{zaritsky04}, or the
Guide Star Catalogue Version 2.3.2 \citep{lasker08}. The
resulting sample consisted of 8628 objects. This initial sample
selection was further refined by imposing a luminosity criterion based
on luminosities calculated from blackbody fits ($L_{\rm bb}$) to the raw photometry. The first luminosity
cut (1000 to 35000 $L_{\rm bb}/$L$_{\odot}$) was 
based on the expected luminosity range of post-AGB stars from evolutionary tracks of
\citet{bloecker95} and was aimed at removing contaminating objects such as massive young stellar objects (YSOs) and the supergiants. This
luminosity cut resulted in a sample of 1517 objects. We gave these
objects a priority 1 when assigning objects for spectroscopic
observations. To relax the selection criteria, we included objects
from \citet{vanaarle11} with $L_{\rm  bb}/$L$_{\odot}$\,$<$\,1000. This luminosity cut resulted in a
further sample of 6823
objects. We gave these
objects a priority 2 when assigning objects for spectroscopic
observations. Finally, the luminosity cut of $L_{\rm bb}/$L$_{\odot}$\,$>$\,35000 was removed to allow the inclusion of
supergiants. This luminosity cut removal resulted in a further sample of 
286 stars. We gave these objects a priority 3 when assigning
objects for observations. 

\begin{figure}
\includegraphics[width=9cm,angle=0]{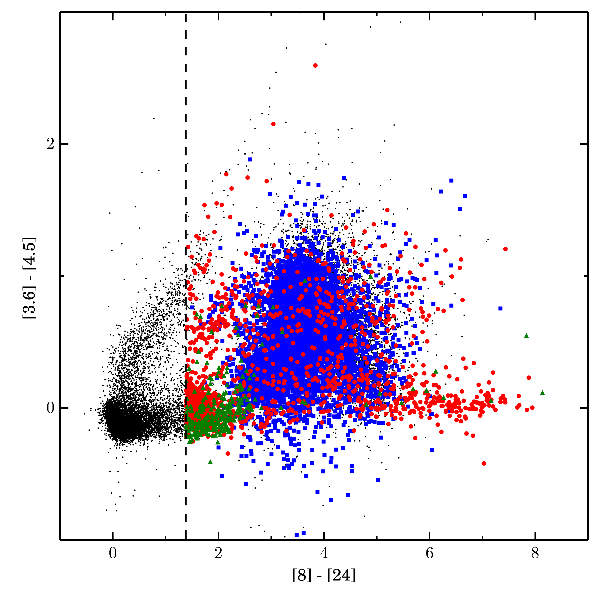}
\caption{The colour$-$colour plot of [8]\,$-$\,[24] versus
[3.6]\,$-$\,[4.5]. The small black dots in the 
background represent the entire field 
LMC population with a valid 24$\mu$m magnitude from the SAGE
catalogue. The red filled circles represent the priority 1 candidates. The blue filled squares represent the priority 2
objects and the green filled triangles represent the priority 3 objects. We note that
some of the priority 1, priority 2 and priority3 objects 
do not show detections at [3.6] or [4.5] microns, and therefore do not
appear in this plot. The region to the right of the black dashed
line defines the selection criterion in the [8]\,$-$\,[24] plane. The
black dots in this region fail one or more of our selection criteria. See text for further details.}
\label{fig:sampleselection}
\end{figure}

Our total initial sample consisted of 8626 objects within which
we expect to find post-AGB and post-RGB
candidates. Note that in the SMC survey (see Paper I), the spectroscopically verified
likely dusty post-RGB stars were discovered from a initial sample selection that was
related to the sample selection of \citet{vanaarle11}. Therefore, we
expect to find similar dusty, evolved, low-luminosity  post-RGB systems in the LMC as
well. 
The positions of the initial sample of objects are marked on the [8]\,$-$\,[24] versus
[3.6]\,$-$\,[4.5] colour$-$colour plot shown in Figure~\ref{fig:sampleselection}. 
The photometric magnitudes along with the blackbody luminosities ($L_{\rm bb}/$L$_{\odot}$) calculated from
blackbody fits of all the 8626 objects in the initial sample can be found in \citet{vanaarle11}. 

\section{Spectroscopic observations}
\label{specobv}

\begin{figure}
\begin{center}
\includegraphics[width=8cm,angle=0]{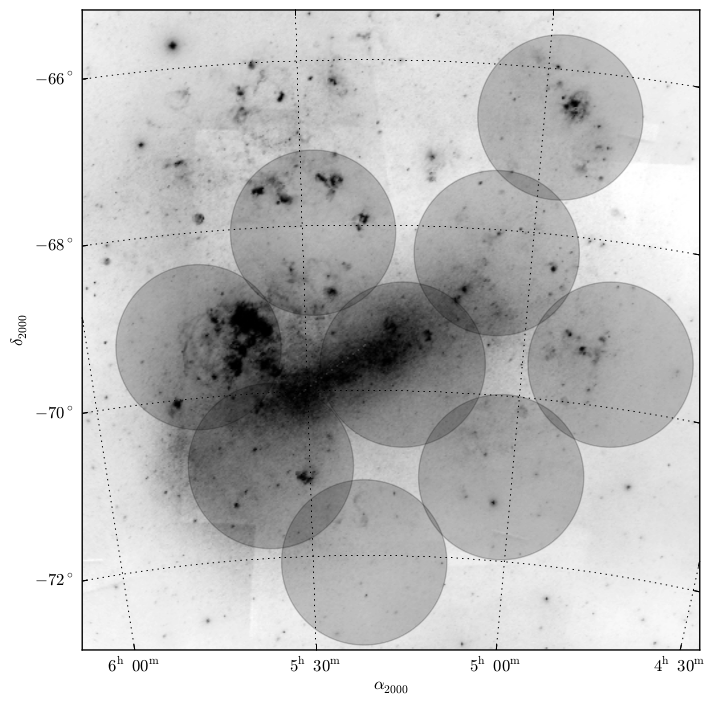} 
\caption{A Digitalised Sky Survey (DSS) image of a 8 degree field of view centered on the LMC. 
The circles represent the observed 2 degree fields in the LMC, with field centers listed in Table~\ref{table:fields}, that were covered in our survey.}
\label{fields}
\end{center}
\end{figure}

We obtained low-resolution spectra of stars selected from the initial sample of objects. The
spectra were taken using the AAOmega double-beam multi-fibre spectrograph \citep{sharp06} mounted 
on the 3.9m Anglo Australian Telescope (AAT) at Siding Spring
Observatory (SSO). AAOmega allows for the 
simultaneous observation of 392 targets 
(including science objects, sky-positions, and fiducial guide stars) over a 
2 degree field using the 2dF fibre positioner \citep{lewis02}. The projected diameter of each fibre is 2 arcsec. 
Within each configuration there is a minimum target 
separation of $\sim$30 arcsec imposed by the 
physical size of the fibres and positioner \citep{miszalski06}. 
The field centers of the LMC observations in our survey are given in 
Table~\ref{table:fields} and they are and shown in Figure~\ref{fields}. A Ne-Ar arc and a quartz lamp flat field exposure were
 taken per field for calibration. We used the 580V grating 
with a central wavelength of 4800\AA\,\, and the 385R grating with a central wavelength of 
7250\AA. This combination provides a maximum spectral coverage of 3700-8800\AA\,\, at a 
resolution of 1300. The AAOmega raw data were reduced using a combination of the AAOmega-2dFDR reduction 
pipeline\footnote{http://www.aao.gov.au/AAO/2df/aaomega/aaomega$\_$software.html} supplied and maintained
by the Anglo-Australian Observatory and 
IRAF\footnote{IRAF is distributed by the National Optical Astronomy Observatory, 
which is operated by the Association of Universities for Research in 
Astronomy (AURA) under cooperative agreement with the National Science
Foundation} routines. The spectral observations and the data reduction
method used in this study is the same as that of the SMC survey (Paper
I) and full details can be found therein. 

\begin{table}
  \caption{The field centers of the LMC observations in our survey and the corresponding exposure times.}
   \centering
  \begin{tabular}{cccc}
  \hline
   Field & RA (2000) & Dec (2000) & Exposure \\
  \hline
   LMC1 & 05 16 48.00 & -69 42 00.0 & 4\,$\times$\,900s \\
   LMC2 & 05 36 00.00 & -70 54 00.0 & 3\,$\times$\,900s \\
   LMC3 & 05 04 48.00 & -68 18 00.0 & 3\,$\times$\,900s \\
   LMC4 & 04 48 00.00 & -69 30 00.0 & 3\,$\times$\,900s \\
   LMC5 & 04 58 12.00 & -66 36 00.0 & 4\,$\times$\,900s \\
   LMC6 & 05 45 00.00 & -69 24 00.0 & 4\,$\times$\,900s \\
   LMC7 & 05 28 48.00 & -68 06 00.0 & 4\,$\times$\,900s \\
   LMC8 & 05 22 12.00 & -72 06 00.0 & 4\,$\times$\,900s \\
   LMC9 & 05 01 48.00 & -71 00 00.0 & 4\,$\times$\,900s \\
  \hline
 \end{tabular}
\label{table:fields}
\end{table}

\begin{figure}
\includegraphics[width=9cm,angle=0]{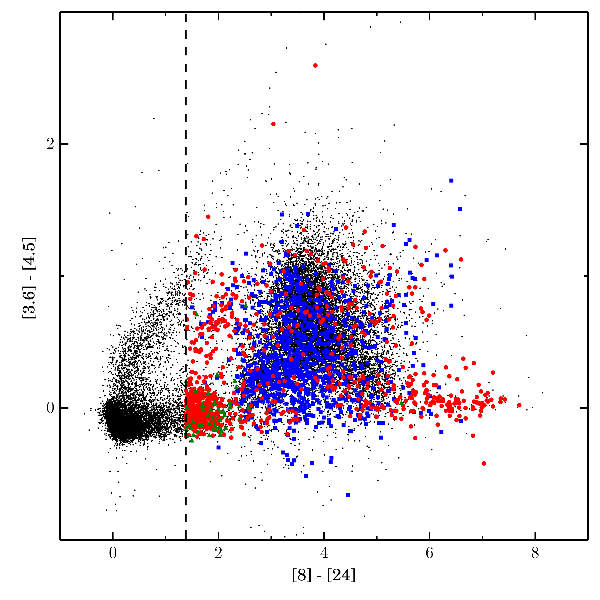}
\caption{The same as Figure~\ref{fig:sampleselection} but for only
  those 2102 objects from the initial selection sample
for which we obtained an optical spectrum. The small black dots in the 
background represent the entire field 
LMC population with a valid 24$\mu$m magnitude from the SAGE catalogue. The red filled circles represent the priority 1 
candidates with an optical spectrum. The blue filled squares represent the priority 2
objects with an optical spectrum and the green filled triangles
represent the priority 3 objects with an optical spectrum. We note that
some of the priority 1, priority 2 and priority 3 objects 
do not show detections at [3.6] or [4.5] microns, and therefore do not
appear in this plot.}
\label{2102sample}
\end{figure}

We obtained a total of 2262 spectra. Not all of the 8626 candidates
could be observed 
due to limited observational time. Moreover, some of the stars from the
initial sample selection were out of the fields that were
observed while some objects were spatially too close for fibre
assignments to be done simultaneously. Furthermore, as shown in
Figure~\ref{fields}, some of the target sources were observed more
than once due to overlap in the observed fields. For such sources, we
either averaged the multiple observations or rejected the observation
with low signal. In the end we were left with spectra of 2102  unique
candidates from the initial sample of 8626 objects. Out of these 2102
objects, 885 were priority 1 objects, 1173 were priority 2 objects and 49 
were priority 3 objects. Figure~\ref{2102sample} shows the positions
of these 2102 objects, marked on the [8]\,$-$\,[24] versus
[3.6]\,$-$\,[4.5] colour$-$colour plot used for our initial sample selection
(see Section~\ref{sampleselection}). 

In Table~\ref{photdatatable}, we provide photometric 
magnitudes for the first five candidates from the group of 2102 unique
candidates for which spectra were obtained. The 
full table which contains the photometry of the 2102 candidates 
is available as online supporting information. As mentioned in Section~\ref{sampleselection}, the selection criteria
for the candidates that we adopted from \citet{vanaarle11} in our survey involve 
optical, near-IR and mid-IR photometry. Photometry in the
$U, V, B, R$ and $I$ bands were added from 
either Massey $U, B, V, R$ CCD survey of the Magellanic Clouds
\citep{massey02}, the LMC stellar catalogue \citep{zaritsky04}, or the
Guide Star Catalogue Version 2.3.2 \citep{lasker08}. The SAGE-LMC survey covers the 
IRAC (3.6, 4.5, 5.8, and 8 $\mu$m) and the MIPS (24.0, 70.0, and 160.0$\mu$m) bands. 
The Spitzer sources have been merged 
with the Two Micron All Sky Survey (2MASS) $J$, $H$, and $K$ bands 
\citep[1.24, 1.66 and 2.16$\mu$m,][]{skrutskie06}. We 
also added WISE photometry in the 3.4, 4.6, 12, and 22 $\mu$m bands 
$W1, W2, W3,$ and $W4$, respectively \citep{wright10}. 
Column (22) of Table~\ref{photdatatable} also gives the 
observed 
luminosity ($L_{\rm ob}$) of the candidates. $L_{\rm ob}$ 
for these sources was obtained by integrating under the 
SED defined by all the photometric bands after correcting for the effects of
Galactic and LMC reddening (similar to the method that was used for the SMC
sources, see Paper I). We used a mean reddening E(B-V) = 0.08 mag for the combined LMC and 
Galactic components \citep{keller06}.

We found that for 556 of the 2102 objects, their
spectra had too low signal (mostly less that 200 counts) such that no
relevant spectral features could be detected. These objects were discarded. 
Hence, at this stage, we were
left with good spectra of 1546 individual objects. 

\begin{landscape}
\begin{table}
\small
\caption{Photometric data. See text for full details. The full table is available online.}
\vspace{-2mm}
\tabcolsep=2pt
\begin{tabular}{rrrrrrrrrrrrrrrrrrrrrr}
\hline
Object Name & RA($^\circ$) & DEC($^\circ$) & $U$ & $B$ & $V$ & $R$ & $I$ & $J$ & $H$ & $K$ & $W1$ & $[3.6]$ & $[4.5]$ & $W2$ & $[5.8]$ & $[8.0]$ & $W3$ & $W4$ & $[24]$ & $[70]$ & $[160]$ \\
$L_{\rm ob}$/L$_\odot$ & $T_{\rm eff,SED}$(K) &&&&&&&&&&&&&&&&&&&\\
\hline
J043617.75-693022.0 & 69.073958 & -69.506111 & 99.999 & 99.999 &
99.999 & 18.34 & 99.999 & 16.789 & 16.112 & 15.377 & 15.016 & 14.876 &
14.289 & 14.641 & 14.563 & 12.667 & 11.282 & 9.687 & 9.38 & 99.999 & 99.999 \\
121 & 3500  & &&&&\\
J043652.36-693356.2 & 69.218167 & -69.565611 & 99.999 & 99.999 &
99.999 & 17.75 & 99.999 & 16.98 & 16.445 & 15.927 & 99.999 & 15.326 &
15.133 & 99.999 & 14.794 & 12.433 & 99.999 & 99.999 & 9.24 & 99.999 & 99.999 \\
118 & 4250  & &&&&\\
J043655.62-694748.1 & 69.23175 & -69.796694 & 99.999 & 99.999 & 99.999
& 18.01 & 99.999 & 16.865 & 16.071 & 15.662 & 15.421 & 15.14 & 15.106 & 15.328 & 14.219 & 11.802 & 11.253 & 9.348 & 9.448 & 99.999 & 99.999 \\
128 & 3750  & &&&&\\
J043702.61-694130.6 & 69.260875 & -69.691833 & 19.738 & 19.025 &
17.327 & 15.54 & 14.313 & 12.573 & 11.668 & 11.376 & 11.353 & 11.237 &
11.324 & 11.335 & 11.107 & 10.786 & 10.185 & 9.253 & 9.307 & 99.999 & 99.999 \\
4011 & 3500  & &&&&\\
J043706.79-694922.6 & 69.278292 & -69.822944 & 18.814 & 18.727 &
17.696 & 15.48 & 99.999 & 15.254 & 14.537 & 14.191 & 13.611 & 14.076 &
14.237 & 13.584 & 13.511 & 12.099 & 11.156 & 8.998 & 9.342 & 99.999 & 99.999 \\
743 & 4750  & &&&&\\
\hline
\end{tabular}
\vspace{-2mm}
\begin{flushleft}
Note: The object name is the SAGE name. The RA and DEC coordinates are given for the 
J2000 epoch. Null magnitudes are listed as 99.999. $L_{\rm
  obs}$/L$_\odot$ is the observed luminosity. 
$T_{\rm eff,SED}$(K) is the photometric temperature obtained by
fitting ATLAS9 atmosphere models \citep{castelli03} to the $B, V, I,$ and $J$ bands corrected for foreground extinction. 
\end{flushleft}
\label{photdatatable}
\normalsize
\end{table}
\end{landscape}

\section{Preliminary spectral classification}
\label{preclass}

Though a mid-IR excess is characteristic of post-AGB and post-RGB stars,
there are other interesting objects with mid-IR excesses such as M
stars, C stars, PNe, background galaxies, and QSOs etc. 
These intermingled contaminants have characteristic spectra which can be
used to identify them. Therefore, we performed a preliminary spectral analysis,
by eye, to categorise the group of 1546 objects with good optical spectra into bins based on the
nature of their spectra. 

We found a group of 290 background galaxies and 39 QSOs that were identified by their large redshifts and the width 
of the emission lines of hydrogen and metallic lines
\citep[e.g.][]{field73, vandenberk01}. Some of the background galaxies and QSOs identified in this study have
been published in \citet{cioni13}. The full sample of background galaxies and QSOs will
be discussed in a following publication. A group of 382 cool M-stars were identified based on 
the presence of 
strong molecular absorption features of titanium oxide (TiO) and vanadium oxide (VO) 
\citep[e.g.][]{kirkpatrick99}. We also identified 
55 cool C-stars characterised by the presence of key molecules such as C$_{\rm 2}$, 
CN, and CH \citep[see][for a review on C stars]{wallerstein98}. 
Based on the presence of an emission-line spectrum characterised by recombination lines 
of hydrogen and helium as well as various collisionally-excited forbidden lines of heavier elements 
such as O, N, C, Ne, and Ar, we were able to identify 123 PNe 
\citep[see][for further details on identifying PNe]{frew10}. From the
123 PNe, 32 objects are likely newly identified PNe, as they have not been
previously classified in the literature. The sample of M-stars, 
C-stars and PNe is presented in Appendix~\ref{mcpne}. Five stars in our sample were identified 
to be stars with TiO bands in emission. These objects are discussed in \citet{wood13} 
and are not considered further here. We also found a group of 69 objects with prominent
emission lines throughout their spectrum. The majority of these objects also
show a broad H$\alpha$ emission line profile mostly along with a UV
continuum. Hot post-AGB/post-RGB stars are likely to have an emission-line spectrum
characterised by a broad H$\alpha$ profile along with weak recombination lines of hydrogen and helium and
various collisionally-excited forbidden lines of heavier elements
\citep[e.g.,][]{vanwinckel03}. The spectra of YSO candidates are also
likely to show a broad H$\alpha$ emission
line profile owing to the disc accretion \citep{natta02,jayawardhana02}. For 15 of these 69 objects 
(J044739.07-692036.5, J044745.05-694048.3,
J045647.06-695024.7, J050224.17-660637.4, J050504.33-674744.9,
J050951.27-684845.2, J051247.94-690307.1, J051338.88-692108.1,
J051451.00-692544.1, J052605.27-683609.4, J052613.39-684715.0,
J052630.65-674036.6, J052707.10-702001.9, 
J052747.62-714852.8 , J053218.75-681731.5) we
were able to clearly detect  FeII emission line features. Their
spectra not only show a 
UV continuum and strong Balmer emission lines, but also the
low-excitation permitted emission lines of singly ionised metals
(e.g., the multiplet 42 of FeII at $\approx$\,4924\AA, 5018\AA, and
5169\AA), 
forbidden emission lines of [FeII] and [OI] and higher
ionisation emission lines (e.g., [OIII] and [HeII]). These spectral
features in combination with a strong infrared excess in 
the spectra of early-type stars is characteristic of the 
B[e] phenomenon \citep{zickgraf00,miroshnichenko07}. Out of the 15 objects, three (J045647.06-695024.7, J051338.88-692108.1 and
J052747.62-714852.8) have been previously classified as B[e] stars by
\citet{zickgraf06}. Therefore, it is likely that
the remaining objects are newly identified B[e] stars in the
LMC. Thus, based on their spectral features, we expect the group of 69 objects
with an emission line spectrum to contain hot
post-AGB/post-RGB stars, hot YSOs and B[e] candidates. We do not carry out any further spectral analysis
on the 69 objects, owing to 
their spectra being dominated by emission lines.   
A detailed analysis of these objects will be presented in following a publication.

After the removal of the above mentioned contaminants, the remaining 
583 objects were carried forward for a detailed spectral analysis, similar to the one
carried out in Paper I. The spectra and IR colours of YSOs are very 
similar to those of post-AGB/post-RGB stars. Therefore, amongst the remaining
sample of 583 objects, 
we expect to find not just post-AGB/post-RGB candidates 
but also YSOs, similar to what we found in our SMC study(see Paper I). 

\begin{figure*}
\begin{center}
\includegraphics[width=18cm,angle=0]{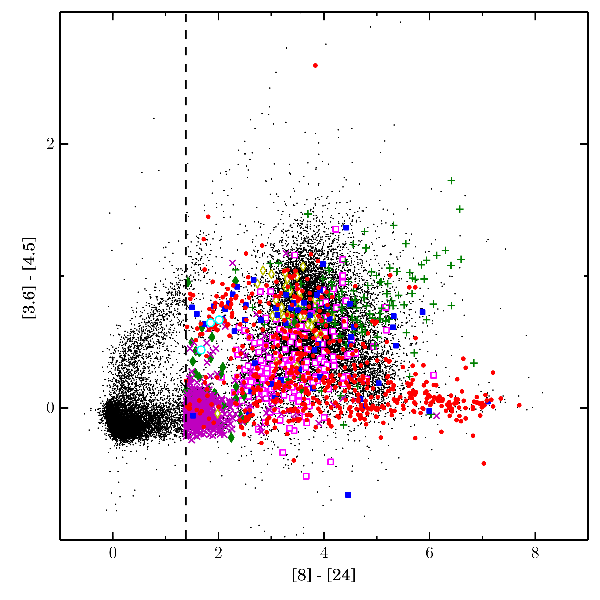} 
\caption{The 1546 sources with good spectra plotted on the
  colour-colour plot of [8]\,$-$\,[24] vs [3.6]\,$-$\,[4.5] with the general spectral classification indicated by symbol type. The small black dots in the 
background represent the entire field 
LMC population with a valid 24$\mu$m magnitude from the SAGE catalogue. The red 
filled-circles represent the 
sample of probable post-AGB/RGB and YSO candidates. The
blue filled-squares represent the sample of objects that have
strong emission lines. The cyan open-circles represent the objects with TiO bands in emission. 
The green plus symbols represent the PNe population. The green filled
diamonds represent 
C-stars, and the magenta crosses represent M-stars. The magenta open squares represent the sample that has been classified as background galaxies. The 
yellow open diamonds represent the sources identified as QSOs. In each
of the groups, objects that do not 
have detections at [3.6] or [4.5] microns do not
appear in this plot.}
\label{sampleselect2}
\end{center}
\end{figure*}

Figure~\ref{sampleselect2} shows all the 1546 sources with good spectra plotted on 
the colour$-$colour plot of [8]\,$-$\,[24] vs [3.6]\,$-$\,[4.5] used for
our sample selection (see Section~\ref{sampleselection}). The
different symbols in Figure~\ref{sampleselect2} represent the nature
of the sources as classified by the preliminary spectral analysis described above. 

\section{Establishing LMC membership of the probable post-AGB/post-RGB and YSO Candidates}
\label{rv}

\begin{figure}
\begin{center}
\includegraphics[width=8cm,angle=0]{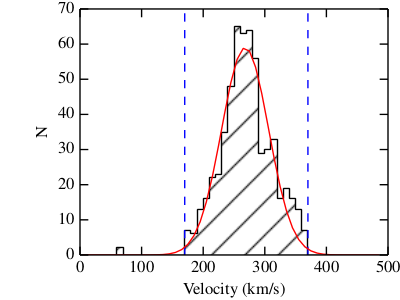}
\caption{Radial velocity histogram for the group of 583 probable post-AGB/post-RGB and YSO candidates. 
The red curve denotes a Gaussian with a mean of 270\,kms$^{-1}$ and a
standard deviation of 40\,kms$^{-1}$. 
The blue dashed lines denote the radial velocity 
interval (defined as a 2.5 sigma deviation from the mean of the
Gaussian fit) used in our study to select stars with a radial velocity
belonging to the LMC.}
\label{fig:rvplot}
\end{center}
\end{figure}

To establish LMC membership of the group of 583 post-AGB/post-RGB/YSO candidates, we 
derived heliocentric radial velocities using the Fourier
cross-correlation technique. For cross-correlation we used only the calcium 
IR triplet (CaT) region from 8400\AA\,\, to 8700\AA\,, 
the Balmer line region from from 3700\AA\,\, to 4000\AA\,, and the
H$\alpha$ region from 6250\AA\,\, to 6450\AA\,. The
procedure is similar to that used for the SMC objects (see Paper I for
full details). For each object we estimated three radial velocities resulting from the 
cross-correlation of the three regions of the spectra with the three templates. 
The adopted heliocentric velocity was chosen to be the one with the minimum 
radial velocity error. Radial velocity errors for each cross-correlation were 
computed by \emph{FXCOR} based on the fitted peak height and the antisymmetric noise as described by 
\citet{tonry79}. A typical error in the adopted radial velocity is 11\,kms$^{-1}$. 
We note that, for some of the objects, instead of the CaT absorption lines, there are CaT lines in 
emission or the Paschen lines. For such objects only two regions (Balmer line region and H$\alpha$ region) 
were used for cross-correlation. This also applies to stars of an earlier spectral type that have 
Paschen lines that lie in the CaT region.

Figure~\ref{fig:rvplot} shows the histogram of the heliocentric
velocities for the 583 targets. A Gaussian fit to the histogram results in an average velocity of 
270\,kms$^{-1}$ and standard deviation 
of 40\,kms$^{-1}$. The estimated average velocity agrees well with the velocity
expected for stars in the LMC \citep[$\approx$\,270\,kms$^{-1}$,][]{vandermarel02, clementini03}. Based
on the expected radial velocity for stars in the LMC and also the shape of the velocity distribution 
of our sample of stars, we found that 581 out of the 583 candidates
lie within the range of 170 to 370\,kms$^{-1}$ (these limits correspond to a 2.5 sigma deviation from the
mean radial velocity of the fitted Gaussian). The 2 stars, J052020.60-693115.4, J052348.94-711201.8, that lie outside this interval have
heliocentric velocities $\approx$\,60\,kms$^{-1}$ and are likely Galactic
objects in the field of view of the LMC. After removing these 2 stars,
we are left with a sample of 581 candidates with
confirmed LMC membership. 

\section{Determination of stellar parameters}
\label{SP}

Deriving stellar parameters such as \teff, \logg\,, [Fe/H], and E($B-V$) is crucial to fully 
characterise the remaining candidates. In the following sub-sections we present 
our systematic procedure to establish a spectroscopically verified
catalogue of post-AGB and post-RGB stars in the LMC.  

\subsection{Spectroscopic analysis}
\label{STP}

We performed a 
detailed spectroscopic analysis of the 581 candidates using a 
fully automated spectral typing pipeline (STP) to simultaneously 
determine the stellar parameters (\teff, \logg, and [Fe/H]) of the candidates. A detailed 
description of our STP is mentioned in Paper I. In brief, our STP matches each
individual observed spectrum to a library of synthetic templates, and
finds the minimum root mean square (RMS) 
deviation over a carefully restricted \teff, \logg, and [Fe/H] grid. 
The synthetic templates were obtained from the Munari synthetic library 
\citep{munari05} which consists of a grid of nearly 60000 spectra, 
based on the local thermodynamical equilibrium (LTE) Kurucz-Castelli
atmosphere models \citep{castelli03}. In the STP, our RMS calculations 
follow a weighting scheme which is based on certain spectral regions that are
sensitive to specific stellar parameters. For instance, the calcium 
IR triplet [CaT] region from 8400\AA\, to 8700\AA\,, serves as a good
metallicity indicator. This helps to break the degeneracy that often plagues 
automated spectral classification algorithms. Since our
spectral resolution is low, the only good metallicity indicator is the
CaT region. Therefore, for those stars whose spectra do not 
contain CaT absorption lines, [Fe/H] is fixed to $-$0.5 which is the
mean metallicity of the LMC \citep[see][and references
therein]{vanderswaelmen13}. Note that for some stars with a low signal, 
the resulting fit from the STP was poor. In these cases, during their 
spectral analysis we fixed the [Fe/H] to provide some constraints to the STP.

 We note that the STP does not consider the H$\alpha$ region, from 6500-6650\AA,
while estimating the minimum RMS deviation owing to the diversity of the observed
H$\alpha$ line profiles that range from
emission to complex line profiles (with both emission and absorption
components). 

\begin{figure}
\begin{center}
\includegraphics[width=7cm,height=15cm,angle=0]{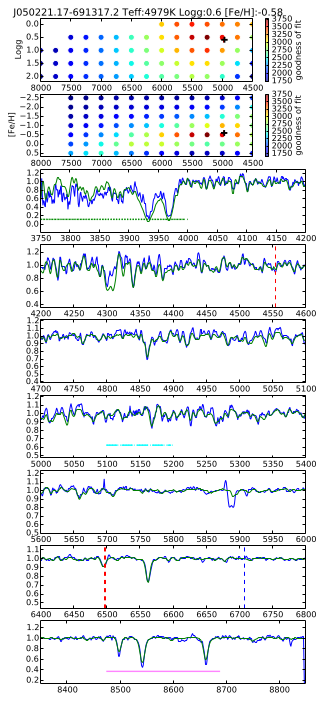}
\caption{An example of the spectral fitting result from the STP for a
  candidate in our sample. The first 
two subplots show the inverse RMS distribution (goodness of fit) in the 
\teff$-$\logg\, space and \teff$-$[Fe/H] space with a black plus representing the 
final interpolated value. In the remaining subplots, the observed spectrum is shown in blue, 
whilst the green line 
represents the best-fitting synthetic spectrum. The green-dotted, cyan-dashed and
pink-solid horizontal lines represents the 
preferentially weighted \teff, \logg\, and [Fe/H] regions (see Paper I for full details). 
The blue dashed vertical line marks the position of the LiI lines at
6707.77\AA\, and 6707.92\AA\,. The single red dashed vertical line 
marks the position of the BaII line at 4554.03\AA\, and the double red
dashed vertical lines mark 
the positions of the BaII line at 6496.89\AA\,, and the BaI line at
6498.76\AA. In this object, the Balmer lines of hydrogen are filled 
in with emission.}
\label{stpplot}
\end{center}
\end{figure}

Figure~\ref{stpplot} shows an example of the fitting result of the STP 
for a candidate in our sample. The plot presents the inverse RMS  distribution (goodness of fit) in the
\teff$-$\,\logg\, space and the \teff\,$-$\,[Fe/H] space, with the interpolated final values marked.
The preferentially weighted spectral 
regions used during the spectral typing process are plotted and important spectral 
features are also indicated. Depending on the strength of the spectral features, lines such 
as the BaII line at 4554.03\AA\,, the BaII line at 6496.89\AA\,, and the BaI line at 6498.76\AA\, 
can be detected. These are indicators of possible 
$s$-process enrichment and are marked on each plot for the identification of the sources in which we can 
identify $s$-process 
enrichment. Similarly, LiI lines at 6707.77\AA\, and 6707.92\AA\, are also marked for identification of 
Li in the candidates (although the two lines are unresolved at the resolution of 
our spectra and their position is indicated by a single blue dashed
vertical line in Figure~\ref{stpplot}.) 

For each of the 581 candidates that were fed into the STP, we were able to derive
\teff, \logg,and [Fe/H] values. Plots similar to Figure~\ref{stpplot}
for all these candidates are available as online supporting
information. 

In Paper I, we present a detailed error estimation of our STP by using
three objects that were then in our study and which were also previously examined using
high-resolution spectra. Currently, we have a sample of seven objects
(including the three objects used in Paper I). Out of these seven objects four
(J050632.10-714229.8,  J051848.84-700247.0,  J053250.69-713925.8 and
J053253.51-695915.1) are LMC objects whose chemical abundances 
have been studied in detail using UVES spectra \citep{vanaarle13},
  and three (J004441.04-732136.4,  J010056.95-715551.4 and
  J010333.89-724405.7) are SMC objects \citep[][De Smedt et al. in
  prep]{desmedt12}. Table~\ref{stperror} gives the estimated \teff,
  \logg, and [Fe/H] values using our STP and the corresponding
  values estimated using high-resolution spectra obtained from the
  literature mentioned above. 
By comparing the two sets of values, we get a mean
RMS difference for each of the stellar parameters as follows: 
$\Delta$\teff\, = 250K, $\Delta$ \logg\, = 0.7, and $\Delta$[Fe/H] = 0.2 dex.

\begin{table*}
  \caption{Determination of error estimates in the derived \teff,
    \logg, and [Fe/H] from the STP based on a
    sample of objects previously studied with high-resolution spectra
    (see text for full details).}
   \centering
  \begin{tabular}{cccccccccc}
  \hline
  Object & $T_{\rm eff,STP}$& \logg$_{\rm STP}$ & [Fe/H]$_{\rm STP}$&
  $T_{\rm eff,high-res}$&\logg$_{\rm high-res}$ & [Fe/H]$_{\rm high-res}$
  & $\Delta$\teff & $\Delta$\logg\, & $\Delta$[Fe/H]\\
  &(K) & & (dex) &(K) & & (dex) & (K) & & (dex)\\
   \hline
  J004441.04-732136.4 & 6168 & 1.0 & -1.0 & 6250 & 0.5 & -1.0 & 82 &
  0.5 & 0.0 \\
  J010056.95-715551.4 & 4295 & 0.0 & -0.9 & 4250 & 1.0 & -1.0 & 45 &
  1.0 & 0.1 \\
  J010333.89-724405.7 & 4621 & 0.0 & -0.9 & 4500 & 1.0 & -0.6 & 121 &
  1.0 & 0.3 \\
  J050632.10-714229.8 & 7614 & 0.5 & -0.4 & 6750 & 0.5 & -1.0 & 864 &
  0.0 & 0.6 \\
  J051848.84-700247.0 & 6015 & 0.0 & -1.0 & 6000 & 0.5 & -1.0 & 15 &
  0.5 & 0.0 \\
  J053250.69-713925.8 & 6073 & 1.0 & -1.1 & 5500 & 0.0 & -1.0 & 573 &
  1.0 & 0.1 \\
  J053253.51-695915.1 & 4698 & 1.5 & -0.7 & 4750 & 2.5 & -0.7 & 52 &
  1.0 & 0.0 \\
  \hline
 \end{tabular}
\label{stperror}
\end{table*}

Additionally, the low-resolution optical AAOmega spectra have a range of signal 
from 100 to 3000 counts. In Paper I, we tested the reliability of our
STP as a function of signal by taking a set of synthetic spectra varying in temperatures from 
3500K to 9500K (the \teff\, region over which we expect most of the post-AGB/RGB and YSO candidates to lie), in 
\logg\,\, from 0.5 to 1.5, with fixed [Fe/H] = -1.0. To these
synthetic spectra we added AAOmega detector
read noise and varying levels of photon noise, resulting in artificial spectra with 
quality equivalent to measured spectra of 100, 500, 1500, and 3000 counts. The artificial spectra were then passed through the 
spectral typing pipeline. These tests resulted in 
mean errors of 250K in \teff, 0.5 in \logg, and 0.5 in [Fe/H], for stars with \teff\, values $\leq$\,10000K. 
From this error estimation exercise and comparison with 
high-resolution studies, we conclude that the expected errors in our derived 
parameters are in the range: 250K to 500K in \teff, 0.5 to 1.0 in \logg, and 0.5 to 1.0 in [Fe/H].
We note that for higher temperatures (\teff\,$>$\,10000K) the grid
spacings of the synthetic templates in \teff\, increase from the 250K
step (which applies to \teff\,$<$\,10000K) to steps of 500K, 1000K, 1500K
and 2500K as \teff\, increases \citep[see][]{munari05}, and therefore accurate error estimations are
not feasible.

\subsection{Reddening estimates}
\label{reddening}

Owing to the large mid-IR excesses displayed by our candidates, the expected reddening in these objects is significant. 
The total reddening, which includes both the interstellar 
and circumstellar reddening, can be determined by estimating the difference between the 
intrinsic colour of the candidate (derived from the \teff\, estimated from the spectrum) and 
the measured colour (derived from the raw photometry based on the
$B$, $V$, $I$ and $J$ magnitudes).  Of the sample of 581 candidates, we
found that 85 objects did not have both $B$ and $I$ magnitudes and
thus there were not enough data points to estimate the value of
E($B-V$). We therefore removed this sample of 85 objects from further
analysis. These objects are presented in Appendix~\ref{nobv}. 

The remaining sample consisted of 496 candidates for which we calculated
the E($B-V)$ by estimating the value of E($B-V)$ that minimised the sum of the 
squared differences between the de-reddened observed and the intrinsic 
$B$, $V$, $I$ and $J$ magnitudes. At longer wavelengths, emission from 
dust can contribute to the observed magnitudes. We used the \citet{cardelli89} extinction law, 
assuming Rv = 3.1. It is possible that the circumstellar extinction law is different from the 
interstellar extinction law but we have not explored this possibility.
The derived E($B-V$) values were used to correct the observed magnitudes for extinction. 
Then the $BVIJ$ fluxes of the best-fit model atmosphere (derived from the STP) were normalised 
to the corrected $BVIJ$ fluxes. For a few stars 
the estimated E($B-V$) was negative which indicates that the \teff\, estimated from the spectra 
is likely to be cooler than the actual \teff\, of the star. For such
stars, we re-estimated the E($B-V$) by adopting \teff\, values
that were increased by 250K or by 500K than the estimated \teff\, values of the stars. Therefore, the
derived total reddening serves as a good check for the \teff\, value
estimated using the STP (see Subsection~\ref{STP}).

The uncertainty in the E($B-V$) estimate 
owing to other errors, such as errors in the photometry and errors in
the derived \teff\, values, are small. 
We estimated the maximum error in E($B-V$) to be the difference between 
E($B-V$) at the estimated \teff, and at \teff\, values of $\pm$\,250. We find that the 
error $\Delta$E($B-V$) $\approx$ 0.2 mag at \teff $\sim$\, 4000K and declines with increasing \teff\, to 
$\Delta$E($B-V$) $\approx$ 0.1 mag at \teff\, $\sim$\, 5000K and
$\Delta$E($B-V$) $\approx$ 0.05 mag at \teff\, $\sim$\, 6500K. 

\subsection{The luminosity of the central star}
\label{lumphot}

For post-AGB/post-RGB and YSO candidates, the central 
star is surrounded by circumstellar dust that is not necessarily spherically symmetric. For such cases, the 
observed luminosity $L_{\rm ob}$ (obtained by integrating the flux under
the observed SED) could either be over-estimated or under-estimated. For this reason 
it is essential to estimate the photospheric luminosity of the central
star $L_{\rm ph}$.  

This photospheric luminosity ($L_{\rm ph}$) can be derived from the bolometric correction for the model atmosphere 
corresponding to each individual candidate normalised to the
de-extincted $V$ magnitude, coupled with the distance modulus to the
LMC. We note that for some objects, we find that the available luminosity $L$$_{\rm ph}$ from 
the central star is too small to account for the 
luminosity $L_{\rm ob}$ derived by integrating the total SED thus resulting in  $L_{\rm ob}$ being 
$\sim$\,1.5\,$-$\,5 times $L_{\rm ph}$. This trend was also found for some of the objects in our 
previous studies \citep[see][]{wood13,kamath14}. This is likely due to
the non-spherically symmetric dust evelope. Alternatively, it could be
due to a substantial flux contribution from another object coincident on the sky with 
the optically observed star or uncertainties in the reddening estimates.

\section{Separation of the post-AGB, post-RGB and YSO candidates}
\label{pagbprgbyso}

Based on the estimated stellar parameters (\teff, \logg, and [Fe/H],
see Subsection~\ref{STP}), the E($B-V$) values (see Subsection~\ref{reddening}) and the derived 
photospheric luminosity $L_{\rm ph}$ (see Subsection~\ref{lumphot})
of the 496 remaining candidates, 
we were able to systematically characterise the sample as follows. 

On analysing their stellar parameters, we found that the sample is more diverse than
expected. Post-AGB/post-RGB stars are expected to have luminosities\,$\lesssim$\,35000\Lsun. Therefore, based on the 
estimated $L_{\rm ph}$ value, we were able to identify a sample of 112 luminous objects with 
$L_{\rm ph}$\,$\geq$\,35000\,\Lsun. These objects are likely to be supergiants or massive main-sequence stars. 
Furthermore, Post-AGB/post-RGB stars are expected to have spectral
types A, F ,G and K \citep{vanwinckel03}, the hotter objects having
already rapidly evolved to the PNe stage. Based on the derived \teff\, values, we were able to identify a 
sample of 68 objects with \teff\,$>$\,10000K. Some of these objects could
very well be hot post-AGB/post-RGB stars or YSO candidates. However,
the spectra of majority of these 68 objects do not show an emission-line spectrum
characterised by weak recombination lines of hydrogen and helium and 
various collisionally-excited forbidden lines of heavier elements
\citep[e.g.,][]{vanwinckel03}, that are characteristic of hot
post-AGB/post-RGB stars. Moreover, hot YSO candidates shows rather broad
H$\alpha$ profiles due to accretion \citep{natta02,jayawardhana02}
while most of our 68 objects do not show such H$\alpha$ emission
line profiles. Therefore, these objects are likely to be main-sequence
stars.

A detailed analysis of the 112 luminous objects with 
$L_{\rm ph}$\,$\geq$\,35000\,\Lsun  and the 68 likely main-sequence
stars with early-A, O and B spectral types will be discussed in a 
following publication. After the removal of these objects, 
the remaining sample consists of 316 objects that are likely to 
be either post-AGB/post-RGB or YSO candidates. 

Disentangling the post-AGB/post-RGB candidates from the YSO candidates is a concern since these 
unrelated objects lie in the same region of the HR diagram with
similar \teff\, and $L_{\rm ph}$ values and they have very similar IR colours and dust excesses. In our SMC survey (see
Paper I), to distinguish between the
post-AGB/post-RGB and YSO objects, we devised a criterion 
based on the \logg\, of the central star.  
At a given luminosity, the mass of a YSO is about 15\,$-$\, 20 times
that of the corresponding post-AGB/post-RGB star  (see
Figure~\ref{hr}), leading to a difference of $\sim$\,1.3 in \logg\, between post-AGB/post-RGB stars 
and YSOs.

To derive the theoretical \logg\, value a star (with a specific
photospheric luminosity, $L_{\rm
  ph}$ and intrinsic temperature, \teff) would 
have during the post-AGB phase, we first derived the mass of the
post-AGB star 
using the photospheric luminosity ($L_{\rm ph}$) and the
luminosity-core mass relation \citep{wood81} for AGB stars. During the post-RGB phase (if $L_{\rm ph}$/\Lsun\,$\leq$\,2500), a
similar procedure can be used 
but using a luminosity-core mass 
relation derived from a fit to the evolutionary tracks of \citet{bertelli08} with $Z$ = 0.004. 
For a given post-AGB/post-RGB star, the stellar mass ($M$/M$_\odot$) is essentially the core mass of the 
progenitor AGB/RGB star. 

Using this estimated mass value along with
the $L_{\rm ph}$ and \teff\, of the object, we calculated the theoretical \logg,\ value the object would have during
the post-AGB/post-RGB phase. Similarly, using the PISA pre-main sequence evolutionary tracks 
\citep{tognelli11} for $Z$ = 0.004 (and an extrapolation to higher masses since the 
maximum mass of the PISA tracks is 7\Msun), we estimated the 
theoretical value of \logg\, the same object would have if it were on the pre-main sequence
evolutionary track. The difference of the two theoretical
values resulted in a difference factor of $\sim$\,1.3 between the post-AGB/RGB stars and the pre-main
sequence stars of a given luminosity and \teff\, (see Paper I for full
details). The errors in our log g estimates for individual stars are typically in the range 
0.5\,$-$\,1.0 (see Subsection~\ref{STP}).  Therefore, we applied this \logg\,
criterion to the remaining sample of 316 objects and we were able to
identify a sample of 162 YSO candidates and 154 post-AGB/post-RGB candidates. 

We note that in determining the 
stellar parameters such as \teff, \logg, and [Fe/H], the \logg\, estimates have the highest 
uncertainty since the \logg\, value least affects the spectra when compared to the \teff\, and [Fe/H]. 
Therefore, despite the criteria used to separate the post-AGB/post-RGB candidates from the YSOs, 
there remains a degree of uncertainty in our classification method. 
Detailed studies based on high-resolution spectra are needed to confirm the 
nature of the individual objects.

We then made the distinction between post-AGB and post-RGB candidates 
using a luminosity criterion. Post-AGB stars were assumed to have $L_{\rm
  ph}$/L$_{\odot}$\, $>$ 2500 and post-RGB stars were assumed to have $L_{\rm ph}$/L$_{\odot}$\, 
$\leq$ 2500. This is based on the 
expected luminosity of the RGB-tip for stars in the LMC and SMC (see
Section~\ref{intro} and Paper I). We found a sample of 35 post-AGB candidates and 119 post-RGB candidates. 

\begin{figure}
\begin{center}
\includegraphics[width=8cm,angle=0]{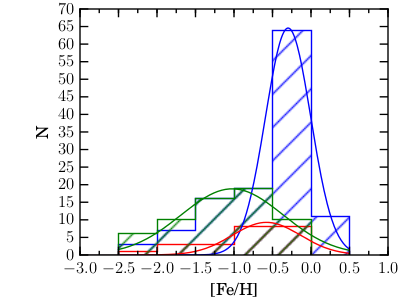}
\caption{Histograms showing the [Fe/H] distribution for the post-AGB, 
post-RGB and YSO candidates. The red 
histogram corresponds to the post-AGB objects, green to the post-RGB
objects, and blue to the YSOs. The red, green and blue curves denote individual 
Gaussian fits to the histograms for the post-AGB, post-RGB and YSO candidates, respectively.}
\label{fehplot}
\end{center}
\end{figure}

Since post-AGB/post-RGB stars are an old to intermediate age population, we expect 
them to be more metal poor than the YSOs. The young LMC 
population has a mean metallicity of 
[Fe/H] $\simeq$ -0.5 \citep[][and references
therein]{vanderswaelmen13}. Figure~\ref{fehplot} shows the [Fe/H] 
distribution for the post-AGB candidates, post-RGB candidates, and YSO candidates\footnote{Note: The post-AGB/RGB and YSO candidates 
for which we imposed a [Fe/H] value of -0.5 
(see Tables~\ref{tab:pagb_param} $-$~\ref{tab:yso_param}) have not 
been considered while plotting the [Fe/H] distribution.}. From Gaussian fits to the histograms (Figure~\ref{fehplot}) we 
find that the post-AGB sample peaks at a [Fe/H] = -0.6 with a standard deviation 
of 0.4 , the post-RGB sample peaks at [Fe/H] = -1.0 with a standard
deviation of 0.7 whilst the YSO sample, peaks at a higher metallcity
of [Fe/H] = -0.3 
with a standard deviation of 0.3. This indicates that, as expected, the post-AGB and
post-RGB evolved objects are indeed more metal poor than the YSO
population. Using the 2-sided
Kolmogorov-Smirnov (KS) test, we find that 
the post-AGB/post-RGB candidates are more metal poor than the YSOs with
high confidence.  The probability of the post-AGB and YSO samples being drawn from the 
same distribution is $P$ $\sim$\, 10$^{-2}$. The probability of the post-RGB and YSO samples to be drawn from the 
same distribution is $P$ $\sim$\, 10$^{-7}$. Finally the probability of the post-AGB and post-RGB samples to be drawn from the 
same distribution is $P$ $\sim$\,0.3, i.e., there is no significant
evidence for a different metallicity distribution for these two
groups. However, there definitely exist distinct metallicity distributions for the evolved and young objects which
strongly supports our separation of post-AGB/post-RGB from the YSO candidates.

\begin{figure*}
\begin{center}
\includegraphics[width=18cm,angle=0]{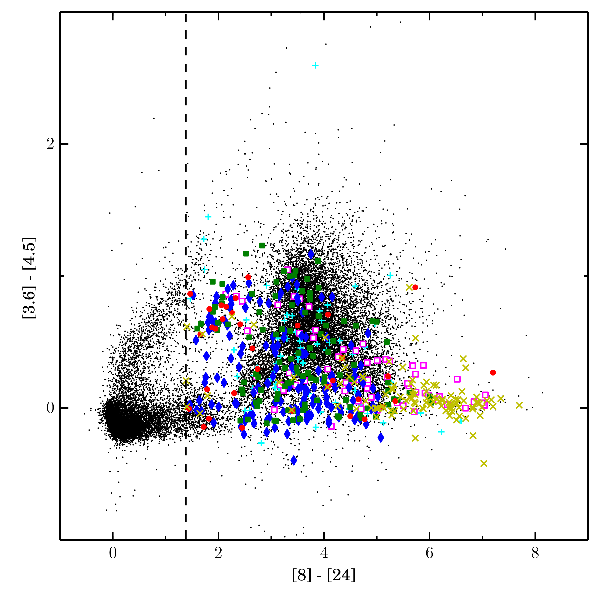} 
\caption{The spectroscopically verified post-AGB, post-RGB and YSO candidates on the 
colour-colour plot of [8]\,$-$\,[24] vs [3.6]\,$-$\,[4.5], represented by red filled-circles, green filled-squares
  and blue filled-diamonds, respectively. 
The yellow cross symbols and the magenta open-squares represent the 112 luminous objects with 
$L_{\rm ph}$\,$\geq$\,35000\,\Lsun and the 68 likely main-sequence
stars with early-A, O and B candidates, respectively. The cyan plus
symbols represent the 85 likely post-AGB, post-RGB and YSO candidates without a detection in the 
$B$ and $I$ bands. In each of the groups, objects that do not have detections at [3.6] or [4.5] microns do not
appear in this plot.}
\label{sampleselect3}
\end{center}
\end{figure*}

Figure~\ref{sampleselect3} shows the positions of the
spectroscopically verified post-AGB, post-RGB, 
and YSO candidates on 
the colour$-$colour plot of [8]\,$-$\,[24] vs [3.6]\,$-$\,[4.5] used for
our sample selection (see
Section~\ref{sampleselection}). Figure~\ref{sampleselect3} also shows 
the positions of the 112 luminous objects with 
$L_{\rm ph}$\,$\geq$\,35000\,\Lsun, the 68 likely main-sequence
stars with early-A, O and B spectral types, and the 85
likely post-AGB, post-RGB and YSO candidates 
without a detection in the $B$ and $V$ bands (see
subsection~\ref{reddening}). 

In Tables~\ref{tab:pagb_param} and~\ref{tab:prgb_param} we list the final sample of 
post-AGB and post-RGB candidates, along with their derived stellar
parameters such as \teff, \logg, [Fe/H], E[$B-V$], stellar mass ($M$/M$_\odot$), the observed
luminosity ($L_{\rm ob}$), the photospheric luminosity ($L_{\rm ph}$) and the estimated radial velocity. Similarly, in 
Table~\ref{tab:yso_param} we present the final sample of YSO
candidates. 

\onecolumn
\begin{landscape}
\begin{ThreePartTable}
\small{
\renewcommand{\arraystretch}{1.0}
\medskip
\tabcolsep=1.0pt
\LTcapwidth=\textwidth
\begin{longtable}{lrrrcrrcllllll>{\scriptsize}l}
\caption{The observational and stellar parameters for the post-AGB candidates.\label{tab:pagb_param}}\\
\hline
Name & $T_{\rm eff}$\,(K) & $\log g$ & [Fe/H] & E($B$-$V$) & ($L_{\rm ob}$/L$_\odot$) & ($L_{\rm ph}$/L$_\odot$) & $V$(mags) & RV\,(km/s) & $M$/M$_\odot$ & SED & H$\alpha$ & FWHM & [OIII] &Previous identifications \\ 
\hline
\endfirsthead
\caption{continued.}\\
\hline
Name & $T_{\rm eff}$\,(K) & $\log g$ & [Fe/H] & E($B$-$V$) & ($L_{\rm ob}$/L$_\odot$) & ($L_{\rm ph}$/L$_\odot$) & $V$(mags) & RV\,(km/s) & $M$/M$_\odot$ & SED & H$\alpha$ & FWHM & [OIII] & Previous identifications  \\ 
\hline
\endhead
\hline
\endfoot
\multicolumn{15}{c}{Candidates with [Fe/H] estimates from spectra}\\ 
\hline
J045119.94-670604.8 & 8280 & 1.0 & -0.4 & 1.11 & 790 & 7921 & 17.0 & $270 \pm 5$ & 0.63 & Shell & e & 5.0 & - & - \\ 
J045526.98-665032.0$^{\rm a}$ & 4267 & 0.5 & -0.6 & 0.02 & 4240 & 2601 & 15.5 & $296 \pm 3$ & 0.54 & Disc & a & - & - & OSARG-AGB$^{23}$ \\ 
J045623.21-692749.0 & 4500 & 0.0 & -1.0 & 0.14 & 6864 & 7131 & 14.6 & $246 \pm 2$ & 0.62 & Disc & a & - & - & LPV$^{4}$,O-AGB$^{29}$,YSO-hp$^{28}$,pagb-disc$^{25}$ \\ 
J045655.23-682732.9$^{\rm a}$ & 3730 & 0.5 & -0.5 & 0.01 & 4116 & 5161 & 15.6 & $238 \pm 2$ & 0.58 & Disc & e & 5.5 & e & Evolved$^{28}$,YSO$^{5}$,pagb-uncer$^{25}$ \\ 
J050221.17-691317.2$^{\rm a}$ & 5229 & 0.5 & -0.6 & 0.02 & 5328 & 4795 & 14.3 & $272 \pm 3$ & 0.58 & Shell & a & - & - & - \\ 
J050632.10-714229.8 & 7614 & 0.5 & -0.4 & 0.26 & 4910 & 7606 & 14.3 & $287 \pm 8$ & 0.62 & Shell & a & - & - & CEP$^{25}$,star$^{5}$,pagb-shell$^{25}$,pagb-$s$$^{26}$ \\ 
J051351.32-704656.4 & 4500 & 0.5 & -0.9 & 0.85 & 1631 & 4130 & 17.4 & $293 \pm 2$ & 0.56 & Disc & a & - & - & LPV$^{4}$,OSARG-RGB$^{23}$,YSO$^{28}$ \\ 
J051418.09-691234.9 & 6112 & 0.5 & -1.6 & 0.14 & 6667 & 4703 & 14.6 & $354 \pm 6$ & 0.57 & Disc & e & 5.5 & - & CEP$^{25}$,EB$^{1}$,Evolved$^{28}$,O-PAGB(RVT)$^{29}$,RVT$^{21,25}$,pagb-disc$^{25}$ \\ 
J051848.84-700247.0 & 6015 & 0.0 & -1.0 & 0.81 & 4477 & 14112 & 15.4 & $250 \pm 3$ & 0.73 & Shell & a & - & - & Evolved$^{28}$,LPV$^{4}$,YSO$^{5}$,SRV-AGB$^{23}$,pagb-shell$^{25}$ \\ 
J051906.86-694153.9 & 5613 & 0.0 & -1.3 & 0.37 & 2052 & 4246 & 15.5 & $270 \pm 1$ & 0.57 & Shell & e & 5.5 & e & PN$^{7}$ \\ 
J052147.95-700957.0 & 4500 & 0.0 & -0.8 & 0.84 & 6350 & 21517 & 15.2 & $262 \pm 3$ & 0.86 & Disc & a & - & - & OTHER(RCB)$^{29}$,RCB$^{22}$,RCB$^{24}$ \\ 
J052204.24-691520.7 & 6179 & 0.5 & -1.9 & 0.04 & 4071 & 3753 & 14.5 & $308 \pm 7$ & 0.56 & Disc & e,a & 2.9 & - & SRV-AGB$^{23}$,YSO$^{28}$ \\ 
J052218.52-695013.3 & 4500 & 0.5 & -0.8 & 0.84 & 5916 & 6303 & 16.9 & $262 \pm 2$ & 0.6 & Disc & a & - & - & YSO$^{28}$ \\ 
J052220.98-692001.5 & 5845 & 0.5 & -0.5 & 0.35 & 3321 & 2698 & 15.8 & $275 \pm 2$ & 0.54 & Shell & a & - & - & LPV$^{4}$,SRV-AGB$^{23}$,YSO-hp$^{28}$ \\ 
J052243.99-693828.0 & 4500 & 0.5 & -1.6 & 0.14 & 13480 & 4870 & 15.0 & $231 \pm 3$ & 0.58 & Disc & e & 5.8 & e & LPV$^{4}$,PN$^{16}$ \\ 
J052315.15-695801.7 & 6061 & 0.5 & -0.8 & 1.17 & 3810 & 7284 & 17.3 & $257 \pm 5$ & 0.62 & Disc & e & 4.3 & e & AGB/post-AGB$^{5}$,Evolved$^{28}$,LPV$^{4,4}$,SRV-AGB$^{23}$ \\ 
J052351.13-680712.2 & 4500 & 0.5 & -0.5 & 0.82 & 11065 & 7834 & 16.7 & $287 \pm 2$ & 0.63 & Disc & a & - & - & YSO$^{29}$,YSO-hp$^{28}$ \\ 
J052557.32-680450.5 & 4500 & 1.0 & -0.7 & 0.27 & 2508 & 2686 & 16.0 & $274 \pm 1$ & 0.54 & Disc & a & - & - & OSARG-RGB$^{23}$ \\ 
J052604.53-685737.6 & 4500 & 1.0 & -0.7 & 0.57 & 2035 & 2929 & 16.9 & $288 \pm 1$ & 0.54 & Disc & a & - & e & LPV$^{4}$,YSO$^{28}$ \\ 
J053116.74-690104.1$\dagger$& 5843 & 0.0 & -1.9 & 1.50 & 5369 & 14696 & 17.6 & $275 \pm 12$ & 0.74 & Disc & e,a & 4.2 & e & LPV$^{4}$,OSARG-AGB$^{23}$ \\ 
J053250.69-713925.8 & 6073 & 1.0 & -1.1 & 0.60 & 4223 & 11056 & 15.1 & $315 \pm 5$ & 0.68 & Shell & a & - & - & LPV$^{4}$,PAGB-$s$$^{26}$,YSO$^{5}$,SRV-AGB$^{23}$,pagb-shell$^{25}$ \\ 
J053530.00-712941.6 & 6012 & 0.5 & -1.7 & 0.81 & 1095 & 4207 & 16.8 & $252 \pm 4$ & 0.57 & Uncer & a & - & - & - \\ 
J054218.24-695249.4$\dagger$ & 5060 & 0.5 & -0.5 & 0.51 & 4625 & 6200 & 15.6 & $258 \pm 2$ & 0.6 & Disc & e & 5.5 & - & Evolved$^{28}$ \\ 
J055122.52-695351.4 & 6237 & 1.5 & -2.5 & 0.24 & 3780 & 3657 & 15.2 & $331 \pm 8$ & 0.56 & Disc & e & 4.9 & - & CEP$^{25}$,O-PAGB(RVT)$^{29}$,RVT$^{21,25}$,star$^{5}$,YSO$^{9}$,pagb-uncer$^{25}$ \\ 
\hline
\multicolumn{15}{c}{Candidates with [Fe/H] = -0.50}\\ 
\hline
J051044.47-691233.0 & 10544 & 2.0 & -0.5 & 0.39 & 2038 & 5340 & 15.6 & $207 \pm 7$ & 0.59 & Disc & e & 7.9 & - & EB$^{1}$,YSO$^{9}$,YSO-hp$^{28}$ \\ 
J052116.15-674511.7 & 8327 & 1.5 & -0.5 & 0.44 & 4084 & 8527 & 14.8 & $248 \pm 4$ & 0.64 & Disc & e,a & 12.0 & - & YSO$^{28}$ \\ 
J052222.95-684101.2 & 10437 & 2.0 & -0.5 & 0.89 & 1755 & 3168 & 17.7 & $288 \pm 12$ & 0.55 & Disc & e & 11.3 & - & YSO$^{5,9,10,29}$,YSO$^{29}$,YSO-hp$^{28}$ \\ 
J052241.52-675750.2 & 8284 & 1.0 & -0.5 & 0.07 & 4232 & 4566 & 14.4 & $293 \pm 15$ & 0.57 & Shell & a & - & - & - \\ 
J052519.48-705410.0 & 8117 & 1.0 & -0.5 & 0.31 & 3219 & 4943 & 15.0 & $237 \pm 3$ & 0.58 & Disc & e,a & 4.8 & - & CEP$^{25}$,O-PAGB(RVT)$^{29}$,RVT$^{21,25}$,pagb-disc$^{25}$ \\ 
J052740.75-702842.0 & 8283 & 1.0 & -0.5 & 0.04 & 4961 & 5123 & 14.1 & $263 \pm 51$ & 0.58 & Shell & e & 9.1 & - & YSO-hp$^{28}$ \\ 
J052947.62-682819.0 & 8274 & 1.0 & -0.5 & 0.10 & 16797 & 3073 & 14.9 & $191 \pm 7$ & 0.55 & Disc & e & 7.6 & e & - \\ 
J053937.58-701544.9 & 8284 & 1.0 & -0.5 & 0.55 & 3577 & 5400 & 15.7 & $301 \pm 6$ & 0.59 & Disc & e & 9.6 & - & Evolved$^{28}$ \\ 
J054013.09-691716.0 & 8281 & 1.0 & -0.5 & 0.12 & 2546 & 2754 & 15.1 & $197 \pm 28$ & 0.54 & Shell & e & 6.0 & e & Evolved$^{28}$ \\ 
J054042.24-695743.0 & 8284 & 1.0 & -0.5 & 0.16 & 4590 & 7634 & 14.1 & $225 \pm 26$ & 0.62 & Disc & e & 6.5 & - & OSARG-RGB$^{23}$,YSO-hp$^{28}$ \\ 
J054246.93-683319.7 & 8313 & 1.0 & -0.5 & 0.35 & 1780 & 3607 & 15.5 & $322 \pm 4$ & 0.56 & Uncer & e,a & 5.7 & - & - \\ 
\end{longtable}}
\begin{tablenotes}
\item[]Notes: ($L_{\rm ob}$/L$_\odot$) is the observed luminosity corrected for foreground extinction, ($L_{\rm ph}$/L$_\odot$) is the photospheric luminosity of the central star, $V$ is the optical V-band magnitude, $M$/M$_\odot$ is the derived mass of the post-AGB candidate (see Section~\ref{pagbprgbyso}). The SED column lists the SED type of the object (see Section~\ref{sedanalysis}). H$\alpha$ and OIII represents the nature of the H$\alpha$ and OIII lines at 6562.8\,\AA and 5001\,\AA, respectively. 'e', 'a', 'e,a' and 'a,e' represent emission, absorption and dominant emission with an absorption component and dominant absorption with an emission component, respectively. '-' indicates no detection. FWHM \textbf{(in \AA)} represents the full width at half maximum of the H$\alpha$ line. The column ''Previous identifications'' gives the result of a positional cross-matching that was done with the following catalogues: $^1$\citet{alcock02}, $^2$\citet{2009AJ....138.1003B}, $^3$\citet{cioni13}, $^4$\citet{fraser08}, $^5$\citet{GC09}, $^6$\citet{groen09}, $^7$\citet{hora08}, $^8$\citet{kontizas01}, $^9$\citet{kozlowski09}, $^{10}$\citet{kozlowski12}, $^{11}$\citet{levato14}, $^{12}$\citet{miszalski11}, $^{13}$\citet{matsuura14}, $^{14}$\citet{morgan03}, $^{15}$\citet{neugent12}, $^{16}$\citet{RP06}, $^{17}$\citet{RP10}, $^{18}$\citet{RP13}, $^{19}$\citet{reid14}, $^{20}$\citet{seale09}, $^{21}$\citet{soszynski08}, $^{22}$\citet{soszynski09b}, $^{23}$\citet{soszynski09}, $^{24}$\citet{tisserand09}, $^{25}$\citet{vanaarle11}, $^{26}$\citet{vanaarle13}, $^{27}$\citet{volk11}, $^{28}$\citet{whitney08}, $^{29}$\citet{woods11}, $^{30}$\citet{yang11}, $^{31}$\citet{zickgraf06}. Catalogue identifications: OSARG-AGB $=$ OGLE small amplitude red giant - AGB star, OSARG-RGB $=$ OGLE small amplitude red giant - RGB star, LPV $=$ long period variable, O-AGB $=$ oxygen-rich AGB, YSO-hp $=$ high probability YSO candidate, pagb-disc $=$ post-AGB with disc-type SED, pagb-shell $=$ post-AGB with shell-type SED, pagb-uncer $=$ post-AGB with uncertain-type SED, YSO $=$young stellar object, Evolved $=$ evolved object, CEP $=$ Cepheid variable, RVT $=$ RV-Tauri, O-PAGB $=$ oxygen-rich post-AGB, PAGB-s $=$ $s$-process enriched post-AGB star, PN $=$ planetary nebula,  SRV-AGB $=$ semi-regular variable AGB star, SRV-RGB $=$ semi-regular variable RGB star, EB $=$ eclipsing binary, RCB $=$ \textbf{R Coronae Borealis star}. See catalogued references for full details of the individual object identification.\\
The superscript 'a' in the first column indicates that the \teff\, value used to estimate E($B-V$) is increased by 250K (see Subsection~\ref{reddening} for further details).\\
The superscript $\dagger$  in the first column indicates that based on their spectra, J053116.74-690104.1 and J054218.24-695249.4 are likely symbiotic objects.\\
\end{tablenotes}
\end{ThreePartTable}
\end{landscape}
\twocolumn

\onecolumn
\begin{landscape}
\begin{ThreePartTable}
\small{
\renewcommand{\arraystretch}{1.0}
\medskip
\tabcolsep=1.0pt
\LTcapwidth=\textwidth
\begin{longtable}{lrrrcrrcllllll>{\scriptsize}l}
\caption{The observational and stellar parameters for the post-RGB candidates.\label{tab:prgb_param}}\\
\hline
Name & $T_{\rm eff}$\,(K) & $\log g$ & [Fe/H] & E($B$-$V$) & ($L_{\rm ob}$/L$_\odot$) & ($L_{\rm ph}$/L$_\odot$) & $V$(mags) & RV\,(km/s) & $M$/M$_\odot$ & SED & H$\alpha$ & FWHM & [OIII] & Previous identifications \\ 
\hline
\endfirsthead
\caption{continued.}\\
\hline
Name & $T_{\rm eff}$\,(K) & $\log g$ & [Fe/H] & E($B$-$V$) & ($L_{\rm ob}$/L$_\odot$) & ($L_{\rm ph}$/L$_\odot$) & $V$(mags) & RV\,(km/s) & $M$/M$_\odot$ & SED & H$\alpha$ & FWHM & [OIII] & Previous identifications\\ 
\hline
\endhead
\hline
\endfoot
\multicolumn{15}{c}{Candidates with [Fe/H] estimates from spectra}\\ 
\hline
J043919.30-685733.4 & 6313 & 1.5 & -2.0 & 0.32 & 106 & 159 & 18.8 & $352 \pm 13$ & 0.3 & Shell & a & - & - & - \\ 
J043948.92-693828.9 & 4928 & 0.5 & -1.6 & 0.23 & 368 & 350 & 17.9 & $212 \pm 5$ & 0.34 & Uncer & a & - & - & - \\ 
J045243.16-704737.3 & 5685 & 0.5 & -1.2 & 0.25 & 2045 & 2284 & 15.7 & $266 \pm 3$ & 0.45 & Disc & e,a & 4.8 & - & YSO$^{9,28}$,pagb-disc$^{25}$ \\ 
J045625.46-695130.7 & 4761 & 1.5 & -0.5 & 0.28 & 490 & 639 & 17.5 & $259 \pm 6$ & 0.37 & Uncer & - & - & - & - \\ 
J045739.78-694854.2 & 5819 & 1.5 & -1.2 & 0.04 & 969 & 142 & 18.1 & $232 \pm 6$ & 0.3 & Uncer & e,a & 3.0 & - & - \\ 
J045755.05-681649.2 & 4927 & 1.5 & -0.3 & 0.2 & 190 & 215 & 18.3 & $231 \pm 25$ & 0.32 & Disc & e & 4.8 & - & - \\ 
J045806.79-694855.8 & 4575 & 1.5 & -0.8 & 0.25 & 434 & 507 & 17.8 & $263 \pm 6$ & 0.36 & Uncer & - & - & - & - \\ 
J045815.58-662507.0 & 5303 & 1.0 & -1.1 & 0.06 & 293 & 236 & 17.7 & $287 \pm 20$ & 0.32 & Shell & e & 5.2 & e & - \\ 
J045840.99-693433.5 & 4500 & 1.0 & -0.5 & 0.2 & 875 & 757 & 17.2 & $243 \pm 8$ & 0.38 & Uncer & e & 4.3 & - & G$^{5}$ \\ 
J050252.50-681705.9 & 5318 & 1.0 & -1.6 & 0.5 & 241 & 374 & 18.6 & $200 \pm 16$ & 0.34 & Uncer & a & - & - & QSO$^{9}$ \\ 
J050304.95-684024.7 & 5586 & 0.5 & -2.3 & 0.01 & 3251 & 2091 & 15.1 & $332 \pm 5$ & 0.44 & Disc & e & 5.2 & - & CEP$^{25}$,EB$^{1}$,O-PAGB(RVT)$^{29}$,RVT$^{21,25}$,S$^{5}$,YSO$^{9}$,pagb-disc$^{25}$ \\ 
J050353.52-684703.8 & 5087 & 0.5 & -1.5 & 0.16 & 287 & 286 & 17.9 & $315 \pm 16$ & 0.33 & Uncer & a & - & - & - \\ 
J050554.76-664708.1 & 4500 & 0.5 & -1.3 & 0.68 & 1394 & 2390 & 17.5 & $275 \pm 7$ & 0.45 & Disc & a & - & - & OSARG-RGB$^{23}$,star$^{5}$ \\ 
J050617.59-690537.7 & 5911 & 1.5 & -1.5 & 0.1 & 367 & 368 & 17.2 & $295 \pm 8$ & 0.34 & Disc & - & - & - & QSO$^{9}$ \\ 
J050738.94-682005.9 & 5420 & 1.5 & -1.0 & 0.13 & 859 & 664 & 16.8 & $321 \pm 3$ & 0.37 & Disc & e,a & 6.4 & - & EB$^{1}$,Evolved$^{28}$,WVir$^{21}$,YSO$^{9}$ \\ 
J050817.90-684315.6 & 5653 & 0.5 & -1.4 & 0.57 & 487 & 1269 & 17.4 & $261 \pm 4$ & 0.41 & Uncer & e & 5.3 & e & - \\ 
J051046.68-690850.0 & 4750 & 1.0 & -0.9 & 0.21 & 1018 & 1094 & 16.7 & $256 \pm 2$ & 0.4 & Shell & e,a & 5.7 & - & OSARG-RGB$^{23}$ \\ 
J051113.66-700357.2 & 8296 & 1.0 & -2.5 & 0.13 & 1453 & 1658 & 15.7 & $280 \pm 4$ & 0.43 & Shell & e & 4.9 & - & - \\ 
J051230.14-682708.9 & 4750 & 1.5 & -1.0 & 0.0 & 123 & 102 & 18.6 & $273 \pm 18$ & 0.28 & Shell & e,a & 3.0 & - & - \\ 
J051240.23-722249.5 & 7220 & 1.5 & -1.9 & 0.04 & 1197 & 677 & 16.3 & $255 \pm 10$ & 0.37 & Disc & a & - & - & G$^{5}$,YSO$^{9}$ \\ 
J051259.43-691223.1 & 4401 & 1.0 & -0.7 & 0.48 & 1077 & 1430 & 17.4 & $335 \pm 2$ & 0.42 & Disc & e & 5.5 & e & EB$^{1}$,Evolved$^{28}$,LPV$^{4}$,OSARG-RGB$^{23}$ \\ 
J051347.57-704450.5 & 4500 & 1.0 & -0.5 & 0.2 & 712 & 840 & 17.1 & $234 \pm 3$ & 0.39 & Shell & a & - & - & - \\ 
J051453.10-691723.5 & 4123 & 0.5 & -0.5 & 0.05 & 2616 & 2171 & 16.0 & $276 \pm 1$ & 0.44 & Uncer & e,a & 4.9 & e & Evolved$^{28}$,O-PAGB$^{29}$, YSO$^{5}$,pagb-uncer$^{25}$ \\ 
J051652.53-702143.2 & 6795 & 1.0 & -0.8 & 0.63 & 283 & 955 & 17.7 & $306 \pm 10$ & 0.39 & Uncer & - & - & - & - \\ 
J051728.71-694246.7 & 5161 & 0.5 & -1.0 & 0.2 & 546 & 487 & 17.4 & $206 \pm 1$ & 0.36 & Shell & e & 5.0 & e & - \\ 
J051752.75-695819.1 & 5280 & 1.0 & -1.1 & 0.09 & 235 & 104 & 18.7 & $226 \pm 10$ & 0.29 & Disc & e & 5.0 & - & YSO/G$^{5}$ \\ 
J051821.10-684922.4 & 6752 & 1.0 & -2.1 & 0.04 & 1912 & 1315 & 15.6 & $296 \pm 9$ & 0.41 & Disc & e,a & 3.5 & - & YSO$^{9,28}$ \\ 
J051916.88-693757.3 & 4500 & 0.0 & -2.5 & 0.06 & 10022 & 210 & 18.3 & $191 \pm 9$ & 0.32 & Shell & e & 5.1 & e & YSO$^{5}$,YSO-P$^{20}$ \\ 
J051920.18-722522.1 & 4500 & 1.5 & -1.5 & 0.13 & 496 & 507 & 17.4 & $177 \pm 4$ & 0.36 & Shell & a & - & - & - \\ 
J051924.80-701152.8 & 4500 & 1.0 & -1.5 & 0.06 & 335 & 335 & 17.7 & $198 \pm 4$ & 0.34 & Uncer & - & - & - & - \\ 
J051942.32-690535.4 & 4500 & 0.5 & -1.0 & 0.12 & 2123 & 1718 & 16.1 & $257 \pm 2$ & 0.43 & Disc & e & 4.2 & - & YSO$^{9,28}$ \\ 
J052034.21-693822.1 & 5992 & 0.5 & -1.2 & 0.02 & 1622 & 588 & 16.4 & $263 \pm 2$ & 0.37 & Uncer & e & 5.2 & e & - \\ 
J052047.95-692948.7 & 5721 & 1.5 & -1.9 & 0.29 & 238 & 430 & 17.7 & $237 \pm 7$ & 0.35 & Shell & e & 5.1 & e & - \\ 
J052115.71-702106.2 & 6185 & 1.0 & -2.5 & 0.46 & 277 & 700 & 17.6 & $218 \pm 9$ & 0.38 & Shell & a & - & - & - \\ 
J052118.97-690103.6 & 4500 & 1.5 & -0.2 & 0.13 & 584 & 554 & 17.3 & $284 \pm 5$ & 0.36 & Disc & e & 6.0 & - & YSO$^{5}$ \\ 
J052135.62-695157.1 & 4500 & 0.5 & -0.8 & 0.32 & 1753 & 2114 & 16.5 & $271 \pm 2$ & 0.44 & Uncer & a & - & - & OSARG-AGB$^{23}$ \\ 
J052203.30-681921.4 & 5430 & 1.5 & -1.0 & 0.44 & 1786 & 1027 & 17.3 & $291 \pm 3$ & 0.4 & Disc & e & 3.2 & - & YSO$^{9,28}$ \\ 
J052411.12-692126.7 & 5387 & 1.0 & -1.7 & 0.0 & 346 & 164 & 17.9 & $224 \pm 10$ & 0.3 & Disc & e & 5.1 & e & YSO$^{9,10}$,YSO-hp$^{28}$ \\ 
J052500.40-721845.3 & 4652 & 1.5 & -2.5 & 0.16 & 163 & 197 & 18.4 & $256 \pm 11$ & 0.31 & Disc & - & - & - & - \\ 
J052546.81-722256.8$^{\rm b}$ & 5000 & 1.0 & -1.9 & 0.04 & 257 & 202 & 17.9 & $250 \pm 6$ & 0.31 & Shell & a & - & - & - \\ 
J052548.17-693700.1 & 4500 & 0.5 & -0.9 & 0.12 & 854 & 818 & 16.9 & $246 \pm 2$ & 0.38 & Uncer & e,a & 5.7 & e & - \\ 
J052819.91-681834.2$^{\rm b}$ & 5000 & 1.5 & -1.9 & 0.04 & 145 & 132 & 18.4 & $276 \pm 5$ & 0.3 & Shell & e & 3.9 & - & - \\ 
J052849.25-674125.3 & 4500 & 1.5 & -1.9 & 0.02 & 193 & 155 & 18.4 & $282 \pm 5$ & 0.3 & Uncer & e,a & 4.9 & e & - \\ 
J053030.78-701805.5 & 4500 & 2.0 & -1.4 & 0.03 & 167 & 130 & 18.6 & $205 \pm 7$ & 0.29 & Disc & e & 3.9 & - & - \\ 
J053045.83-705016.5 & 4500 & 1.0 & -0.8 & 0.18 & 952 & 1170 & 16.7 & $227 \pm 2$ & 0.41 & Shell & e & 5.5 & e & PN$^{7,16}$ \\ 
J053206.81-702101.9 & 4656 & 1.0 & -1.3 & 0.07 & 286 & 289 & 17.7 & $222 \pm 3$ & 0.33 & Uncer & - & - & - & - \\ 
J053253.54-714703.7$^{\rm b}$ & 6310 & 1.0 & -1.0 & 0.02 & 789 & 621 & 16.3 & $267 \pm 10$ & 0.37 & Disc & a & - & - & AGN$^{10}$,YSO$^{9}$ \\ 
J053514.93-680010.8 & 4500 & 0.5 & -1.1 & 0.27 & 1169 & 1440 & 16.7 & $256 \pm 2$ & 0.42 & Uncer & a & - & - & - \\ 
J053528.64-683746.2 & 4500 & 1.5 & -1.9 & 0.07 & 120 & 130 & 18.7 & $254 \pm 7$ & 0.29 & Uncer & e & 2.0 & e & - \\ 
J053547.36-700625.4 & 4054 & 0.5 & -0.5 & 0.16 & 1343 & 1639 & 16.7 & $282 \pm 1$ & 0.43 & Disc & a & - & - & OSARG-AGB$^{23}$ \\ 
J053628.41-701256.1 & 5244 & 2.0 & -0.6 & 0.09 & 288 & 347 & 17.4 & $257 \pm 3$ & 0.34 & Uncer & a & - & - & - \\ 
J053851.10-713301.2 & 5250 & 1.5 & -1.2 & 0.07 & 106 & 92 & 18.8 & $264 \pm 10$ & 0.28 & Shell & a & - & - & - \\ 
J053930.60-702248.5 & 4315 & 0.5 & -0.9 & 0.14 & 248 & 294 & 18.2 & $262 \pm 2$ & 0.33 & Shell & - & - & e & - \\ 
J054024.49-692950.4 & 5343 & 1.0 & -1.4 & 0.67 & 453 & 1080 & 18.0 & $262 \pm 11$ & 0.4 & Shell & e & 5.2 & e & - \\ 
J054054.35-701318.6 & 4822 & 1.5 & -0.5 & 0.05 & 488 & 387 & 17.3 & $280 \pm 4$ & 0.35 & Disc & e & 6.1 & - & YSO$^{5}$,QSO$^{9}$ \\ 
J054115.40-711248.9 & 5037 & 1.0 & -1.5 & 0.28 & 332 & 372 & 18.0 & $259 \pm 4$ & 0.34 & Uncer & e & 5.0 & - & - \\ 
J054144.88-712351.7 & 5262 & 2.0 & -0.7 & 0.03 & 125 & 91 & 18.6 & $252 \pm 5$ & 0.28 & Disc & e & 5.6 & - & AGN$^{10}$,QSO$^{9}$ \\ 
J054228.75-693021.2 & 5547 & 0.5 & -2.5 & 0.07 & 1123 & 1107 & 16.0 & $365 \pm 17$ & 0.4 & Disc & e & 5.5 & e & - \\ 
J054249.59-705015.2 & 4500 & 1.0 & -0.5 & 0.26 & 609 & 850 & 17.3 & $227 \pm 2$ & 0.39 & Uncer & a & - & - & - \\ 
J054306.57-713038.3 & 5843 & 1.5 & -0.9 & 0.06 & 193 & 84 & 18.7 & $296 \pm 27$ & 0.28 & Shell & a & - & - & - \\ 
J054555.68-705730.3 & 3902 & 0.0 & -1.0 & 0.14 & 1973 & 2147 & 16.4 & $239 \pm 1$ & 0.44 & Disc & a & - & - & LPV$^{4}$,OSARG-RGB$^{23}$ \\ 
\hline
\multicolumn{15}{c}{Candidates with [Fe/H] = -0.50}\\ 
\hline
J044049.73-684419.0 & 8607 & 2.5 & -0.5 & 0.72 & 129 & 511 & 18.8 & $210 \pm 21$ & 0.36 & Uncer & a & - & - & - \\ 
J044931.84-691551.6 & 7655 & 1.5 & -0.5 & 0.1 & 297 & 111 & 18.4 & $265 \pm 20$ & 0.29 & Disc & e & 5.5 & e & - \\ 
J045058.16-671634.4 & 7625 & 2.5 & -0.5 & 0.21 & 250 & 200 & 18.1 & $212 \pm 16$ & 0.31 & Disc & a & - & - & QSO$^{9}$ \\ 
J045339.50-672833.5 & 7734 & 1.0 & -0.5 & 0.16 & 588 & 715 & 16.6 & $282 \pm 9$ & 0.38 & Disc & e,a & 5.0 & - & QSO$^{9}$ \\ 
J045539.61-681202.9 & 7625 & 1.5 & -0.5 & 1.19 & 212 & 1609 & 18.9 & $293 \pm 20$ & 0.42 & Uncer & - & - & - & - \\ 
J045555.15-712112.3 & 10000 & 2.0 & -0.5 & 0.09 & 454 & 191 & 18.2 & $240 \pm 15$ & 0.31 & Disc & e & 5.3 & - & YSO$^{5}$ \\ 
J045636.68-710226.2 & 9054 & 2.0 & -0.5 & 0.7 & 134 & 524 & 18.8 & $255 \pm 23$ & 0.36 & Disc & a & - & - & QSO$^{9,10}$ \\ 
J045736.84-705127.2 & 10000 & 2.0 & -0.5 & 0.07 & 80 & 88 & 18.9 & $292 \pm 34$ & 0.28 & Disc & a & - & - & QSO$^{9,10}$ \\ 
J045745.46-683724.1 & 7626 & 3.0 & -0.5 & 0.1 & 121 & 118 & 18.4 & $276 \pm 24$ & 0.29 & Disc & a & - & - & QSO$^{9}$ \\ 
J045836.92-701120.1 & 8000 & 1.0 & -0.5 & 0.12 & 277 & 95 & 18.7 & $261 \pm 32$ & 0.28 & Disc & a & - & - & G$^{5}$,QSO$^{9,10}$ \\ 
J045923.34-703157.3 & 9111 & 1.5 & -0.5 & 0.17 & 2601 & 470 & 17.3 & $266 \pm 17$ & 0.36 & Disc & e & 5.9 & - & YSOs$^{5}$ \\ 
J050009.35-673506.7 & 8000 & 1.0 & -0.5 & 0.08 & 307 & 137 & 18.1 & $223 \pm 18$ & 0.3 & Disc & a & - & - & QSO$^{9}$ \\ 
J050117.17-675033.6 & 8000 & 1.0 & -0.5 & 0.9 & 214 & 1177 & 18.4 & $290 \pm 23$ & 0.41 & Uncer & a & - & - & - \\ 
J050149.27-664935.4 & 7000 & 1.0 & -0.5 & 0.89 & 274 & 2030 & 17.7 & $250 \pm 16$ & 0.44 & Uncer & a & - & - & G$^{5}$ \\ 
J050204.10-655017.7 & 9948 & 2.0 & -0.5 & 0.06 & 145 & 83 & 19.0 & $301 \pm 20$ & 0.28 & Disc & e & 4.6 & - & - \\ 
J050257.89-665306.3 & 4586 & 0.5 & -0.5 & 0.12 & 267 & 177 & 18.5 & $254 \pm 22$ & 0.31 & Disc & - & - & - & - \\ 
J050338.50-671754.4 & 7625 & 2.5 & -0.5 & 0.53 & 367 & 379 & 18.4 & $272 \pm 7$ & 0.34 & Disc & e & 5.5 & e & QSO$^{9}$,YSO-hp$^{28}$ \\ 
J050452.48-701201.0 & 8250 & 1.0 & -0.5 & 0.01 & 820 & 443 & 16.7 & $276 \pm 8$ & 0.35 & Uncer & e & 5.4 & e & - \\ 
J050458.28-703102.4 & 10000 & 2.0 & -0.5 & 0.45 & 310 & 1187 & 17.3 & $249 \pm 9$ & 0.41 & Uncer & e & 5.2 & e & - \\ 
J050503.65-660841.9 & 7500 & 1.0 & -0.5 & 0.77 & 174 & 809 & 18.3 & $258 \pm 12$ & 0.38 & Uncer & a & - & - & - \\ 
J050504.15-684440.8$\dagger$ & 8332 & 1.0 & -0.5 & 0.2 & 1232 & 1346 & 16.1 & $280 \pm 13$ & 0.41 & Disc & e & 10.2 & - & YSO$^{9}$ \\ 
J050515.28-713403.8 & 4864 & 1.0 & -0.5 & 0.35 & 202 & 290 & 18.6 & $196 \pm 11$ & 0.33 & Uncer & a & - & - & G$^{5}$ \\ 
J050627.11-713800.0 & 6260 & 2.0 & -0.5 & 0.58 & 208 & 353 & 18.7 & $251 \pm 11$ & 0.34 & Uncer & a & - & - & - \\ 
J050723.82-661510.3 & 5576 & 1.0 & -0.5 & 0.42 & 211 & 299 & 18.5 & $264 \pm 12$ & 0.33 & Uncer & a & - & - & - \\ 
J050954.58-684947.3 & 7658 & 1.0 & -0.5 & 0.05 & 2178 & 113 & 18.2 & $170 \pm 14$ & 0.29 & Uncer & e & 5.3 & e & YSO$^{5}$,YSO-P$^{20}$ \\ 
J051012.81-685535.9 & 7250 & 0.5 & -0.5 & 0.33 & 2501 & 354 & 17.8 & $233 \pm 10$ & 0.34 & Uncer & e & 5.3 & e & YSO$^{5}$ \\ 
J051101.61-705221.4 & 4500 & 1.0 & -0.5 & 0.11 & 729 & 547 & 17.3 & $251 \pm 2$ & 0.36 & Disc & a & - & - & Evolved$^{28}$,S$^{5}$,YSO$^{9}$ \\ 
J051228.19-690755.8 & 8289 & 1.0 & -0.5 & 0.17 & 2923 & 1164 & 16.2 & $203 \pm 22$ & 0.41 & Disc & e & 5.4 & - & C-PAGB$^{29}$,C-PPNe$^{27}$,YSO$^{5}$,YSO$^{28}$ \\ 
J051412.13-702026.0 & 7626 & 3.0 & -0.5 & 0.11 & 173 & 117 & 18.4 & $244 \pm 14$ & 0.29 & Disc & a & - & - & QSO$^{9,10}$ \\ 
J051503.27-703351.0 & 5794 & 2.0 & -0.5 & 0.13 & 3204 & 145 & 18.3 & $314 \pm 31$ & 0.3 & Disc & e & 6.5 & - & YSO$^{5}$,YSO-P$^{20}$,YSO$^{28}$ \\ 
J051627.12-692623.0 & 7699 & 1.0 & -0.5 & 0.08 & 753 & 263 & 17.4 & $306 \pm 13$ & 0.33 & Disc & e & 8.0 & e & YSO$^{9}$,YSO-hp$^{28}$ \\ 
J051751.41-713507.3 & 8124 & 2.0 & -0.5 & 0.15 & 129 & 85 & 18.9 & $238 \pm 16$ & 0.28 & Disc & a & - & - & QSO$^{9,10}$ \\ 
J052009.57-683800.5 & 7681 & 1.5 & -0.5 & 0.46 & 2123 & 2058 & 16.4 & $301 \pm 7$ & 0.44 & Disc & e,a & 12.6 & e & YSO$^{9}$,YSO-hp$^{28}$ \\ 
J052034.42-700033.1 & 7734 & 1.0 & -0.5 & 0.37 & 708 & 781 & 17.1 & $243 \pm 14$ & 0.38 & Disc & e & 7.7 & e & EB$^{1}$,YSO$^{9}$,YSO-hp$^{28}$ \\ 
J052107.90-714104.4 & 10000 & 2.5 & -0.5 & 0.08 & 126 & 114 & 18.7 & $355 \pm 42$ & 0.29 & Disc & e,a & 6.2 & - & - \\ 
J052115.92-693226.0 & 9875 & 2.0 & -0.5 & 0.03 & 684 & 607 & 16.7 & $243 \pm 12$ & 0.37 & Disc & e & 4.6 & e & - \\ 
J052128.28-681324.9 & 7675 & 1.5 & -0.5 & 0.05 & 373 & 113 & 18.2 & $255 \pm 32$ & 0.29 & Uncer & - & - & - & - \\ 
J052245.11-713610.2$\dagger$ & 8289 & 1.0 & -0.5 & 0.66 & 5539 & 1413 & 17.5 & $186 \pm 10$ & 0.42 & Disc & e & 6.8 & e & YSO$^{5}$,YSO-P$^{20}$,YSO-hp$^{28}$ \\ 
J052300.15-681120.8 & 8291 & 1.0 & -0.5 & 0.28 & 997 & 1541 & 16.2 & $266 \pm 15$ & 0.42 & Shell & e & 9.2 & e & - \\ 
J052603.31-684220.0 & 8290 & 1.0 & -0.5 & 0.02 & 1236 & 1366 & 15.5 & $319 \pm 11$ & 0.41 & Shell & e,a & 5.1 & e & EB$^{1}$ \\ 
J052834.16-715558.1$\dagger$ & 9964 & 2.0 & -0.5 & 0.31 & 2788 & 1148 & 16.9 & $335 \pm 13$ & 0.4 & Disc & e & 7.2 & - & AGB/post-AGB$^{5}$,YSO$^{9}$ \\ 
J052902.80-682321.4 & 7641 & 2.0 & -0.5 & 0.06 & 95 & 79 & 18.7 & $196 \pm 16$ & 0.27 & Uncer & - & - & - & - \\ 
J053102.33-682952.6 & 8297 & 1.0 & -0.5 & 0.06 & 528 & 528 & 16.7 & $284 \pm 17$ & 0.36 & Shell & e,a & 4.9 & e & - \\ 
J053130.65-714448.2$\dagger$ & 8294 & 1.0 & -0.5 & 0.24 & 2155 & 968 & 16.6 & $246 \pm 7$ & 0.39 & Disc & e & 8.1 & - & - \\ 
J053305.75-673844.0 & 10000 & 2.0 & -0.5 & 0.06 & 187 & 116 & 18.6 & $322 \pm 19$ & 0.29 & Uncer & e,a & 5.3 & e & - \\ 
J053431.48-683513.7 & 8322 & 1.0 & -0.5 & 0.01 & 4973 & 532 & 16.5 & $250 \pm 11$ & 0.36 & Disc & e & 6.2 & e & YSO$^{5}$,YSO-P$^{20}$,YSO$^{9}$,YSO-hp$^{28}$ \\ 
J053452.18-695649.6 & 7665 & 2.0 & -0.5 & 0.04 & 123 & 123 & 18.1 & $259 \pm 47$ & 0.29 & Shell & e & 4.7 & e,a & - \\ 
J053804.72-695918.5 & 7655 & 1.0 & -0.5 & 0.06 & 941 & 128 & 18.1 & $175 \pm 8$ & 0.29 & Disc & e & 7.7 & e & G/YSO$^{5}$,RRLab$^{1}$,YSO$^{9}$,YSO-hp$^{28}$ \\ 
J053811.63-680824.9 & 4935 & 2.0 & -0.5 & 0.02 & 99 & 81 & 18.8 & $311 \pm 15$ & 0.28 & Disc & - & - & - & - \\ 
J053907.26-693514.5 & 8292 & 1.0 & -0.5 & -0.0 & 2065 & 1182 & 15.6 & $192 \pm 9$ & 0.41 & Uncer & e & 5.2 & e & HII$^{12}$,PN(probable)$^{16}$,HII$^{19}$ \\ 
J053925.37-674653.8 & 10000 & 2.0 & -0.5 & 0.02 & 1055 & 970 & 16.2 & $330 \pm 23$ & 0.39 & Disc & - & - & - & G/YSO$^{5}$ \\ 
J054034.77-683228.2 & 8250 & 1.0 & -0.5 & 0.15 & 845 & 906 & 16.4 & $240 \pm 32$ & 0.39 & Uncer & e & 5.0 & e & - \\ 
J054038.64-702800.5 & 8129 & 1.0 & -0.5 & 0.35 & 1922 & 249 & 18.4 & $293 \pm 45$ & 0.32 & Disc & e & 6.8 & e & PN/YSO$^{5}$,YSO$^{9}$,YSO-hp$^{28}$ \\ 
J054038.99-694850.2 & 8286 & 1.0 & -0.5 & 0.06 & 1716 & 2095 & 15.2 & $319 \pm 16$ & 0.44 & Shell & e & 5.2 & e & - \\ 
J054051.04-700712.9 & 5278 & 2.5 & -0.5 & 0.1 & 78 & 83 & 19.0 & $272 \pm 10$ & 0.28 & Uncer & e & 4.2 & e & - \\ 
J054551.75-702005.0 & 6533 & 2.0 & -0.5 & 0.58 & 157 & 342 & 18.7 & $289 \pm 8$ & 0.34 & Disc & - & - & - & - \\ 
J054849.92-682918.6 & 10500 & 2.0 & -0.5 & 1.09 & 237 & 2139 & 18.8 & $270 \pm 31$ & 0.44 & Uncer & e,a & 3.8 & - & - \\ 
J055102.44-685639.1 & 7625 & 1.0 & -0.5 & 0.94 & 452 & 815 & 18.9 & $352 \pm 39$ & 0.38 & Disc & e,a & - & - & G$^{5}$,YSO-hp$^{28}$ \\ 
\end{longtable}}
\begin{tablenotes}
\item[]Notes: As for Table~\ref{tab:pagb_param}.\\
The column ''Previous identifications'' gives the result of a
positional cross-matching that was done with the catalogues mentioned
in Table~\ref{tab:pagb_param}. Catalogue identifications: YSO $=$young
stellar object, YSO-P $=$ YSO with PAH emission features, YSO-hp $=$ high probability YSO candidate,  
pagb-disc $=$ post-AGB with disc-type SED, pagb-uncer $=$ post-AGB with uncertain-type SED, G $=$ background galaxy, QSO $=$ quasi-stellar object, Evolved $=$
evolved object, PN $=$ planetary nebula, OSARG-RGB $=$ OGLE small amplitude red giant - RGB star, OSARG-AGB $=$ OGLE small amplitude red giant - AGB star, C-PPNe $=$ carbon-rich proto-planetary nebula, EB $=$ eclipsing binary, 
RRLab - RR Lyrae, C-PAGB $=$ carbon-rich post-AGB, RVT $=$ RV-Tauri, O-PAGB $=$ oxygen-rich post-AGB, LPV $=$ long period variable, AGN $=$ active galactic nuclei, WVir $=$ W Virginis star, HII $=$ HII region. See catalogued references for full details of the individual object identification.\\
The superscript 'b' in the first column indicates that the \teff\, value used to estimate E($B-V$) is increased by 500K (see Subsection~\ref{reddening} for further details).\\
$\dagger$J050504.15-684440.8 , J052245.11-713610.2, J052834.16-715558.1 and J053130.65-714448.2 show FeII emission line features similar to that of B[e] stars (see Section~\ref{preclass}).
\end{tablenotes}
\end{ThreePartTable}
\end{landscape}
\twocolumn

\onecolumn
\begin{landscape}
\begin{ThreePartTable}
\small{
\renewcommand{\arraystretch}{1.0}
\medskip
\tabcolsep=1.0pt
\LTcapwidth=\textwidth
\begin{longtable}{lrrrcrrclllll>{\scriptsize}l}
\caption{The observational and stellar parameters for the YSO candidates.\label{tab:yso_param}}\\
\hline
Name & $T_{\rm eff}$\,(K) & $\log g$ & [Fe/H] & E($B$-$V$) & ($L_{\rm ob}$/L$_\odot$) & ($L_{\rm ph}$/L$_\odot$) & $V$(mags) & RV\,(km/s) & H$\alpha$ & FWHM & [OIII] & Previous identifications \\ 
\hline
\endfirsthead
\caption{continued.}\\
\hline
Name & $T_{\rm eff}$\,(K) & $\log g$ & [Fe/H] & E($B$-$V$) & ($L_{\rm ob}$/L$_\odot$) & ($L_{\rm ph}$/L$_\odot$) & $V$(mags) & RV\,(km/s) & H$\alpha$ & FWHM & OIII & Previous identifications\\ 
\hline
\endhead
\hline
\endfoot
\multicolumn{13}{c}{Candidates with [Fe/H] estimates from spectra}\\ 
\hline
J044535.72-685840.4 & 5313 & 3.0 & -1.0 & 0.31 & 224 & 329 & 18.2 & $210 \pm 16$ & - & - & - & - \\ 
J044920.32-690900.0 & 4500 & 1.0 & -0.3 & 0.5 & 4169 & 6564 & 15.8 & $263 \pm 2$ & - & - & - & - \\ 
J044937.42-682924.9 & 5484 & 3.0 & -0.1 & 0.15 & 233 & 138 & 18.5 & $290 \pm 12$ & - & - & - & - \\ 
J045009.84-701320.9 & 4500 & 1.5 & -0.1 & 0.38 & 798 & 1116 & 17.3 & $227 \pm 3$ & a & - & - & - \\ 
J045145.31-692837.1 & 4500 & 2.0 & -0.7 & 0.61 & 1888 & 3173 & 16.9 & $277 \pm 3$ & e & 4.0 & e & LPV$^{4}$,OSARG-RGB$^{23}$ \\ 
J045216.81-713421.5 & 7486 & 4.0 & -0.7 & 0.7 & 118 & 457 & 18.8 & $278 \pm 9$ & a & - & - & - \\ 
J045253.76-664029.2 & 4416 & 1.5 & -0.4 & 0.45 & 1805 & 2579 & 16.6 & $262 \pm 4$ & a & - & - & YSO$^{28}$ \\ 
J045330.56-695150.7 & 4500 & 1.5 & -0.1 & 0.43 & 2074 & 2774 & 16.5 & $253 \pm 1$ & - & - & - & OSARG-RGB$^{23}$ \\ 
J045424.37-685240.6 & 5756 & 3.0 & -0.0 & 0.11 & 267 & 274 & 17.6 & $245 \pm 12$ & e & 5.6 & - & QSO$^{9}$ \\ 
J045445.18-680656.4 & 7007 & 3.5 & -1.4 & 0.83 & 335 & 1342 & 18.1 & $312 \pm 13$ & a & - & - & G$^{5}$ \\ 
J045500.50-675115.7 & 4500 & 2.0 & -0.6 & 0.23 & 577 & 634 & 17.5 & $243 \pm 11$ & a & - & - & - \\ 
J045552.51-664517.2 & 3944 & 1.0 & -0.5 & 0.22 & 2250 & 2504 & 16.4 & $211 \pm 3$ & a & - & - & SRV-AGB$^{23}$ \\ 
J045601.49-683958.2 & 7627 & 3.5 & -0.6 & 0.12 & 332 & 251 & 17.6 & $226 \pm 18$ & e & 5.4 & e & - \\ 
J045607.04-712422.3 & 4500 & 1.5 & -0.4 & 0.3 & 1141 & 1567 & 16.7 & $249 \pm 1$ & a & - & - & - \\ 
J045638.17-670352.3 & 4500 & 1.5 & -1.7 & 0.06 & 949 & 978 & 16.5 & $359 \pm 13$ & a & - & - & - \\ 
J045639.01-694846.2 & 4500 & 2.0 & -0.9 & 0.16 & 363 & 414 & 17.7 & $238 \pm 7$ & a & - & - & - \\ 
J045749.15-715407.5 & 6271 & 3.0 & -1.6 & 0.28 & 421 & 812 & 17.0 & $254 \pm 16$ & a & - & - & QSO$^{9}$ \\ 
J045815.38-684658.5 & 7187 & 3.5 & -1.6 & 0.6 & 167 & 421 & 18.6 & $257 \pm 13$ & e & 5.3 & e & QSO$^{9}$ \\ 
J045908.22-703254.1 & 4500 & 2.5 & -1.0 & 0.59 & 2843 & 5455 & 16.3 & $236 \pm 3$ & e & 5.9 & - & LPV$^{4}$,LPV-C$^{1}$,SRV-AGB$^{23}$ \\ 
J045913.17-705029.9 & 4500 & 1.5 & -0.5 & 0.89 & 2601 & 5316 & 17.2 & $235 \pm 2$ & e & 5.8 & - & LPV$^{4}$,LPV-C$^{1}$,SRV-AGB$^{23}$ \\ 
J050017.38-680143.4 & 7168 & 3.0 & -2.2 & 0.53 & 465 & 405 & 18.4 & $211 \pm 17$ & e,a & 3.5 & e & - \\ 
J050022.04-713802.2 & 5113 & 3.0 & -0.2 & 0.1 & 347 & 428 & 17.3 & $278 \pm 2$ & a & - & - & - \\ 
J050022.37-691413.6 & 5855 & 2.5 & -0.6 & 0.96 & 942 & 4707 & 17.2 & $241 \pm 6$ & e & 3.9 & - & - \\ 
J050055.66-720013.5 & 5019 & 3.0 & -0.5 & 0.35 & 430 & 554 & 17.8 & $259 \pm 3$ & a & - & - & QSO$^{9}$, G$^{5}$ \\ 
J050150.72-700450.0 & 5120 & 3.0 & -0.2 & 0.37 & 540 & 882 & 17.3 & $233 \pm 8$ & e & 5.4 & e & - \\ 
J050201.73-673419.2 & 6423 & 3.0 & -0.9 & 0.17 & 422 & 584 & 16.9 & $345 \pm 8$ & e & 5.8 & - & - \\ 
J050217.28-685428.5 & 6195 & 3.5 & -0.3 & 0.14 & 85 & 89 & 18.9 & $211 \pm 9$ & a & - & - & - \\ 
J050232.06-690501.5 & 8211 & 2.0 & -0.0 & 0.22 & 2654 & 4879 & 14.7 & $341 \pm 9$ & a & - & - & G$^{5}$ \\ 
J050252.54-674848.6 & 4922 & 2.0 & -0.6 & 0.3 & 427 & 560 & 17.6 & $244 \pm 4$ & a & - & - & - \\ 
J050305.55-712846.3 & 4747 & 2.5 & -1.1 & 0.1 & 166 & 114 & 18.8 & $261 \pm 19$ & a & - & - & QSO$^{9}$ \\ 
J050335.39-700347.2 & 4500 & 1.0 & -0.4 & 0.19 & 2848 & 3332 & 15.5 & $256 \pm 1$ & a & - & e & - \\ 
J050431.87-701037.0 & 5771 & 3.0 & -2.5 & 0.13 & 646 & 146 & 18.5 & $327 \pm 15$ & e & 5.3 & e & YSO$^{5}$ \\ 
J050503.87-683210.7 & 7631 & 3.5 & -1.6 & 0.79 & 166 & 550 & 18.9 & $263 \pm 10$ & e & 3.8 & - & - \\ 
J050524.47-670125.4 & 5145 & 2.5 & -0.1 & 0.39 & 519 & 799 & 17.4 & $246 \pm 5$ & a & - & - & - \\ 
J050527.24-684844.5 & 5096 & 2.5 & -0.5 & 0.3 & 481 & 949 & 17.0 & $226 \pm 4$ & a & - & - & - \\ 
J050600.08-683714.0 & 4750 & 1.0 & -0.8 & 0.55 & 3491 & 5701 & 16.0 & $260 \pm 1$ & - & - & - & OSARG-AGB$^{23}$ \\ 
J050651.48-713030.2 & 5093 & 3.5 & -1.2 & 0.06 & 96 & 78 & 19.0 & $252 \pm 16$ & a & - & - & - \\ 
J050718.33-690742.9 & 7206 & 1.5 & -1.1 & 0.05 & 10075 & 8348 & 13.6 & $329 \pm 9$ & e,a & 3.0 & - & YSO-hp$^{28}$,pagb-uncer$^{25}$ \\ 
J050732.94-682925.4$^{\rm b}$ & 7910 & 3.0 & -1.3 & 0.01 & 512 & 415 & 16.8 & $270 \pm 12$ & e,a & 3.7 & e & - \\ 
J050735.59-701546.4 & 5169 & 3.5 & -0.2 & 0.15 & 95 & 100 & 18.9 & $282 \pm 20$ & e,a & 5.0 & - & - \\ 
J050830.51-692237.4$\dagger$ & 8062 & 2.5 & -1.7 & 0.09 & 8725 & 5090 & 14.3 & $287 \pm 6$ & - & - & - & C$^{25}$,O-PAGB$^{29}$,SRV-AGB$^{23}$,YSO-hp$^{28}$,pagb-uncert$^{25}$ \\ 
J050953.41-701833.8 & 4500 & 1.5 & -0.4 & 0.34 & 628 & 866 & 17.5 & $261 \pm 1$ & a & - & - & - \\ 
J051200.28-683819.4 & 4698 & 2.0 & -0.0 & 0.55 & 1947 & 3763 & 16.4 & $264 \pm 2$ & a & - & - & OSARG-RGB$^{23}$ \\ 
J051255.84-721644.5 & 4500 & 2.5 & -1.8 & 0.1 & 182 & 188 & 18.4 & $203 \pm 6$ & - & - & - & - \\ 
J051315.46-711415.0 & 4500 & 1.5 & -0.5 & 0.32 & 976 & 1377 & 16.9 & $238 \pm 2$ & a & - & - & - \\ 
J051443.20-685034.5 & 4500 & 2.0 & -0.6 & 0.85 & 2394 & 6036 & 17.0 & $300 \pm 4$ & e & 5.7 & - & LPV$^{4}$,LPV-C$^{1}$,SRV-AGB$^{23}$,YSO$^{28}$ \\ 
J051450.04-692735.3 & 5850 & 2.5 & -0.5 & 0.29 & 1809 & 2488 & 15.8 & $260 \pm 3$ & e,a & 4.6 & - & CEP$^{25}$,Evolved$^{28}$,LPV$^{4}$,pagb-uncertain$^{25}$ \\ 
J051550.59-684105.2 & 6036 & 2.0 & -0.4 & 0.71 & 764 & 2434 & 17.1 & $299 \pm 3$ & - & - & - & - \\ 
J051624.89-690000.8 & 6837 & 2.0 & -0.7 & 0.02 & 3761 & 3590 & 14.4 & $335 \pm 7$ & a & - & - & YSO-hp$^{28}$,pagb-shell$^{25}$ \\ 
J051810.34-695714.1 & 4500 & 2.0 & -0.1 & 0.4 & 1624 & 3127 & 16.3 & $212 \pm 2$ & - & - & - & OSARG-RGB$^{23}$ \\ 
J051844.30-695357.0 & 6474 & 3.0 & -1.6 & 0.02 & 517 & 519 & 16.6 & $289 \pm 5$ & e & 4.6 & - & - \\ 
J051845.23-700534.5 & 6152 & 2.0 & -1.2 & 1.01 & 4651 & 3930 & 17.5 & $306 \pm 10$ & e & 5.9 & - & C-PAGB$^{29}$,C-PPNe$^{27}$,LPV$^{4,4}$,S$^{5}$,SRV-AGB$^{23}$,YSO-hp$^{28}$ \\ 
J051845.47-690321.8 & 5860 & 1.5 & -0.8 & 0.11 & 4001 & 3664 & 14.8 & $261 \pm 3$ & e & 5.6 & - & CEP$^{25}$,O-PAGB(RVT)$^{29}$,RVT$^{21,25}$,YSO$^{28}$,pagb-disc$^{25}$ \\ 
J051917.29-693147.3 & 4500 & 1.5 & -0.3 & 0.23 & 3800 & 3787 & 15.5 & $283 \pm 2$ & - & - & - & YSO$^{5}$,YSO-hp$^{28}$,pagb-uncer$^{25}$ \\ 
J051917.83-714359.4 & 4500 & 3.0 & -0.3 & 0.31 & 953 & 1376 & 16.9 & $273 \pm 2$ & a & - & - & - \\ 
J051922.13-683901.9 & 4500 & 2.5 & -0.2 & 0.34 & 1416 & 1921 & 16.6 & $273 \pm 2$ & e & 3.6 & - & - \\ 
J052106.33-711755.7 & 5866 & 3.5 & -1.6 & 0.3 & 122 & 167 & 18.8 & $237 \pm 16$ & e & 5.1 & e & - \\ 
J052141.52-710909.2 & 4678 & 2.5 & -0.6 & 0.22 & 1300 & 1452 & 16.4 & $228 \pm 2$ & a & - & - & YSO$^{9}$ \\ 
J052151.46-673455.8 & 4500 & 2.0 & -0.6 & 0.26 & 387 & 383 & 18.1 & $257 \pm 3$ & a & - & - & - \\ 
J052159.63-710859.1 & 4943 & 3.0 & -2.5 & 0.0 & 218 & 211 & 17.8 & $256 \pm 10$ & e,a & 3.1 & - & - \\ 
J052202.91-694127.8 & 4895 & 1.5 & -0.5 & 0.13 & 2362 & 1814 & 15.8 & $291 \pm 2$ & e & 5.4 & - & YSO$^{9}$ \\ 
J052211.94-690420.5 & 4500 & 1.5 & -0.5 & 0.29 & 972 & 1046 & 17.1 & $217 \pm 3$ & e,a & 4.3 & - & LPV$^{4}$,YSO$^{28}$ \\ 
J052300.83-700431.1 & 7645 & 3.5 & -0.3 & 0.04 & 394 & 477 & 16.7 & $328 \pm 19$ & e & 7.9 & - & Be$^{10}$,QSO$^{9}$ \\ 
J052317.90-692330.0 & 7626 & 3.5 & -0.0 & 0.71 & 314 & 1583 & 17.4 & $286 \pm 4$ & e,a & 4.3 & e & - \\ 
J052338.74-700054.2 & 5956 & 2.5 & -0.6 & 0.33 & 441 & 804 & 17.1 & $273 \pm 3$ & e & 4.7 & e & - \\ 
J052410.77-705414.5 & 4500 & 1.5 & -0.3 & 0.28 & 870 & 1101 & 17.0 & $283 \pm 5$ & a & - & - & - \\ 
J052417.75-710708.7 & 5093 & 2.0 & -0.5 & 0.34 & 444 & 681 & 17.5 & $296 \pm 3$ & a & - & - & - \\ 
J052449.29-693320.4 & 5000 & 3.0 & -0.5 & 0.29 & 372 & 542 & 17.6 & $262 \pm 8$ & e & 4.6 & - & QSO$^{9}$,red$^{10}$ \\ 
J052507.64-702219.7 & 4436 & 2.0 & -0.3 & 0.35 & 873 & 1170 & 17.2 & $243 \pm 3$ & a & - & - & OSARG-RGB$^{23}$ \\ 
J052526.01-703523.9 & 4500 & 2.0 & -0.4 & 0.31 & 857 & 1111 & 17.1 & $250 \pm 2$ & a & - & - & - \\ 
J052533.79-674240.7 & 4982 & 3.0 & -0.7 & 0.2 & 117 & 139 & 18.8 & $288 \pm 21$ & - & - & - & - \\ 
J052559.42-671932.9 & 4500 & 2.0 & -0.1 & 0.36 & 1173 & 1650 & 16.9 & $279 \pm 2$ & e & 2.5 & e & - \\ 
J052602.26-690306.2 & 4500 & 2.0 & -1.6 & 0.15 & 204 & 212 & 18.4 & $257 \pm 5$ & e & 2.4 & e & - \\ 
J052608.03-704232.9 & 4500 & 2.0 & -0.5 & 0.38 & 1815 & 2826 & 16.3 & $232 \pm 2$ & e,a & 4.7 & e & G$^{5}$,OSARG-RGB$^{23}$ \\ 
J052622.28-704131.2 & 6357 & 2.5 & -0.7 & 0.04 & 4993 & 712 & 16.3 & $281 \pm 5$ & e,a & 3.8 & e & OSARG-RGB$^{23}$ \\ 
J052709.99-683333.2 & 4500 & 2.5 & -1.6 & 0.08 & 215 & 235 & 18.1 & $285 \pm 3$ & - & - & e & - \\ 
J052715.55-701425.0 & 5116 & 3.0 & -0.2 & 0.08 & 379 & 321 & 17.5 & $259 \pm 2$ & a & - & e & - \\ 
J052725.04-671245.3 & 4888 & 3.0 & -0.3 & 0.29 & 491 & 711 & 17.3 & $269 \pm 2$ & - & - & e & - \\ 
J052728.46-702359.9 & 4500 & 1.5 & -0.4 & 0.08 & 707 & 899 & 16.6 & $248 \pm 2$ & a & - & - & G$^{5}$ \\ 
J052838.12-683101.8 & 4500 & 2.0 & -0.4 & 0.37 & 810 & 1231 & 17.2 & $257 \pm 2$ & - & - & e & OSARG-AGB$^{23}$ \\ 
J052909.56-684258.3 & 4728 & 2.0 & -1.0 & 0.3 & 739 & 993 & 17.1 & $277 \pm 2$ & - & - & e & - \\ 
J052917.10-703422.1 & 5447 & 3.0 & -1.8 & 0.43 & 221 & 374 & 18.4 & $326 \pm 21$ & e & 5.3 & e & - \\ 
J052959.77-705545.9 & 4500 & 1.5 & -0.5 & 0.23 & 589 & 725 & 17.3 & $287 \pm 2$ & e & 5.5 & e & - \\ 
J053030.10-675018.0 & 4500 & 2.5 & -0.4 & 0.25 & 649 & 802 & 17.3 & $316 \pm 2$ & - & - & - & - \\ 
J053153.96-673845.7 & 4777 & 2.5 & -0.6 & 0.3 & 2894 & 3438 & 15.7 & $286 \pm 2$ & e & 5.4 & e & - \\ 
J053249.63-704640.8 & 4500 & 1.5 & -0.3 & 0.15 & 1654 & 1446 & 16.3 & $245 \pm 1$ & a & - & - & YSO$^{9}$,red$^{10}$ \\ 
J053253.51-695915.1 & 4698 & 1.5 & -0.7 & 0.1 & 1391 & 1422 & 16.1 & $234 \pm 2$ & a & - & - & CEP$^{25}$,PAGB-s$^{26}$,YSO-hp$^{28}$,pagb-disc$^{25}$ \\ 
J053445.96-704737.0 & 5489 & 3.0 & -0.6 & 0.33 & 177 & 310 & 18.2 & $226 \pm 5$ & a & - & - & - \\ 
J053505.62-691923.2 & 5142 & 2.5 & -0.5 & 0.4 & 3298 & 5011 & 15.5 & $222 \pm 4$ & e & 5.3 & e & Evolved$^{28}$,SC$^{5}$,pagb-uncert$^{25}$ \\ 
J053546.56-702810.4 & 6216 & 2.5 & -0.5 & 0.03 & 537 & 449 & 16.8 & $276 \pm 5$ & e & 5.4 & - & G$^{5}$,QSO$^{9}$ \\ 
J053551.32-702400.4 & 5378 & 3.0 & -0.4 & 0.37 & 273 & 480 & 17.9 & $223 \pm 6$ & - & - & - & - \\ 
J053553.70-704844.7 & 4500 & 2.0 & -0.3 & 0.24 & 660 & 810 & 17.2 & $219 \pm 3$ & a & - & - & - \\ 
J053605.89-695802.6 & 6030 & 1.5 & -1.5 & 0.03 & 5569 & 4499 & 14.3 & $316 \pm 7$ & a & - & - & YSO$^{28}$,pagb-disc$^{25}$ \\ 
J053614.00-695543.8 & 4712 & 1.5 & -0.5 & 0.52 & 1694 & 2259 & 16.9 & $263 \pm 2$ & - & - & - & Evolved$^{28}$,YSO$^{5}$ \\ 
J053618.84-672909.4 & 4500 & 2.5 & -1.4 & 0.08 & 443 & 418 & 17.5 & $324 \pm 3$ & e,a & 2.6 & e & - \\ 
J053625.85-702834.7 & 7623 & 3.5 & -0.7 & 0.05 & 295 & 363 & 17.0 & $316 \pm 6$ & a & - & - & - \\ 
J053637.40-704709.2 & 4500 & 2.5 & -1.6 & 0.04 & 301 & 223 & 18.0 & $240 \pm 7$ & a & - & - & - \\ 
J053648.05-700722.4$^{\rm b}$ & 7578 & 3.5 & -1.6 & 0.01 & 144 & 63 & 18.8 & $251 \pm 13$ & e & 2.9 & - & QSO$^{9}$ \\ 
J053712.04-702131.2 & 7273 & 3.5 & -0.6 & 0.02 & 393 & 434 & 16.7 & $319 \pm 8$ & e,a & 4.4 & - & G$^{5}$,QSO$^{9}$ \\ 
J053720.22-702628.1 & 4500 & 1.5 & -0.4 & 0.33 & 1641 & 2445 & 16.3 & $262 \pm 1$ & a & - & - & OSARG-AGB$^{23}$ \\ 
J053802.32-675552.9 & 4500 & 1.5 & -1.0 & 0.04 & 783 & 827 & 16.6 & $308 \pm 2$ & - & - & - & - \\ 
J053808.67-680522.6 & 5154 & 3.0 & -0.6 & 0.12 & 631 & 591 & 16.9 & $272 \pm 3$ & - & - & - & Evolved$^{28}$,QSO$^{9}$ \\ 
J053918.05-704841.8 & 5519 & 3.0 & -1.0 & 0.12 & 181 & 100 & 18.8 & $244 \pm 4$ & e & 5.9 & - & QSO$^{9}$,red$^{10}$ \\ 
J053951.46-695703.9 & 4576 & 1.5 & -0.5 & 0.64 & 1031 & 1321 & 18.0 & $265 \pm 1$ & e & 4.8 & e & LPV$^{4}$,YSO$^{9,28}$ \\ 
J054008.64-704650.7 & 4500 & 2.0 & -1.4 & -0.0 & 364 & 295 & 17.6 & $298 \pm 6$ & - & - & - & - \\ 
J054044.70-700916.3 & 7250 & 4.0 & -1.6 & 0.04 & 257 & 101 & 18.4 & $200 \pm 10$ & e & 5.2 & - & YSO$^{28}$ \\ 
J054128.10-701501.7 & 4500 & 1.5 & -0.4 & 0.97 & 670 & 2185 & 18.5 & $277 \pm 2$ & e & 3.4 & - & YSO-hp$^{28}$ \\ 
J054131.78-705614.5 & 4760 & 2.0 & -0.1 & 0.35 & 472 & 735 & 17.6 & $285 \pm 2$ & a & - & - & LPV$^{4}$ \\ 
J054153.85-710903.7 & 4500 & 2.0 & -0.1 & 0.34 & 529 & 402 & 18.3 & $275 \pm 3$ & e & 6.1 & - & YSO$^{9,28}$,red$^{10}$ \\ 
J054247.71-695304.2 & 6076 & 2.5 & -0.6 & 0.83 & 689 & 2383 & 17.5 & $273 \pm 4$ & e & 5.1 & e & - \\ 
J054247.76-713738.2 & 5221 & 2.5 & -0.3 & 0.11 & 167 & 174 & 18.2 & $283 \pm 2$ & a & - & - & - \\ 
J054312.86-683357.1 & 5103 & 1.5 & -1.9 & 0.02 & 3085 & 1381 & 15.8 & $312 \pm 6$ & a & - & - & CEP$^{25}$,Evolved$^{28}$,RVTau$^{21,25}$,pagb-disc$^{25}$ \\ 
J054432.64-704330.7 & 4965 & 2.5 & -0.0 & 0.41 & 777 & 1293 & 17.0 & $228 \pm 2$ & a & - & - & - \\ 
J054440.60-713520.5 & 4500 & 1.5 & -0.2 & 0.69 & 2430 & 5138 & 16.6 & $252 \pm 2$ & - & - & - & LPV$^{4,4}$,OSARG-RGB$^{23}$ \\ 
J054442.05-712024.1 & 5055 & 2.0 & -0.6 & 0.06 & 363 & 412 & 17.2 & $257 \pm 2$ & a & - & - & - \\ 
J054535.99-692859.9 & 4669 & 3.0 & -0.4 & 0.48 & 277 & 397 & 18.6 & $309 \pm 12$ & e & 5.3 & e & - \\ 
J054623.31-704325.8 & 5287 & 2.5 & -0.3 & 0.3 & 401 & 529 & 17.6 & $256 \pm 5$ & a & - & - & - \\ 
J054718.55-705859.7 & 4500 & 1.0 & -0.8 & 0.67 & 1857 & 3515 & 17.0 & $264 \pm 1$ & a & - & - & Evolved$^{28}$,LPV$^{4}$,OSARG-RGB$^{23}$ \\ 
J054901.96-690322.9 & 4500 & 2.0 & -0.2 & 0.53 & 1955 & 3159 & 16.7 & $286 \pm 2$ & a,e & - & - & OSARG-RGB$^{23}$ \\ 
J055219.64-693454.1 & 4973 & 2.5 & -0.9 & 0.47 & 490 & 922 & 17.6 & $329 \pm 6$ & a & - & - & - \\ 
\hline
\multicolumn{13}{c}{Candidates with [Fe/H] = -0.50}\\ 
\hline
J045329.79-693211.9 & 6965 & 3.5 & -0.5 & 0.04 & 303 & 78 & 18.7 & $193 \pm 23$ & e & 6.6 & e & YSO$^{28}$ \\ 
J045330.21-690316.6 & 7628 & 3.0 & -0.5 & 0.13 & 518 & 472 & 17.0 & $235 \pm 21$ & e & 7.6 & - & G/YSO$^{5}$,YSO$^{9,28}$ \\ 
J045442.81-654637.9 & 7626 & 3.0 & -0.5 & 0.31 & 284 & 188 & 18.5 & $310 \pm 10$ & a & - & - & YSO$^{9}$,YSO-hp$^{28}$ \\ 
J045625.05-673708.6 & 9176 & 3.5 & -0.5 & 0.01 & 218 & 214 & 17.6 & $256 \pm 19$ & a & - & - & - \\ 
J045746.14-670206.2 & 7626 & 3.5 & -0.5 & 0.77 & 152 & 599 & 18.7 & $273 \pm 16$ & a & - & - & - \\ 
J045750.97-654135.0 & 8916 & 3.0 & -0.5 & 0.42 & 183 & 282 & 18.6 & $267 \pm 17$ & a & - & - & QSO$^{9}$ \\ 
J045804.32-715634.7 & 4998 & 2.5 & -0.5 & 0.55 & 338 & 639 & 18.2 & $214 \pm 18$ & a & - & - & - \\ 
J045835.46-675614.1 & 7420 & 3.0 & -0.5 & 0.33 & 720 & 1272 & 16.5 & $290 \pm 7$ & e,a & 2.9 & - & YSO$^{9,28}$ \\ 
J050350.12-655953.9 & 7625 & 3.0 & -0.5 & 1.19 & 1745 & 15066 & 16.5 & $255 \pm 10$ & a & - & - & - \\ 
J050403.92-680048.6 & 7064 & 3.0 & -0.5 & 0.37 & 783 & 765 & 17.2 & $265 \pm 20$ & e & 8.6 & e & YSO$^{5,9}$,YSO-hp$^{28}$ \\ 
J050422.62-684555.3 & 7625 & 3.0 & -0.5 & 1.25 & 2150 & 18467 & 16.5 & $249 \pm 2$ & a & - & - & OSARG-RGB$^{23}$,SG$^{5}$ \\ 
J050538.12-693002.2 & 6784 & 2.5 & -0.5 & 0.5 & 315 & 463 & 18.1 & $270 \pm 13$ & a & - & - & - \\ 
J050835.91-711730.7 & 7625 & 2.5 & -0.5 & 1.33 & 4330 & 4010 & 18.4 & $231 \pm 24$ & e & 6.0 & - & AGB/post-AGB$^{5}$,C-PPNe$^{27}$,LPV$^{4}$,YSO$^{28}$ \\ 
J050919.91-682842.1 & 7643 & 3.0 & -0.5 & 0.2 & 1856 & 776 & 16.6 & $288 \pm 29$ & e & 5.7 & - & YSO$^{5}$ \\ 
J051031.10-713013.0 & 6059 & 3.0 & -0.5 & 0.88 & 203 & 784 & 18.9 & $209 \pm 23$ & a & - & - & - \\ 
J051113.48-681644.6 & 7628 & 2.5 & -0.5 & 0.89 & 2997 & 13134 & 15.7 & $234 \pm 6$ & e & 6.9 & - & LPV$^{4}$ \\ 
J051543.32-683340.5 & 7627 & 3.0 & -0.5 & 0.39 & 79 & 162 & 18.9 & $228 \pm 14$ & a & - & - & AGN$^{10}$ \\ 
J051726.80-692132.0 & 9948 & 2.0 & -0.5 & 0.06 & 8314 & 13365 & 13.5 & $310 \pm 21$ & a,e & - & e & YSO$^{5}$ \\ 
J051812.13-715655.2 & 7630 & 3.5 & -0.5 & 0.08 & 213 & 156 & 18.0 & $230 \pm 14$ & e & 5.2 & - & QSO$^{9}$ \\ 
J051829.85-711023.3 & 5752 & 3.0 & -0.5 & 0.09 & 131 & 82 & 18.9 & $294 \pm 31$ & a & - & - & - \\ 
J051831.46-685926.3 & 11000 & 2.5 & -0.5 & 0.02 & 5179 & 6286 & 14.3 & $315 \pm 8$ & e & 7.6 & e & YSO$^{9,28}$ \\ 
J052000.01-693617.7 & 7623 & 2.5 & -0.5 & 0.81 & 5048 & 10289 & 15.7 & $286 \pm 4$ & e,a & 4.1 & e & Evolved$^{28}$, YSO$^{5}$ \\ 
J052429.44-693723.7 & 8876 & 2.0 & -0.5 & 0.06 & 7132 & 9647 & 13.6 & $333 \pm 9$ & a & - & - & pagb-shell$^{25}$ \\ 
J052520.76-705007.5 & 6742 & 2.5 & -0.5 & 0.52 & 3349 & 6558 & 15.3 & $277 \pm 23$ & a & - & - & C-PPNe$^{27}$,LPV$^{4}$,OSARG-RGB$^{23}$, PN/YSO$^{5}$,pagb-shell$^{25}$ \\ 
J052549.58-715627.7 & 7640 & 3.5 & -0.5 & 0.26 & 389 & 247 & 18.1 & $211 \pm 11$ & e & 8.8 & - & - \\ 
J052700.36-711854.7 & 7626 & 3.0 & -0.5 & 0.55 & 419 & 615 & 18.0 & $270 \pm 27$ & e & 8.1 & - & G/YSO$^{5}$,QSO$^{9}$,YSO-hp$^{28}$ \\ 
J052705.32-713800.2 & 7372 & 3.5 & -0.5 & 0.14 & 448 & 203 & 17.9 & $282 \pm 30$ & e,a & 5.1 & - & YSO/G$^{5}$,YSO$^{28}$ \\ 
J052731.18-684352.0 & 7158 & 3.5 & -0.5 & 0.06 & 210 & 196 & 17.7 & $299 \pm 22$ & e,a & 13.1 & e & - \\ 
J052924.07-702724.7 & 5111 & 3.0 & -0.5 & 0.04 & 92 & 95 & 18.7 & $244 \pm 28$ & e & 6.7 & e & - \\ 
J053020.07-714933.9 & 7625 & 2.5 & -0.5 & 0.97 & 269 & 1824 & 18.1 & $302 \pm 38$ & a & - & - & - \\ 
J053050.58-681210.3 & 7642 & 3.0 & -0.5 & 0.05 & 621 & 318 & 17.1 & $306 \pm 13$ & e & 2.7 & - & YSO$^{5}$ \\ 
J053244.62-715347.7 & 9576 & 3.0 & -0.5 & 0.72 & 148 & 816 & 18.4 & $234 \pm 22$ & - & - & - & - \\ 
J053337.09-711046.1 & 4578 & 1.5 & -0.5 & 0.24 & 764 & 738 & 17.3 & $274 \pm 2$ & a & - & - & YSO/G$^{5}$,YSO$^{9}$,red$^{10}$ \\ 
J053501.15-695717.6 & 5113 & 2.5 & -0.5 & 0.16 & 271 & 339 & 17.7 & $249 \pm 4$ & e & 5.5 & e & - \\ 
J053944.35-710350.6 & 7650 & 3.5 & -0.5 & 0.04 & 496 & 370 & 17.0 & $266 \pm 12$ & e & 9.5 & - & YSO$^{9,28}$ \\ 
J054000.07-713941.2 & 7624 & 3.0 & -0.5 & 1.15 & 1294 & 9754 & 16.9 & $243 \pm 1$ & e,a & 3.7 & - & LPV$^{4}$,OSARG-RGB$^{23}$ \\ 
J054033.54-703240.8 & 7634 & 3.0 & -0.5 & 0.33 & 2064 & 169 & 18.7 & $220 \pm 21$ & e & 7.4 & - & YSO$^{5}$,PN$^{16,28}$,YSO$^{29}$ \\ 
J054037.01-695951.2 & 7045 & 3.0 & -0.5 & 0.06 & 514 & 351 & 17.1 & $340 \pm 22$ & e & 5.5 & e & YSO$^{28}$ \\ 
J054503.85-682915.2 & 7626 & 3.5 & -0.5 & 1.04 & 531 & 1231 & 18.8 & $183 \pm 30$ & - & - & - & G$^{5}$,YSO$^{28}$ \\ 
J054540.61-693453.3 & 7626 & 3.0 & -0.5 & 1.34 & 1013 & 7855 & 17.7 & $271 \pm 3$ & e & 5.1 & e & YSO$^{9,28}$ \\ 
J054530.64-685610.2 & 7626 & 3.0 & -0.5 & 0.46 & 442 & 233 & 18.8 & $197 \pm 33$ & e & 4.6 & e & - \\ 
J055133.18-691633.7 & 7628 & 3.0 & -0.5 & 0.43 & 192 & 286 & 18.4 & $347 \pm 32$ & a,e & 3.5 & - & G$^{5}$,QSO$^{9}$ \\ 
\end{longtable}}
\begin{tablenotes}
\item[]Notes: As for Table~\ref{tab:pagb_param}.\\
The column ''Previous identifications'' gives the result of a
positional cross-matching that was done with the catalogues mentioned
in Table~\ref{tab:pagb_param}. Catalogue identifications: OSARG-AGB $=$ OGLE small amplitude red giant - AGB star, OSARG-RGB $=$ OGLE small amplitude red giant - RGB star, LPV $=$ long period variable, YSO $=$young
stellar object, YSO-hp $=$ high probability YSO candidate, 
pagb-disc $=$ post-AGB with disc-type SED, pagb-shell $=$ post-AGB with shell-type SED, pagb-uncer $=$ post-AGB with uncertain-type SED,  G $=$
background galaxy, Evolved $=$ evolved object, CEP $=$ Cepheid variable, RVT $=$ RV-Tauri, C-PPNe $=$ carbon-rich proto-planetary nebula, Be $=$ Be star,  O-PAGB $=$ oxygen-rich post-AGB, C-PAGB $=$ carbon-rich post-AGB, PN $=$
planetary nebula,  SRV-AGB $=$ semi-regular variable AGB star, SRV-RGB $=$ semi-regular variable RGB star, QSO $=$ quasi-stellar
object, red $=$ stars with red spectra, AGN $=$ active galactic nuclei, C $=$ C-star. See catalogued references for full details of the individual object identification.\\
The superscript 'b' in the first column indicates that the \teff\, value used to estimate E($B-V$) is increased by 250K (see Subsection~\ref{reddening} for further details).\\
$\dagger$Based on the spectra of J050830.51-692237.4, it is likely a YSO candidate with a C-star in the line-of-sight (see Section~\ref{specanalysis}).
\end{tablenotes}
\end{ThreePartTable}
\end{landscape}
\twocolumn

For each of the spectroscopically verified post-AGB, post-RGB and YSO candidates, we performed a
positional cross-matching with the most relevant previous studies. The
results of the positional cross-matching are listed in the last column
of Tables~\ref{tab:pagb_param},~\ref{tab:prgb_param} and
~\ref{tab:yso_param} for the post-AGB, post-RGB and YSO candidates,
respectively. 
In Table~\ref{tab:pagb_param} (see footnote) we list all
the previous studies used for the positional cross-matching. We find
that, in many cases, the
positional cross-matching gives a similar evolutionary nature for the
objects. However, in some cases, mostly when the positional cross-matching
is performed with studies that use a purely photometric classification
\citep[e.g.,][]{whitney08}, we find that the evolutionary nature of the object
was misclassified. This goes to show that spectroscopic
characterisation of objects is crucial to confirm their evolutionary
nature. The positional cross-matching also helps to test our 
\logg\, criterion for the separation of post-AGB/post-RGB and YSO
objects. We find that for majority of the cases, the nature of the
objects the nature of the
object from our spectroscopic analysis matched
well with other studies that are based either on high-resolution
optical spectroscopic studies \citep[e.g.,][]{vanaarle13} or studies
using SST spectra \citep[e.g.,][]{woods11}. 

However, in some cases there exists an inconsistency in the
  nature of the object when compared to studies using SST spectra. 
For  instance, \citet{woods11} have classified J051228.19-690755.8 as
a  C-rich PAGB star. \citet{volk11} classify this object as a C-rich
proto-planetary nebula. Furthermore, they estimate the luminosity 
of the object to be 4200\,$\pm$\,550\,\Lsun. In our study, 
J051228.19-690755.8 has been classified as a post-RGB star 
based on its \logg\, value and its photospheric luminosity 
($L_{\rm ph}$\,=\,1164\,\Lsun) (see Table~\ref{tab:prgb_param}). The
observed luminosity of this object is $L_{\rm  ob}$\,=\,2932\,\Lsun. This 
object could be a post-AGB star if its estimated reddening value is
too low. A good high-resolution spectrum is required to obtain a more
accurate estimate of the reddening.

The relevance of our
\logg\, criterion can be illustrated by considering one of the objects in our
sample: J053253.51-695915.1. This object is one of the four objects considered in a detailed chemical
abundance study of candidate $s$-process enriched post-AGB objects, using
high-resolution UVES spectra \citep{vanaarle13}. \citet{vanaarle13} 
find that though J053253.51-695915.1 shows a mild
$s$-process enrichment, there is an anti-correlation between the
strength of the neutron irradiation and the efficiency of the third dredge-up. However, observationally,
post-AGB stars have always shown a strong correlation between the two
parameters \citep{vanwinckel00,reyniers04}. \citet{vanaarle13}
attribute this anti-correlation to the possible extrinsic nature of
J053253.51-695915.1 i.e. it obtained its $s$-process enhancement by
mass transfer from an AGB star in a binary system. However, in our study, based on our
\logg\, criterion, we classify this object as a
YSO candidate (see Table~\ref{tab:yso_param}), indicating that the mild $s$-process enhancement comes
from the initial composition of the LMC
\citep{vanderswaelmen13}. 

The detailed spectroscopic analysis in our study is performed
using low-resolution spectra (R\,$\approx$\,1300). Since the
classification of post-AGBs/post-RGBs from the YSOs is based on the
estimated \logg\, value. it is possible that some of the objects have
been misclassified, especially when the signal-to-noise ratio of the
spectra are low (either due to faint optical fluxes or
poor observing conditions). For instance, \citet{volk11} classify
J051845.23-700534.5 ($V$\,=\,17.5 mags), J050835.91-711730.7 ($V$\,=\,18.4 mags) and J052520.76-705007.5 ($V$\,=\,15.3 mags) as C-rich proto-planetary
nebulae. J051845.23-700534.5 and  J050835.91-711730.7 show the
21$\mu$m feature, that is especially prominent in the spectra of some
carbon-rich proto-planetary nebulae and also the
30$\mu$m feature which which is commonly observed in the spectra of
carbon stars and generally attributed to MgS \citep[see][and
references therein]{volk11}. J052520.76-705007.5 shows the 21$\mu$m
feature and only a weak trace of the 30$\mu$m feature. In our study,
we classify all the three objects as YSOs (see
Table~\ref{tab:yso_param}) based on their spectroscopically derived
\logg\, values. On visually inspecting the low-resolution spectra of
these 3 objects, we find that the spectra of J051845.23-700534.5 and
J050835.91-711730.7 are rather noisy and of low signal (which is
likely due to their faint optical magnitudes).  For
J052520.76-705007.57, the spectroscopically determined \teff\, for
this object is $\approx$ 7000K. The spectrum of this star does not
contain CaT absorption lines but instead contains the Paschen
lines. For such objects without the CaT absorption lines and with
\teff\, values in the range of 6000K\,$-$\,8000K, the Balmer lines in
the wavelength region of 3750\AA\,-\,3950\AA\, are used to determine the
\logg\, (see Paper I for full details). For J052520.76-705007.5, this region of the spectrum is rather
noisy. Therefore it is likely that the spectroscopically estimated
\logg\, values for these three objects are uncertain beyond the
estimated error of 0.5\,$-$\,1.0 resulting in a possible misclassification of their nature. Note that the spectral fitting for these three objects (as well as all the other objects in our study) can be found in the online supporting information

The importance of a detailed spectroscopic analysis such as that
presented in this study can be further illustrated by comparing the results of
our spectral analysis to the sample of 70 post-AGB candidates from 
\citet{vanaarle11} that were identified on the basis of a visual 
spectral-type classification only. Of the 2102 objects in our study for
which we obtained optical spectra, 37 were in common with the 70
objects classified as post-AGB
candidates by \citet{vanaarle11}. Table~\ref{elskamath} lists these 37
objects along with the nature of the object as classified in our study. We find that, though some of the
object classifications from the two studies match, majority of the 37 objects are not
necessarily post-AGB candidates, as classified by \citet{vanaarle11}. They cover a wide range of object
classes such as post-RGBs, YSOs, luminous supergiants and likely
main-sequence stars. Therefore, our systematic spectral analysis provides
a more detailed classification of the nature of the objects. 

\begin{table}
  \caption{The nature of the 37 objects that were common to our study
    and that of \citet{vanaarle11}. In \citet{vanaarle11}, these 37 objects were classified as post-AGB stars
    based on a visual spectral-type classification.}
  \label{elskamath}
  \begin{tabular}{lll}
  \hline
   Object & Nature of the object based on our study \\
  \hline
J045243.16-704737.3 & post-RGB \\
J045444.13-701916.1 & Reject \\
J045543.20-675110.1 & C star \\
J045615.77-682042.3 & M star \\
J045623.21-692749.0 & post-AGB \\
J045655.23-682732.9 & post-AGB \\
J050304.95-684024.7 & post-RGB \\
J050431.84-691741.4 & Main-sequence star \\
J050632.10-714229.8 & post-AGB \\
J050718.33-690742.9 & YSO \\
J050733.83-692119.9 & C star \\
J050830.51-692237.4 & YSO \\
J051418.09-691234.9 & post-AGB \\
J051450.04-692735.3 & YSO \\
J051453.10-691723.5 & post-RGB \\
J051624.89-690000.8 & YSO \\
J051845.47-690321.8 & YSO \\
J051848.84-700247.0 & post-AGB \\
J051917.29-693147.3 & YSO \\
J052340.49-680528.2 & Luminous object \\
J052429.44-693723.7 & YSO \\
J052519.48-705410.0 & post-AGB \\
J052520.76-705007.5 & YSO \\
J052722.11-694710.1 & Luminous object \\
J052836.59-685829.9 & M star \\
J053011.67-710559.7 & Luminous object \\
J053250.69-713925.8 & post-AGB \\
J053253.51-695915.1 & YSO \\
J053416.45-695740.4 & M star \\
J053444.17-673750.1 & QSO \\
J053505.62-691923.2 & YSO \\
J053605.89-695802.6 & YSO \\
J054312.86-683357.1 & YSO \\
J055122.52-695351.4 & post-AGB \\
  \hline
 \end{tabular}
\end{table}

\subsection{The newly discovered post-RGB objects}
\label{prgb}

The spectroscopic analysis along with the determination of the
photospheric luminosities ($L_{\rm ph}$) have revealed a new population of evolved,
dusty ''post-RGB'' objects with $L_{\rm
  ph}$\,$\lesssim$\,2500\Lsun. This class of objects was discovered in
our SMC survey (see Paper I) where 41 new
dusty post-RGB
objects were presented. The present study has confirmed the existence
of 
dusty post-RGB stars, by revealing 119 such candidates in the LMC. So far, in the
Galaxy, these objects have not been identified, owing to the unknown distances and
hence unknown luminosities. However, it is likely that some of
the suppossed Galactic post-AGB objects are indeed post-RGB objects. 

The post-RGB objects have stellar
parameters (\teff, \logg, [Fe/H]), IR colours and SED types similar to those of post-AGB stars. However, their
luminosities (and hence masses and radii) are much lower than that expected for post-AGB
stars. Furthermore, we expect that they should be in binary
systems, since single-star mass loss that 
occurs during the RGB phase is insufficient to remove the 
H-rich envelope and produce a dusty post-RGB star
\citep[e.g. ][]{vw93}. Therefore, the only way large amounts of mass
loss followed by evolution off the RGB can
occur is via binary interaction \citep[e.g.,][]{han95a,heber09}. 

To confirm the evolutionary nature of these objects and establish 
evolutionary links to their possible precursors and progeny, in the
near-future we intend to carry out a radial velocity
monitoring program to verify their likely binary nature. Furthermore, we
also intend to pursue a chemical abundance study to
investigate their chemical patters. 

\section{Classification of spectral energy distributions}
\label{sedanalysis}

\begin{figure*}
\begin{center}
\includegraphics[width=18cm]{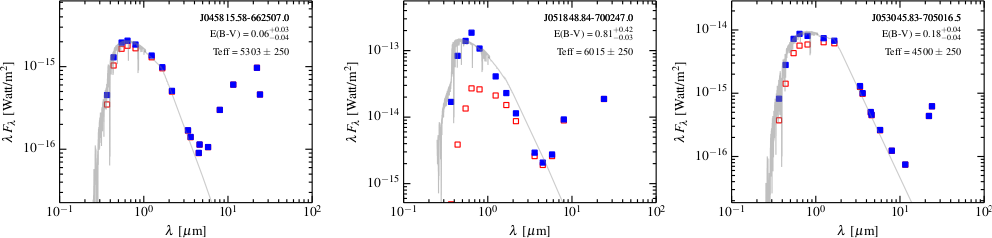}
\end{center}
\caption{Example SEDs of the post-AGB/post-RGB candidates classified
  as shell sources. The red open squares represent the original broadband 
photometry. The blue filled squares represent the dereddened broadband 
photometry. Up to a wavelength of 10500\AA, we over-plot (grey solid-line) the 
flux-calibrated Munari synthetic spectrum which is estimated to have
the best-fit to the observed spectra (see Section~\ref{STP}). From
10500\AA\, onwards we over-plot the low-resolution flux distribution
from the corresponding appropriate ATLAS9 atmospheric model
\citep{castelli04}. The SED plots also show the name of the individual object, the estimated E(B-V) 
value with error bars (see Section~\ref{reddening}) and the estimated
\teff\, value.}
\label{pagbsed_eg1}
\end{figure*}

\begin{figure*}
\begin{center}
\includegraphics[width=18cm]{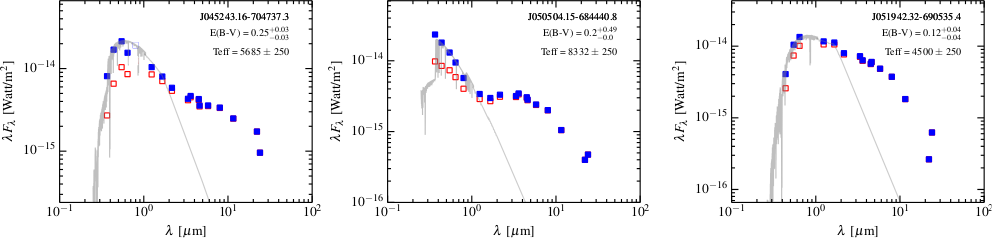}
\end{center}
\caption{Example SEDs of the post-AGB/post-RGB candidates classified
  as disc sources. The symbols used in the plots are same as that
  mentioned in Figure~\ref{pagbsed_eg1}. }
\label{pagbsed_eg2}
\end{figure*}

\begin{figure*}
\begin{center}
\includegraphics[width=18cm]{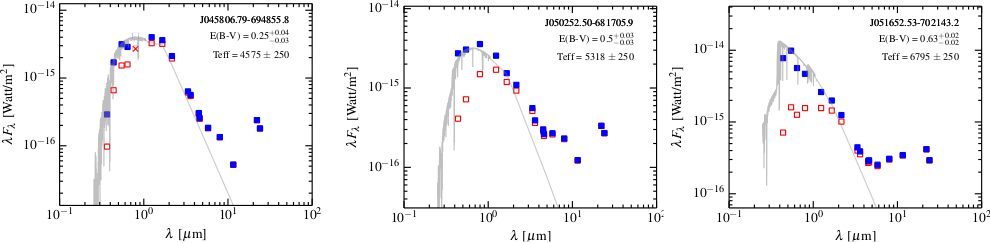}
\end{center}
\caption{Example SEDs of the post-AGB/post-RGB candidates that are
  classified as 'uncertain'. The symbols used in the plots are same as that
  mentioned in Figure~\ref{pagbsed_eg1}. }
\label{pagbsed_eg3}
\end{figure*}

Post-AGB/post-RGB and YSO candidates consist of two components: 
the central star and the circumstellar environment. SED analysis of the Galactic
post-AGB stars revealed two highly 
distinct populations: the shell-sources with double peaked SEDs \citep[probably arising from
single stars,][]{vanwinckel03}, 
and disc-sources with a near-IR excess indicative of hot dust and
circumstellar discs probably arising from 
binary stars \citep[][]{deruyter06,vanwinckel07,gielen09,vanwinckel09,dermine13}. The
post-AGB/post-RGB stars in the SMC also showed these two
populations. We performed a preliminary SED analyses to classify the
LMC post-AGB and post-RGB candidates into shell- and disc-souces, based on the visual
inspection of the position
of the peak of their dust excess. In shell-sources, the dust 
emission peaks at wavelengths greater than 10$\mu$m \citep[as shown by
radiative transfer models of the 
well known expanding shell source HD161796, where the peak of the dust
SED is at around 30$\mu$m, see][for details]{min13}. On the other
hand, for disc-sources the peak of the dust SED lies around 10\,$\mu$m and in some cases even bluer 
\citep{deruyter06,gielen11}. Figure~\ref{pagbsed_eg1} and
Figure~\ref{pagbsed_eg2} show a few examples of 
post-AGB/post-RGB candidates that 
we classified as shell-sources and disc-sources, respectively. We note that for a
group of objects, we were unable to classify them based on their SEDs
and we call these objects
'uncertain-sources'. Figure~\ref{pagbsed_eg3} shows a few 
examples of SEDs of post-AGB/post-RGB candidates that we
classify as uncertain.  Of the 35 post-AGB objects, we find
a group of 10 shell-sources, 23 disc-sources and 2 are of uncertain
SED type. Similarly, of the 119 
post-RGB objects, we find a group of 23 shell-sources, 56 disc-sources
and 40 uncertains. 

\begin{figure*}
\begin{center}
\begin{minipage}{240pt}
\resizebox{\hsize}{!}{\includegraphics[clip=true,width=9cm]{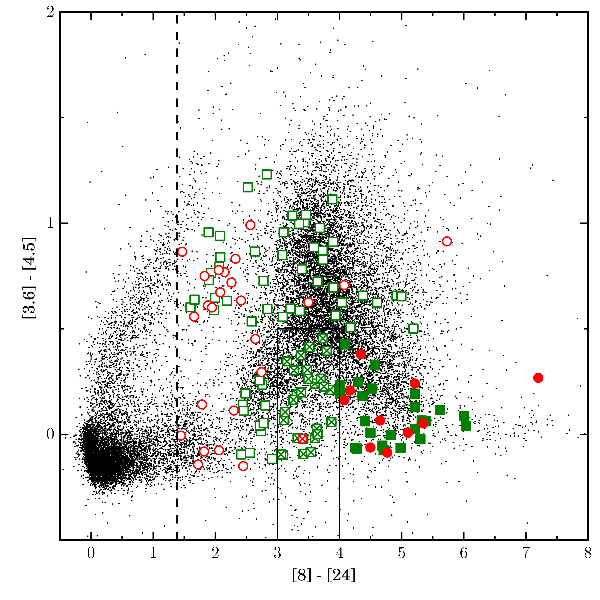}}
\end{minipage}
\hspace{0.1cm}
\begin{minipage}{240pt}
\resizebox{\hsize}{!}{\includegraphics[clip=true,width=9cm]{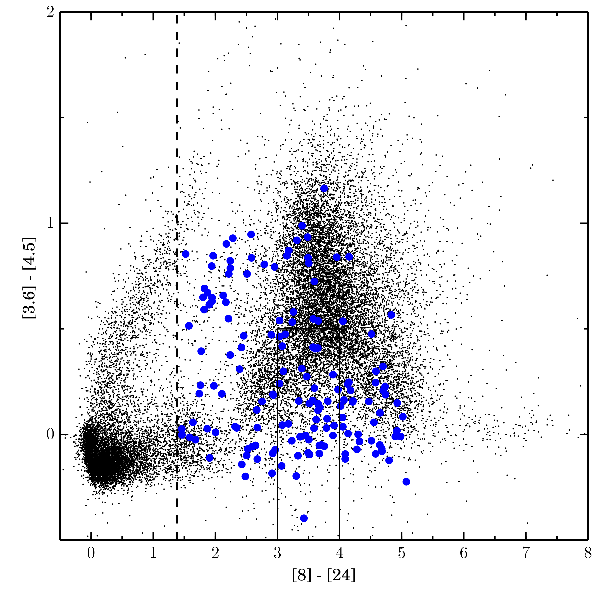}}
\end{minipage}
\end{center}
\caption{Left panel: Same as Figure~\ref{fig:sampleselection}, but 
  indicating the visual SED-based classification as disc-, shell- or
  uncertain-sources for the post-AGB (red circles) and post-RGB
  candidates (green squares). Disc sources are represented as open symbols, shells as filled symbols and 
uncertain sources with enclosed-crosses. The region 3.0\,$<$\,[8]\,$-$\,[24]\,$<$\,4.0 and
[3.6]\,$-$\,[4.5]\,$<$\,0.5 is marked with black solid lines and is
where the uncertain-sources lie. 
Right panel: Same as
Figure~\ref{fig:sampleselection} 
but for the  YSO candidates, represented as blue 
filled circles. The uncertain-sources region is transferred from the
left panel to the right panel for comparison purposes only.}
\label{sedclassification}
\end{figure*}

To find a more quantitative method for classifying disc- and
shell-sources, we explore the position of the sources in the [8]\,$-$\,[24] vs [3.6]\,$-$\,[4.5] 
colour$-$colour plot used previously. We note that a large [3.6]\,$-$\,[4.5] indicates the presence of a
near-IR excess due to hot dust surrounding the central star. 
This is indicative of a stable dust structure close to the central
star, as these objects have
photospheres too hot to be in a dust producing phase. A large 
[8]\,$-$\,[24] colour indicates the presence of colder dust in the
system. In Figure~\ref{sedclassification},  we show 
the post-AGB and post-RGB candidates on a [8]\,$-$\,[24] vs [3.6]\,$-$\,[4.5] 
colour$-$colour plot indicating their visual SED-based classification as discs, shells or uncertain
sources. We find that the majority of shell-sources show a
[8]\,$-$\,[24]\,$>$\,4.0. These shell-sources show only 
a mild [3.6]\,$-$\,[4.5] excess with
 [3.6]\,$-$\,[4.5]\,$<$\,0.5. Most disc-sources have a clear near-IR excess with
 [3.5]\,$-$\,[4.5]\,$>$\,0.5.  Some disc-sources, however, have
 [3.6]\,$-$\,[4.5]\,$<$\,0.5 but in combination with a
 [8.0]\,$-$\,[24]\,$<$\,3.0, indicating that hot dust dominates their SEDs. Finally the uncertain-sources are found to occupy the
 region between the disc- and shell-sources. This region is
 characterised by 3.0\,$<$\,[8]\,$-$\,[24]\,$<$\,4.0 and
 [3.6]\,$-$\,[4.5]\,$<$\,0.5.

We find that approximately half the post-RGB
sample are classified as disc-sources while the remainder do not show
disc-type SEDs.  This is inconsistent with the
suggested formation channel in which the RGB evolution of \textit{all} post-RGB stars is terminated
prematurely by a binary interaction which also results in the creation
of a dusty disc.  In fact, we find that 23 objects are classified as
shell-sources within our classification scheme. Since binary
interaction is, to our knowledge, the only mechanism that can prematurely terminate the
RGB evolution, we think that the objects with shell-type SEDs in our
study are not exclusively single stars but may also be stars in binary systems with
likely evolved dusty disc components that peak at longer
wavelengths. Furthermore, we find that the majority of the sources
classified as 'uncertain', are post-RGB stars and therefore we expect
these objects to also have dusty discs. This also shows that our SED based
classification is a rather approximate classification and does not
translate directly to the single or binary nature of these objects,
especially since their SEDs are limited to 24$\mu$m in most cases. To
be able to clearly differentiate between shell-type and disc-type
SEDs, we would need SEDs which extend to longer wavelengths. 
For each of the post-AGB and post-RGB objects, their SED classifications
("disc", "shell" or  ''uncertain'') are given in the SED column of
Tables~\ref{tab:pagb_param} and ~\ref{tab:prgb_param}.

In the right panel of Figure~\ref{sedclassification} we also plot the YSO candidates 
(as blue filled circles) though the classification scheme for YSOs are different 
compared to that of post-AGB/post-RGB candidates. A SED based YSO
classification is beyond the scope of this study.
We note that the SEDs of the post-AGB, post-RGB and YSO candidates can be found in
Appendix~\ref{seds} and are also available as online supporting
information.

\section{Features in the stellar spectra}
\label{specanalysis}

Spectra can provide a plethora of information about the nature of an object. To further investigate the post-AGB, post-RGB and YSO
candidates, we inspected their spectra individually.

We find in some cases prominent emission lines in the Balmer sequence
(especially the H$\alpha$ line at 6562.8\AA), forbidden lines of oxygen ([OIII]), sulphur ([SII]), 
nitrogen ([NII]), the HeI lines, the CaT line or the Paschen
lines. This indicates either that the star is of an early spectral type capable of exciting
circumstellar gas or in those cases when the line profiles are narrow then there is 
un-associated ISM nebulosity in line-of-sight to the candidate. 

The most prominent diagnostic is certainly the H$\alpha$ line. We find
that in many of our post-AGB and post-RGB stars, the line profile is dominated by a
strong emission component along with a weaker absorption component, similar to that 
seen in the post-AGB stars studied by \citet{vandesteene00-pr}. 
Even with the low spectral resolution and low signal-to-noise ratio of our
spectra, we can resolve the H$\alpha$ emission line profile which
indicates the presence of strong winds. \citet{vandesteene00} explain that the emission component originates close to the central star 
and the absorption component originates from the fast wind itself or in the region
where the fast wind sweeps up the slow wind. 
In some rare cases we find that the H$\alpha$ line profile is dominated by an
absorption component, along with a weaker emission component which is likely to be indicative of
on-going mass-loss in these objects.  Furthermore, in some of the objects
with a H$\alpha$ emission line profile, an absorption core is also detected and
vice-versa. 
The H$\alpha$ line profile of the 162 YSO candidates also show similar
trends to that of the post-AGB/post-RGB objects explained above.  
\citet{natta02} and \citet{jayawardhana02} show that the H$\alpha$ emission and
forbidden line emission in YSOs indicate disc 
accretion. 

In Tables~\ref{tab:pagb_param}, ~\ref{tab:prgb_param} and
~\ref{tab:yso_param}, we have summarised the line profile of the
H$\alpha$ line as well as the forbidden oxygen [OIII] line of the
final sample of post-AGB, post-RGB and YSO candidates,
respectively. The tables also list the full width at half maximum of the
H$\alpha$ for an emission line profile, as an indication of whether the line is spectrally
resolved or not. For broad lines, the FWHM is a good proxy of the H-alpha gas outflow
velocity. Considering the resolution of the spectrograph
(R\,$\approx$\,1300), the outflow velocity must be larger than $\sim$\,115kms$^{-1}$ (FWHM of
5\AA) to be resolved. We find that for the majority of the objects, the
H$\alpha$ line is resolved, indicating a large outflow velocity. 

Four of the post-RGB candidates (J050504.15-684440.8 ,
J052245.11-713610.2, J052834.16-715558.1 and J053130.65-714448.2), show FeII emission line
features in their spectra. These features are characteristic of B[e]
stars (see Section~\ref{preclass}).

A significant element with respect to post-AGB stars is barium. 
The presence of enhanced Ba lines in the stellar spectrum indicates a 
\emph{s}-process enriched post-AGB object.  
For low- to intermediate-mass stars, a significant amount of \emph{s}-process 
nucleosynthesis takes place prior to the post-AGB phases of 
stellar evolution. Therefore we expect to detect the presence of 
$\emph{s}$-process elements for post-third dredge-up objects. 

On visually inspecting the spectra of the post-AGB candidates, we were unable to identify 
barium in the majority of cases. This could be due to the low-resolution of our spectra. 
However, for strongly $s$-process enriched stars, we were able to detect the presence of the strong 
BaII line at 4554.03\AA. We found that 6 out of 35 stars
(J050632.10-714229.8, J051906.86-694153.9, J052147.95-700957.0,
J052220.98-692001.5, J052243.99-693828.0, J052604.53-685737.6) show the presence of the BaII line at
4554.03\AA. J050632.10-714229.8 and J053250.69-713925.8  has been previously identified, 
from abundance studies with high resolution spectra, as \emph{s}-process enriched post-AGB stars 
by \citet{vanaarle13}. Two of the post-RGB stars (J051453.10-691723.5 and
J054034.77-683228.2) also the presence of the BaII line at
4554.03\AA. J051453.10-691723.5  has $L_{\rm ob}$\,$=$\,2616\Lsun\, and
$L_{\rm ph}$\,$=$\,2171\Lsun, close to the post-AGB--post-RGB dividing
line of 2500\Lsun\, so this star is probably a genuine post-AGB star. On
the other hand, J054034.77-683228.2 has $L_{\rm ob}$\,$=$\,845\Lsun\, and $L_{\rm
  ph}$\,$=$\,906\Lsun\, so it seems to be an unambiguous post-RGB star. 
In this case, since post-RGB stars have not yet gone
through the AGB phase, it is likely that the object is extrinsically
enriched by binary mass transfer putting in the category of Ba-stars.

Five of the YSO candidates
(J050252.54-674848.6, J051845.47-690321.8, 
J053030.10-675018.0, J053153.96-673845.7, J053625.85-702834.7) show the presence of the BaII line at
4554.03\AA, indicating $s$-process enrichment. This could imply that either these objects are
post-AGB objects that have been misclassified YSOs or that the strong detection at 4554.03\AA\,s is not
real. 

Another element of interest is lithium, which can be detected by the presence of the 
LiI line at 6708 \AA. Lithium is abundant in the parent molecular cloud but it is 
destroyed in the stellar interior at relatively 
low temperatures ($\sim$ 2\,$\times$\,10$^{6}$K). If these interior temperatures are reached when 
the star is convective, Li will be depleted at the stellar surface during the 
pre-main sequence phase. During the evolution beyond the main sequence, lithium is 
further decreased owing to the first and second-dredge up processes that occur during the 
red-giant phase of evolution and the early-AGB phase of evolution \citep{karakas03b}. 
However, in massive stars ($>$\, 4\Msun) during the thermally pulsing AGB phase, lithium can be created by hot bottom burning 
\citep{boothroyd95,lattanzio96}. We searched for the presence of lithium in the stellar photospheres of 
both the post-AGB/RGB and YSO candidates by visually inspecting the
spectra. We could not detect the LiI (6708 \AA) line in absorption in any of the post-AGB and post-RGB
candidates. This is as expected since current evolutionary models for
these initial mass ranges (calculated based on their luminosities) do not predict an 
enhanced Li abundance.

We detected the LiI (6708 \AA) line in absorption in for 11 out of 162
YSO candidates (J044535.72-685840.4, J044920.32-690900.0,
J045009.84-701320.9, 
J045145.31-692837.1, J045424.37-685240.6, J050422.62-684555.3,
J052141.52-710909.2, 
J053050.58-681210.3, J053337.09-711046.1,
J053614.00-695543.8, J054153.85-710903.7) indicating that these latter objects are 
probably early stage YSOs or massive YSOs. We note again that the 
low-resolution of the spectra could possibly affect the 
number of identifications.

\begin{figure*}
\begin{center}
\includegraphics[width=18cm]{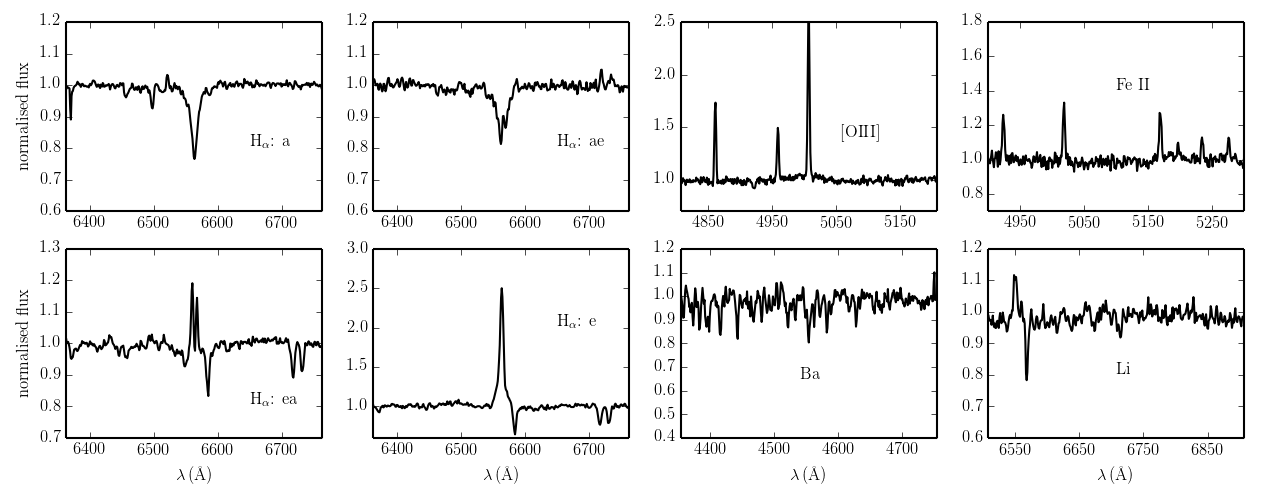}
\end{center}
\caption{A small part of the continuum-normalised spectrum for a few sample stars that show the different H$\alpha$ line
profiles, the forbidden [OIII] emission line, the region with FeII emission lines, the barium
line (4554.03\AA ) and the Li absorption line (6708 \AA). The spectral
feature shown is labelled in each of the panels (see text for further details). In the case of the
H$\alpha$ line, the nature of the line profile is also mentioned. See Tables~\ref{tab:pagb_param},\,~\ref{tab:prgb_param},\,and\,~\ref{tab:yso_param}, for more details.}
\label{spec_eg}
\end{figure*}

Figure~\ref{spec_eg} shows examples of the different H$\alpha$ line
profile that the stars in our sample display. Also shown in
Figure~\ref{spec_eg} are selected parts of the continuum-normalised spectrum for a few sample objects that
show the forbidden [OIII] emission line (at 5001\AA), the region with the multiplet
42 of FeII (at $\approx$\,4924\AA, 5018\AA, and 5169\AA), the barium
line (at $\approx$\,4554\AA ) as well as the Li absorption line (at $\approx$\,6708 \AA). 

A peculiar object in our sample is J050830.51-692237.4 (see
Table~\ref{tab:yso_param}), which has been classified as a YSO candidate based on
our \logg\, criterion. On analysing the spectrum of this object, we find
that it shows a hot component that corresponds to an A-type
star as well as a cooler component with molecular bands of carbon, corresponding to a
C-star. Based on the radial velocity estimates of the two components
of the spectra, we find that both the components have LMC
velocities. The hot  component has a heliocentric radial velocity of 
$\approx$\,220\,kms$^{-1}$ while the cool component shows
$\approx$\,290\,kms$^{-1}$. The spectral features and the estimated radial
velocities of the two components were also confirmed using UVES
spectrum of this object that was obtained as a part of our
high-resolution follow-up studies. \citet{woods11} have analysed the SST IRS spectra of this
object and have found that the circumstellar environment of this
object is clearly O-rich but with very large crystallinity, which
indicates that the dust is in the form of a disc and therefore 
associated with an object in a binary system (with the
C-star as it's companion) or a YSO.  The former scenario is very
unlikely as it would require a binary system with a post-AGB star and
an AGB star or a luminous main-sequence star and a AGB star, which is
rather uncommon. Furthermore, from the STP, 
the estimated \logg\, value points to a YSO evolutionary
nature for J050830.51-692237.4. Therefore, it is likely that this
object is YSO with a O-rich
disc and in the line-of-sight of the object is a non-related
C-star. 

\section{Evolutionary stages of the post-AGB, post-RGB and YSO candidates}
\label{hrtext}

\begin{figure*}
\begin{center}
\begin{minipage}{500pt}
\resizebox{\hsize}{!}{\includegraphics[clip=true,width=9cm]{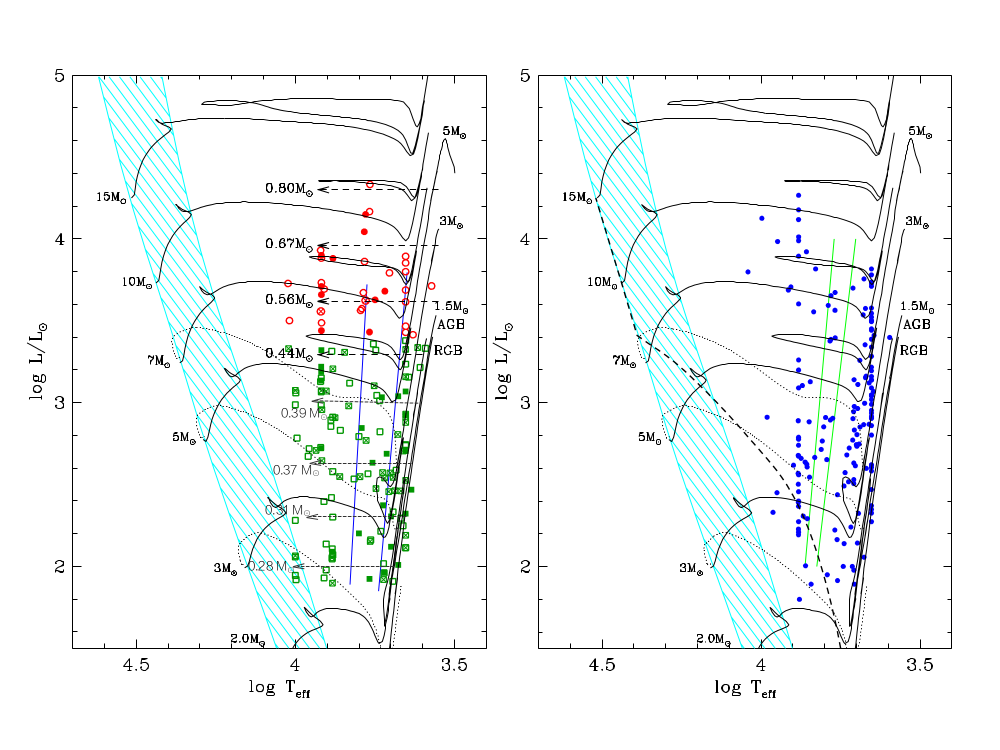}}
\end{minipage}
\end{center}
\caption{The HR diagram for the sample of the post-AGB and post-RGB candidates 
(left panel) and YSO candidates (right panel). In the left panel, 
the red circles represent the post-AGB candidates and the green squares represent the 
post-RGB candidates. The open symbols represent disc sources and the
filled symbols 
represent the shell sources.  
The enclosed-crosses represent those post-AGB/post-RGB candidates for which 
the nature of the SED is uncertain. In the right panel, the blue filled circles 
represent the YSO candidates. Each plot shows the main sequence as a cyan cross-hatched region. 
In both the plots the black solid lines represent evolutionary tracks starting from the main-sequence and 
the black dotted lines represent pre-main sequence evolutionary tracks. The black dashed arrows on 
the HR diagram for the 
post-AGB/post-RGB candidates schematically represents the
post-AGB/post-RGB evolutionary tracks, each labelled by the current
stellar mass. Also shown 
on this plot is the empirical OGLE instability strip for the Population 
II Cepheids represented with blue vertical lines. 
In the right panel, the thick black dashed line in right panel is the 
birth-line and the green vertical lines represent the Cepheid instability strip. See text 
for further details.}
\label{hr}
\end{figure*}

To understand the evolutionary stage of the post-AGB, post-RGB and YSO candidates, we show their positions 
in the HR diagram in Figure~\ref{hr}. The left panel shows the
post-AGB and post-RGB 
population. We note that the \teff\, values are those 
derived from the spectral fitting and the luminosities plotted are 
the photospheric luminosities ($L_{\rm ph}$).

Each plot shows the main sequence as a cyan cross-hatched region. Evolutionary 
tracks starting from the main sequence and continuing up to the AGB-tip according to the tracks of 
\citet{bertelli08,bertelli09} are shown as black solid lines. Note that these tracks 
use a synthetic AGB calculation adopting unusual mass loss rates, and almost certainly terminate at too 
low a luminosity.  The plots also show the PISA pre-main sequence (PMS) 
evolutionary tracks \citep[black dotted lines:][]{tognelli11} 
up to the maximum 
computed mass of 7\Msun. A metallicity $Z$ = 0.004 was selected for both sets of evolutionary tracks. 
The masses of the evolutionary tracks are marked on the plots with the 
PMS and main-sequence masses marked on the left side of the plots and RGB-tip masses marked 
on the right side of the plots. The positions of the RGB and AGB are also marked.

In the figure showing the post-AGB and post-RGB candidates, post-AGB and post-RGB evolutionary 
tracks are shown schematically (black dashed arrows). The masses for the post-AGB 
evolutionary tracks are from \citet{vw94} for $Z$ = 0.004. The post-RGB evolutionary track masses 
are estimated from the RGB luminosity-core mass relation of the \citet{bertelli08} models 
with $Z$ = 0.004. In Section~\ref{pagbprgbyso}, we estimate the masses
of  the post-AGB and post-RGB candidates. The post-AGB candidates (with $L_{\rm
  phot}$\,$\gtrsim$\,2500\Lsun) have masses of
$\sim$\,0.45\,$-$\,0.8\,\Msun and the post-RGB candidates (with ($L_{\rm
  phot}$\,$\approx$\,100\,$-$\,2500\Lsun) have masses
of $\sim$\,0.28\,$-$\,0.45\,\Msun.

In the HR diagram of the post-AGB/post-RGB candidates, the blue vertical lines shows the empirical 
OGLE instability strip for the Population 
II Cepheids presented in \citep{soszynski08}, since post-AGB/post-RGB evolutionary tracks cross the Population 
II Cepheids instability strip.  In the HR diagram showing the YSO candidates, the green vertical 
lines on this plot denotes the Cepheid instability strip from \citet{chiosi93}. Also 
shown in the HR diagram showing the YSO candidates, is the birthline 
(thick black dashed line in right panel of Figure~\ref{hr}), which may be considered as the 
dividing line between the obscured protostellar and the observable pre-main sequence stage of stellar evolution. 
The location of the birthline depends highly on the 
mass accretion rate, with higher accretion rates shifting the line to the right. A mass accretion rate of 10$^{-5} \Msun$/yr (used for the birthline in right panel of Figure~\ref{hr}) 
represents the typical value for stars in the 
mass range from few tenths of a solar 
mass to about 10\Msun\, \citep{stahler83,palla93}.

The YSOs, with masses in the range of $\sim$\,3\,$-$\,10\,\Msun (see
Section~\ref{pagbprgbyso} and Figure~\ref{hr}), are rather luminous ($L_{\rm
  phot}$\,$>$\,200\Lsun) and lie on the cool side of the usually adopted birthline in the
HR-diagram. We also find that the majority of YSOs 
lie to the right of the birthline so they should not be visibly
detectable. A similar trend was observed for the
YSOs identified in the SMC (see Paper I), where we concluded that the discrepancy may be due 
to the assumption of symmetric and spherical dust 
shells in the birth line modelling (with asymmetries, it may be possible to see the central star through a region of low 
extinction), or too high an assumed accretion rate since the birthline depends on the mass-accretion rates \citep{palla93}. 
A low mass accretion rate could move the 
birthline to lower values of \teff\, so that our stars could become visible. A group of 
massive pre-main sequence stars 
similar to the Galactic Herbig AeBe stars was found in the LMC by \citet{lamers99} and these are also located 
above the traditional birth line used for the Galactic
sources. \citet{lamers99} suggested that this could be 
due to either a shorter accretion timescale for Galactic Herbig AeBe stars due to lower metallicity in the LMC, or a lower dust-to-gas ratio in the LMC, 
again owing to the lower metallicity. Therefore for the LMC, a higher birth line for YSOs could be expected in the HR diagram. 

\section{Completeness of the survey}
\label{completeness}

\begin{table*}
  \caption{A breakdown of the number of objects with respect to their assigned priorities, as analysis has 
proceeded.}
  \centering
  \label{samplevo2}
  \begin{tabular}{lcccc}
  \hline
   Stage & Total objects & Priority 1 objects & Priority 2 objects & Priority 3 objects \\
  \hline
   Initial photometric selection for the whole LMC & 8628 & 1517 & 6823 & 286 \\
   Objects with optical spectra & 2102 & 883 & 1170 & 49 \\
   Objects rejected due to poor spectra & 556 & 102 & 449 & 5 \\
   Objects retained with good optical spectra & 1546 & 781 & 721 & 44 \\
   Objects classified as M stars & 382 & 333 & 10 & 39 \\
   Objects classified as C stars & 55 & 51 & 3 & 1 \\
   Objects classified as PN & 123 & 56 & 67 & 0 \\
   Objects classified as red-shifted galaxies & 290 & 26 & 264 & 0 \\
   Objects classified as QSOs & 39 & 2 & 37 & 0 \\
   Objects with TiO in emission & 5 & 5 & 0 & 0 \\
   Objects with strong emission lines and an UV continuum  & 69 & 35 & 32 & 2 \\
   Sample of probable post-AGB/post-RGB and YSO candidates & 581 & 271 & 308 & 2 \\
   with confirmed LMC membership &  &  &  &  \\
   for which we carry out detailed spectral analysis &  &  &  &  \\
   Objects without $B,V$ magnitudes & 85 & 12 & 73 & 0 \\
   Objects with \teff\,$>$\,12000\,K & 68 &34 & 34 & 0 \\
   (likely hot main-sequence objects)  &  &  &  &  \\
   Objects with $L_{\rm ph}$\,$>$\,35000\Lsun & 112 & 105 & 5 & 2 \\
   (likely luminous super-giants)  &  &  &  &  \\
   Final number of post-AGB candidates & 35 & 34 & 1 & 0\\
   Final number of post-RGB candidates & 119 & 34 & 85 & 0\\
   Final number of YSO candidates & 162 & 52 & 110 & 0 \\
  \hline
 \end{tabular}
\end{table*}

This study is aimed at identifying optically visible post-AGB and RGB candidates in the LMC. 
However, the survey obviously has its limitations and does not
catalogue all the post-AGB and post-RGB stars.  
In this section we have listed the limitations that govern this survey and also provide the 
extent of the completeness of our survey.

To obtain a complete sample of post-AGB/post-RGB candidates in the
LMC, we used the selection criteria of \citet{vanaarle11} and these selection 
criteria restrict our search to post-AGB/post-RGB candidates with an
excess at 24$\mu$m. Older post-AGB/post-RGB 
stars with expanding shells for which the excess is undetectable at
24$\mu$m will not be selected. For instance, the LMC counterparts of the hot Galactic post-AGB stars studied by 
\citet{gauba04} will not be selected since their 24$\mu$m fluxes would
be below the detection threshold. Furthermore, we require that all the 
selected candidates have $V$\,$<$\,20 mags, which only selects those
post-AGB/post-RGB stars that are 
optically visible, therefore omitting those that are faint or heavily dust
enshrouded. 

Based on the mid-IR selection criteria, we started with a sample of 8628
objects, out of which 1517 objects were given a priority 1, 6823 were given a priority 2 and 
286 were given a priority 3. We note that the priorities were 
assigned based on the selection criteria used and the likelihood of finding post-AGB/post-RGB 
candidates with that selection 
criteria (see Section~\ref{sampleselection}). In Table~\ref{samplevo2} we give a breakdown of the number of objects with respect to their assigned priorities 
as the analysis has proceeded. We performed a
low-resolution optical survey that covered 
a large part of LMC, as 
shown in Figure~\ref{fields}. From the initial selected sample of 8628
objects, we obtained spectra of 2102 objects.  
Excluding the 556 objects that 
had a poor-quality spectrum with low signal ($<$\,200 counts), we were left with a sample of 1546 unique spectra.
Taking into consideration the candidates for which we did not obtain spectra and the objects that we rejected from the spectral analysis due to 
poor quality of their spectrum, we estimate the completeness of the
survey (defined as the fraction of all possible candidates that were
observed) to be 1546/8626\,$\approx$\,20$\%$. However, when only 
considering the priority 1 objects, we find that we were able to study
and assign candidature for 781 out of 
the 1517 objects, resulting in $\approx$\,50$\%$ completeness. Similarly we estimate the completeness of the study of the priority 2 and priority 3 
objects to be $\approx$\,10$\%$ and $\approx$\,15$\%$,
respectively. 

We can provide a rough estimate 
of the number of optically visible post-AGB and post-RGB candidates we expect to find in the unstudied sample 
(see Table~\ref{samplevo2}). Out of the 781 priority 1 
objects, 34 turned out to be post-AGB candidates. So for the remaining
unstudied sample of 1517$-$781\,$=$\,736 priority 1 candidates we can expect 
$\approx$\,32 more post-AGB candidates in the whole LMC. One out of the 721 
priority 2 objects turned out to be post-AGB candidates. So from the remaining unstudied sample of 
6102 priority 2
candidates, we can expect $\approx$\,8 more post-AGB
candidates. None of the priority 3 candidates are post-AGB stars and
therefore we do not expect to find any post-AGB stars from the
remaining unstudied sample of priority 3 objects. Similarly, based on
the numbers of post-RGB objects, in the LMC we expect to
find $\approx$\,32 additional post-RGB candidates from the priority 1
unstudied sample and  $\approx$\,719 post-RGB candidates from the
priority 2 unstudied sample. This implies that it might be possible to identify a 
further group of $\approx$\, 40 optically visible 
post-AGB candidates and $\approx$\,751 optically visible post-RGB
candidates from the total unstudied group of initially selected 
candidates in the LMC. We note that the number of observed and
expected post-RGB candidates with priority 2 
is much larger than the objects with priority 1, because while
assigning priority for spectroscopic observations, priority 1 was
assigned to objects with luminosities in-between 1000\Lsun and
35000\Lsun, while priority 2 was assigned to objects with luminosities
$<$\,1000\Lsun, where it is more likely to find post-RGB stars. 
Since priority 3 is assigned to those objects with luminosities
$>$\,35000\Lsun, as expected, we do not find any post-RGB objects with
priority 3. 

As part of the future work, we intend to complete the survey by
observing the unstudied sample of objects. We note that
out of the 556 objects that we rejected based on low signal spectrum,
456 objects were faint with $V$, $I$ or $R$ band magnitudes greater than
17\,mags. For these objects a larger exposure time than that used obtain
the current spectrum or a larger
telescope would be required to obtain a more suitable spectrum. For the remaining 101 out of the
556 objects, it is likely that they were either observed in
poor observing conditions or subject to a bad fibre
alignment. 

We note that the above estimated completeness applies only to
optically visible post-AGB and post-RGB stars. We do not take into
account the fainter or heavily enshrouded objects.

\section{Estimating Post-AGB and post-RGB Evolutionary Rates}
\label{evolrate}

To be able to estimate the evolutionary rate, a complete
  sample of post-AGB  and post-RGB stars is required. During the
  post-AGB phase, for a particular luminosity, \teff\, is determined by the mass of the hydrogen 
envelope \citep{schoenberner81,vw94}. The rate of evolution in \teff\, is 
therefore determined by the luminosity (which determines the rate of consumption of 
the hydrogen envelope by nuclear burning) and by the mass loss rate 
(in the case of single post-AGB stars) and the mass re-accretion rate
(in the case binary of post-AGB and post-RGB stars). 
This mass loss rate/mass accretion rate is essentially completely 
unknown. Based on the available sample of post-AGB and post-RGB candidates from this study, 
we now try to estimate this mass loss/accretion rate by 
determining the numbers of stars in a given \teff\, interval along the
post-AGB and post-RGB tracks and comparing those numbers to the number of 
stars in the top magnitude of the RGB, where the duration of evolution is known.

To estimate the lifetime on the top one magnitude of the RGB, we used the 
\citet{bertelli08} evolutionary tracks. They show that 
stars in the mass range 1.0 to 1.8 \Msun\, and LMC-like 
metallicity take $\sim$\,3$\times$\,10$^{6}$ years to traverse the 
top magnitude of the RGB. Subsequent AGB evolution through the 
same luminosity range takes $\sim$\,1$\times$\,10$^{6}$ years. 
Hence, the total time spent by a low mass star in the luminosity range corresponding 
to the top one magnitude of the RGB is $\sim$\,4$\times$\,10$^{6}$ years.

In order to estimate the observed number of stars on the top 
magnitude of the RGB in the nine fields LMC\,1-9, we followed the prescriptions in 
\citet{nie12}.  Stars in the SAGE \citep{meixner06,blum06} catalog were plotted in the 
$J$, $J-$[3.6] diagram and those in a parallelogram coinciding 
 with the top 1 magnitude of the RGB were selected. The 
parallelogram has sides $J$\,=\,13.9, $J$\,=\,14.9, 
$J-$[3.6]\,=\,3.25 $-$ 0.17$J$ and $J-$[3.6]\,=\,3.75 $-$ 0.17$J$.
We find the numbers of stars n1, n2, 
n3, n4, n5, n6, n7, n8 and n9 in each of LMC1, LMC2, LMC3, LMC4, LMC5,
LMC6, LMC7, LMC8 and LMC9 are 34198, 19727, 16258, 6686, 4881, 11480,
10859, 7346 and 7491, respectively, with an error of approximately 5\% \citep{nie12}.
The total number of stars in the magnitude interval corresponding
to the top mag of the RGB in all 9 LMC fields is thus 118927.

Next we estimated the lifetimes of the stars in the early part of the
post-AGB and post-RGB phase. If we assume that post-AGB stars, whether they leave the AGB by single 
 star mass loss or binary interaction, have all passed through the top 
 magnitude of the RGB then we can easily derive their average post-AGB lifetime. 
If there are N$_{\rm AGB}$ post-AGB stars $(\log L \ga 3.4)$ in all 9 LMC fields in a 
 certain evolutionary phase (say from the AGB to \teff\,=\,10000\,K i.e. 
 $\sim$3.6\,$<$\,$\log$\teff\,$<\,\sim$4.0), then the lifetime of these stars in that 
 evolutionary phase is 4\,$\times$\,10$^{6}$\,*\,N$_{\rm AGB}$/118927
 years. From our analysis, we find a total of N$_{\rm AGB}$ = 35 
which results in a post-AGB lifetime of 1177 yrs. For post-RGB stars, 
the calculation is not so easy. \citet{nie12} find that $\sim$4\% of red giants evolving up the RGB produce post-RGB stars when they fill their Roche lobes before reaching the RGB tip. Most of the post-RGB stars are produced during the top two magnitudes of the RGB where most of our post-RGB stars are observed. 
The median luminosity of these stars is $\log$\,L/\Lsun\,$\sim$\,2.8. In this case, 
 if N$_{\rm RGB}$ post-RGB stars are observed, then their average lifetime is 
 4\,$\times$\,10$^{6}\,*\,($N$_{\rm RGB}/0.04)/118927$ years. From our analysis, we find a 
total of N$_{\rm RGB}$ = 119 post-RGB candidates, which results in a
post-RGB lifetime of 100061 yrs.

The estimated 
lifetime assumes that the samples of post-AGB and post-RGB candidates in the fields LMC 1 $-$ 9 is
complete. However, as mentioned in Section~\ref{completeness}, our survey is not entirely complete and 
we can expect to find an additional sample of 40 post-AGB candidates
and 751 post-RGB candidates. To illustrate the impact of this addition, we recalculated the
lifetimes and we estimate a post-AGB lifetime of 2522 yrs 
and a post-RGB lifetime of 731541 yrs. 

We need to compare the above post-AGB lifetime to that of
  post-AGB stars without external mass loss during the post-AGB life. 
Following the procedures in Paper I, we
  estimated evolutionary times for a post-AGB to traverse the interval
  3.7 $< \log$\,\teff $<$ 4.0 where most of the observed post-AGB
  stars lie and also for a post-RGB star to traverse from
  $\log$\,\teff(RGB)\,+\,0.05 to $\log$\,\teff\,=\,4.0 by consuming
  the hydrogen-rich envelope.  The lifetime for post-AGB
evolution without post-AGB mass loss from $\log$ \teff\,=\,3.7 to 4.0 
and for an intermediate luminosity star ($\log$L/\Lsun\,=\,3.8) was
estimated to be 9800 yrs. From the observations of the post-AGB stars
in the LMC, we find an observational lifetime of 2522 yrs for 
the post-AGB stars. Formally, the numbers suggest  that for the 
post-AGB stars, some mass loss is required to hasten the evolution. 
This mass loss rate is M$_{\odot}$\,$\approx$\,2.7\,$\times$\,10$^{-7}$ \Msun/yr, but the
uncertainties are very large and we do not consider the post-AGB
binaries (with mass-accretion).

Using the procedures in Paper I, we find that the post-RGB evolution time
without mass loss from $\log$ \teff(RGB)\,+\,0.05 to $\log$ \teff = 4.0 and for an intermediate luminosity
 star ($\log$L/\Lsun=3.0) is 109000 years (the envelope mass is assumed to
be consumed by hydrogen burning only).
From the observations of
 post-RGB stars in the LMC, we find an observational
 lifetime of 731541 years which is greater than the theoretically predicated lifetime
therefore suggesting mass-accretion on the post-RGB, as expected for
these likely binary objects.   The mass accretion rate formally required to slow
down the evolution to give the observed number of post-RGB stars 
is M$_{\odot}$\,$\approx$\,1.0\,$\times$\,10$^{-8}$ \Msun/yr.

Finally, we note that our calculations of
  lifetimes do not include the possibility that some objects in the post-AGB and post-RGB
phases could be heavily dust
  enshrouded and therefore not observed by us.  If there are significant numbers
of these objects then our estimate observational lifetimes will be underestimated,
our post-AGB mass loss rates will be overestimated and our post-RGB accretion
rates will be slightly underestimated.

\section{Summary and conclusions}
\label{end}

In the LMC, we have identified a sample of 35 well characterised high probability
optically visible post-AGB candidates with spectroscopically determined stellar
parameters (\teff, \logg, [Fe/H] and E[$B-V$]) spanning a wide range
in luminosities. These objects have spectral types between A and K and being an evolved 
class of objects, they have a lower metallicity
([Fe/H]\,$\approx$\,$-$1.0) than the mean present-day LMC
metallicity. Their spectral energy distributions and their locations on the [8]$-$[24] vs [3.6]$-$[4.5]
colour-colour plot allowed us to distinguish between shell-sources,
(with a double-peaked SED) and disc-sources (with an SED that is
indicative of hot circumstellar material).  Out of the 35 post-AGB
candidates, we found a group of 10 shell-sources and 23
disc-sources. For the remaining 2 sources, we were 
unable to classify them based on their SEDs. The low-resolution
spectra of the post-AGB objects revealed a diversity in their H$\alpha$ line
profiles. Some objects showed a line profile with a
strong emission component along with a weaker absorption component,
indicative of strong stellar winds, while other objects showed a strong absorption component with a weaker
emission component, indicative of on-going weak mass loss. We were also able to identify the presence of barium for 6
candidates, which is an expected product of the nucleosynthesis during
the AGB phase of evolution. 

The determination of the
photospheric luminosity ($L_{\rm ph}$) confirmed the existence of
optically visible dusty post-RGB stars by revealing 119 such candidates. This class of
objects was discovered in our SMC survey (see Paper I). These objects 
have mid-IR excesses and stellar parameters similar to those of post-AGB stars 
(late-G to late-A spectral types, low log g values, and low
metallicities with [Fe/H]\,$\approx$\,$-$1.0). 
However, their luminosities ($\approx$\,100\,-\,2500\Lsun), and hence
masses and radii, are much lower than those expected for post-AGB stars. We expect these objects to be a
result of mass loss induced via binary interaction that terminates
their RGB phase. Detailed radial velocity monitoring studies are required to confirm the true nature of all these
sources. Based on their SEDs, we were able to classify the 119
post-RGB stars into groups of 23 shell-sources and 56
disc-sources. For the remaining 40 sources we were
unable to classify them based on their SEDs. The low-resolution spectra
of the post-RGB candidates also showed similar H$\alpha$ profiles as
the post-AGB stars. Furthermore, we were able to identify the presence
of barium in two of these objects. One of these objects,
J051453.10-691723.5, is likely to be a post-AGB star since its central
star luminosity is close to the post-AGB--post-RGB dividing
line of $\approx$\,2500\Lsun. The second object, J054034.77-683228.2, is a
low-luminosity star and since post-RGB stars have not yet gone through
the AGB phase, it is likely that this objects falls in the category of
Ba-stars, which are objects that are extrinsically
enriched by binary mass transfer. 

This study has also resulted in a large sample of luminous YSOs, since YSOs are
also surrounded by large amounts of circumstellar material (and display a large IR excess) and 
are present in the luminosity range occupied by post-AGB and post-RGB stars. We
used a \logg\, criterion to disentangle the YSOs since at a given luminosity, the
mass of a YSO is about 15\,$-$\,20 times that 
of the corresponding post-AGB/post-RGB star, leading to a difference
of ∼ 1.3 in log g between them.  This resulted in a sample of 162 high probability YSO candidates, majority
of which are newly identified in this study. These objects have temperatures ranging  
between 4000K and 9000K, high surface gravities, and a mean metallicity [Fe/H]\,$\approx$\,$-$0.60, which agrees well with the average present-day LMC metallicity. From the position of these YSO candidates on the HR diagram, we 
were able to infer that they have masses of
$\sim$\,3$-$\,10\,\Msun. The YSO candidates showed H$\alpha$ emission
and forbidden line emission indicative of disc 
accretion in YSOs. We were also able to identify the presence of Li in
eleven candidates. 

Our detailed spectroscopic analysis combined with our robust \logg\, criterion has provided a systematic characterisation of the
objects in our study and in some cases revealed the true nature of the
objects. For instance, J050830.51-692237.4 was previously classified as a
oxygen-rich post-AGB star by \citet{woods11} based on IRS spectra. The
study by \citet{vanaarle11} classified this objects as a
C-star. However, based
on the \logg\, value of this object and
inspection of its spectrum and SED, we find that this object is more
likely to be a YSO with a O-rich disc and in
the line-of-sight of the object is a non-related C-star. Another object, J053253.51-695915.1, which
is a low-luminosity object that shows a mild
$s$-process enrichment, was classified as a $s$-process rich post-AGB
star by \citet{vanaarle13}. However, this object has too low a luminosity to be a post-AGB star and
based on the \logg\, value of the object, our studies reveal that this
object is likely a YSO, indicating that the mild $s$-process enhancement comes
from the initial composition of the LMC
\citep{vanderswaelmen13}. 

We note that the classification of post-AGB, post-RGB and YSO candidates depend
significantly on their estimated stellar parameters (\teff, \logg,
[Fe/H], E($B-V$) and luminosity). Since our study is performed 
using low-resolution spectra, this study results in spectroscopically
verified, likely optically bright post-AGB, post-RGB and YSO
candidates. To confirm their nature, follow-up studies with high-resolution spectra are essential.

Other interesting by-products of this survey include a group of 69 hot objects whose spectra show
emission lines and in some cases, a significant UV continuum. These
objects are likely to be either hot post-AGB, post-RGB or luminous
YSO candidates. Fifteen of these objects are likely to be B[e] star candidates, out
of which 12 are newly discovered. 

This study has also resulted in the discovery of a significant number contaminants. They are: M-stars, C-stars 
and PNe (presented in Appendix~\ref{mcpne}), a group of QSOs and 
red-shifted galaxies (to be presented in a following publication), a
group of stars with TiO band emission \citep{wood13}, a group of
luminous supergiants and a group of hot main-sequence stars (to be presented in a following publication).

We note that, due to limitations introduced by the selection criteria, our study 
is restricted to optically visible post-AGB and post-RGB stars of spectral type
A $-$ K, in the LMC. The completeness of this survey is $\approx$\,20$\%$ since we were not able to obtain spectra all of the
candidates from within the initially selected sample of candidates. Some 
of the candidates with optical spectra were rejected as their spectra
were of poor quality due to the faintness of the targets (majority of
which had $V$mag\,$>$\,17mags) combined with the low resolution 
of our spectra ($\approx$ 1300). Based on the current final sample of 
post-AGB/RGB candidates (of A $-$ K) in the LMC, we expect to find 
approximately an additional 40 optically visible post-AGB and 751
optically visible post-RGB candidates
from the remaining unstudied initial sample selection of objects. We
note that the estimated completeness and the expected numbers of
remaining post-AGB/post-RGB stars only apply to optically visible 
objects and do not include objects that are faint ($V$\,$>$\,20) or heavily dust enshrouded.

We also used the numbers of the post-AGB and post-RGB objects, to
study the evolutionary rates as well as the mass loss/mass accretion
rates. Though the uncertainties are large, we found that the number of post-AGB stars require stellar evolution models with
some mass-loss and the number of post-RGB stars suggests re-accretion
of gas.

This study has provided a rich sample of post-AGB stars and a new
population of dusty post-RGB stars that cover a
wide range of luminosities and hence masses. Our on-going and future work 
includes investigating the chemical properties of these objects using
high-resolution optical UVES spectra. The known luminosities and chemical abundances of these objects will provide
excellent constrains for AGB nucleosynthesis models. We also intend to
perform a long-term radial velocity monitoring of these objects,
especially the newly discovered dusty post-RGB sample. This will help
establish their likely binary nature and a possibility to probe this
unexplored phase of stellar evolution. Additional future work aims at
characterising in detail the diverse 
samples of contaminants with IR excesses. 

\section*{Acknowledgments}
DK acknowledges support of the FWO grant G.OB86.13.
HVW and DK acknowledge support of the KU Leuven contract
GOA/13/012. PRW acknowledges support from Australian Research Council Discovery
Project DP120103337. We thank the Australian Astronomical Observatory
for allowing us to use the observatory facilities and our AAT support astronomer, 
Dr. Paul Dobbie, who was very helpful during our observing run. We thank the AAO Service 
Program, especially Dr. Sarah Brough and Dr. Daniel Zucker, our service observers, for 
observing three of the LMC fields. We would also like to thank the
referee for his/her useful comments and suggestions.

\bibliography{mnemonic,devlib}

\appendix
 
\section[]{Tables of the PN, C-stars, and M-stars in our sample}
\label{mcpne}

Here, we present the list of C-stars (Table~\ref{tab:C}), M-stars
(Table~\ref{tab:M}) and the PNe (Table~\ref{tab:PN}). For the PNe
objects listed in Table~\ref{tab:PN}, we separate the objects into the
previously catalogued PNe and the newly identified likely PN
candidates, identified based on the visual inspection of their
spectra, as mentioned in Section~\ref{preclass}. In
Table~\ref{tab:PN}, we also perform a positional cross-matching of all
the objects to the most relevant previous studies (see
Table~\ref{tab:pagb_param} for a list of catalogues using which the
positional cross-matching was performed). Based on the result of the
positional cross-matching, out of the 123 PNe in our study, 91 objects
have been previously catalogued as PNe, while the remaining 32
objects have not been catalogued as PNe. Some of these objects have been
classified as YSOs or QSOs, based on either photometric
classifications \citep[e.g.,][]{GC09,whitney08} or inspection of Spitzer Space
Telescope (SST) spectra \citep[e.g.,][]{seale09}. Therefore, we refer to these
objects as uncatalogued planetary nebula candidates. A follow-up detailed
spectral analysis is required to confirm the PN nature of these
objects. 

\begin{table*}
\scriptsize{
\caption{The C-stars in our sample}
\begin{tabular}{ccccc}
\hline
Name & Name & Name & Name & Name \\ 
\hline
J044114.11-700109.5 & J050040.18-682911.6 & J051028.50-694704.4 & J052046.78-690124.1 & J052805.91-700753.4 \\ 
J044542.06-700352.8 & J050252.33-690252.6 & J051121.80-693456.0 & J052224.83-693837.4 & J052943.72-700913.9 \\ 
J044918.46-695314.5 & J050403.80-675408.0 & J051255.63-704551.1 & J052248.93-714312.1 & J053004.33-722253.4 \\ 
J044941.97-683821.1 & J050508.30-701242.0 & J051307.99-683311.8 & J052337.00-700905.4 & J053030.25-683807.7 \\ 
J045111.42-710001.0 & J050527.05-690643.6 & J051320.59-703928.3 & J052346.76-684653.0 & J053128.44-701027.1 \\ 
J045144.79-661757.0 & J050536.98-701734.4 & J051405.00-682758.4 & J052419.63-710424.9 & J053556.87-690045.0 \\ 
J045241.94-713137.5 & J050538.97-684732.1 & J051435.67-713837.0 & J052500.49-694811.8 & J053738.67-712124.0 \\ 
J045543.20-675110.1 & J050558.23-680923.6 & J051734.05-692609.4 & J052619.82-681602.1 & J054124.63-694806.5 \\ 
J045645.54-675048.6 & J050606.51-690302.7 & J051807.82-685355.9 & J052714.36-712731.8 & J054216.19-695636.6 \\ 
J045731.30-663232.0 & J050733.83-692119.9 & J052010.84-684212.3 & J052747.03-691913.4 & J054406.22-701342.1 \\ 
J045902.04-692102.1 & J050839.80-681205.6 & J052040.41-691909.2 & J052755.27-695828.8 & J054418.29-705320.3 \\ 
\hline
\label{tab:C}
\end{tabular}}
\end{table*}
\begin{table*}
\scriptsize{
\caption{The M-stars in our sample.}
\begin{tabular}{ccccc}
\hline
Name & Name & Name & Name & Name \\ 
\hline
J043702.61-694130.6 & J050350.14-691729.1 & J051606.79-690602.1 & J052351.92-691809.7 & J053328.48-721349.9 \\ 
J043758.24-695545.7 & J050414.10-671614.3 & J051607.12-693705.4 & J052354.13-701005.9 & J053333.78-695537.2 \\ 
J044317.72-693742.3 & J050420.11-662753.0 & J051615.59-690622.9 & J052419.31-693849.1 & J053337.18-680600.5 \\ 
J044417.48-693741.3 & J050421.14-702222.0 & J051616.53-685705.4 & J052422.04-683842.5 & J053338.71-711854.5 \\ 
J044606.86-701913.7 & J050426.60-703759.7 & J051622.18-704235.7 & J052430.35-703616.3 & J053340.13-695815.0 \\ 
J044821.86-663834.4 & J050502.46-692837.2 & J051625.87-685754.7 & J052432.86-685445.6 & J053342.83-710722.0 \\ 
J044833.34-693115.4 & J050517.48-673033.8 & J051643.69-703906.7 & J052443.32-675844.3 & J053344.53-705743.3 \\ 
J045002.86-684734.1 & J050518.18-690449.1 & J051705.07-694423.3 & J052452.09-691108.1 & J053411.73-712701.5 \\ 
J045112.96-695019.3 & J050556.33-701854.3 & J051708.96-693221.2 & J052452.65-720156.5 & J053416.45-695740.4 \\ 
J045120.54-692913.9 & J050607.00-693558.6 & J051709.20-702024.4 & J052502.42-691737.5 & J053437.38-700832.4 \\ 
J045134.09-690451.0 & J050626.65-694437.0 & J051709.43-695927.8 & J052508.01-683511.2 & J053521.76-685016.3 \\ 
J045148.45-691903.3 & J050630.69-683014.9 & J051709.90-714402.7 & J052511.58-700414.7 & J053548.07-703146.6 \\ 
J045156.93-701151.5 & J050639.42-672927.1 & J051745.44-690719.3 & J052511.64-685929.5 & J053549.96-715324.8 \\ 
J045159.94-700725.8 & J050639.50-695956.6 & J051745.54-680133.4 & J052527.69-711808.1 & J053610.81-680030.7 \\ 
J045226.45-683437.4 & J050647.80-702659.6 & J051747.18-681842.6 & J052529.08-685021.2 & J053625.70-683945.2 \\ 
J045314.77-691217.9 & J050652.85-684112.5 & J051753.19-694427.0 & J052534.67-700254.6 & J053646.72-713725.9 \\ 
J045328.26-690859.6 & J050657.66-682947.6 & J051802.26-700534.8 & J052535.52-683612.1 & J053654.42-703746.7 \\ 
J045331.78-695616.8 & J050703.08-685626.7 & J051809.11-695001.3 & J052544.83-690448.8 & J053655.60-681124.6 \\ 
J045337.03-672625.8 & J050705.91-673429.2 & J051811.85-693551.4 & J052547.91-701255.6 & J053704.92-675313.8 \\ 
J045346.73-691548.8 & J050708.96-672618.7 & J051818.38-685735.5 & J052615.92-675110.1 & J053751.09-704823.6 \\ 
J045412.87-701708.4 & J050712.74-682622.0 & J051822.17-711150.4 & J052620.10-693902.1 & J053757.05-700511.8 \\ 
J045439.40-690436.4 & J050738.30-701838.0 & J051834.00-694917.3 & J052623.55-695225.7 & J053804.59-703218.0 \\ 
J045456.74-662940.4 & J050744.83-695501.3 & J051846.47-723844.8 & J052624.49-722051.5 & J053807.90-700137.8 \\ 
J045503.97-685822.0 & J050759.35-683925.8 & J051846.97-715111.3 & J052628.20-690757.6 & J053822.62-710950.8 \\ 
J045511.91-693030.7 & J050803.57-692111.6 & J051850.32-723624.5 & J052630.28-693921.6 & J053840.63-695909.9 \\ 
J045516.04-691912.1 & J050935.85-720039.4 & J051852.64-712545.2 & J052631.33-674627.3 & J053845.51-703937.1 \\ 
J045525.77-693837.9 & J050937.01-713612.5 & J051857.56-693459.9 & J052645.52-700332.9 & J053857.96-712409.2 \\ 
J045529.99-692910.5 & J050951.35-691711.2 & J051903.24-693955.2 & J052653.23-690438.3 & J053910.19-695916.8 \\ 
J045541.81-692624.2 & J050957.27-692033.2 & J051904.18-692318.0 & J052659.39-715011.7 & J053925.59-705106.5 \\ 
J045548.25-692406.5 & J051006.54-711340.2 & J051907.05-680302.2 & J052708.87-694334.5 & J053938.30-705623.5 \\ 
J045615.77-682042.3 & J051032.63-675848.3 & J051922.25-684931.2 & J052710.28-691617.5 & J053939.25-705724.5 \\ 
J045617.44-705942.4 & J051040.59-704809.2 & J051939.42-692527.3 & J052712.54-691300.4 & J053955.21-705328.9 \\ 
J045620.06-694000.6 & J051045.38-693052.3 & J051953.27-692733.6 & J052714.72-690755.7 & J054011.30-693003.0 \\ 
J045626.51-692350.6 & J051053.27-702646.3 & J051953.74-692253.7 & J052716.32-700205.0 & J054026.90-703202.9 \\ 
J045627.67-661111.0 & J051106.94-693443.2 & J051953.89-691014.9 & J052728.20-700426.5 & J054057.50-711125.5 \\ 
J045628.26-694037.0 & J051117.62-695809.8 & J052001.20-694729.9 & J052735.26-704437.7 & J054101.93-704311.0 \\ 
J045652.11-664633.9 & J051133.51-694854.1 & J052008.84-674749.9 & J052738.58-692843.9 & J054114.56-713236.0 \\ 
J045656.46-692405.3 & J051134.56-700039.1 & J052010.69-702215.0 & J052743.11-695306.3 & J054130.47-705801.7 \\ 
J045654.12-655918.7 & J051136.14-690317.9 & J052010.76-702322.5 & J052745.02-713228.2 & J054141.68-694043.9 \\ 
J045659.81-684952.4 & J051144.12-692641.8 & J052014.17-684737.9 & J052754.00-702437.0 & J054159.06-694301.3 \\ 
J045717.06-685806.3 & J051146.79-691813.2 & J052018.93-692253.0 & J052802.15-700140.3 & J054203.64-701326.7 \\ 
J045718.17-680534.7 & J051147.36-691714.6 & J052020.24-694607.0 & J052821.82-701037.7 & J054211.08-713513.8 \\ 
J045721.08-692340.6 & J051147.38-685631.9 & J052024.96-685308.2 & J052825.40-702303.1 & J054254.38-700807.4 \\ 
J045727.05-664408.8 & J051154.70-690941.9 & J052032.32-721348.8 & J052836.59-685829.9 & J054257.21-703948.8 \\ 
J045732.39-713808.5 & J051213.54-683922.8 & J052041.11-723858.0 & J052842.48-693755.1 & J054305.81-712205.6 \\ 
J045735.16-692927.8 & J051228.72-723956.2 & J052041.72-715126.9 & J052847.68-680658.6 & J054310.43-704232.0 \\ 
J045800.99-670457.4 & J051234.05-691716.2 & J052044.36-715649.6 & J052924.89-684331.4 & J054316.25-714431.9 \\ 
J045817.98-691328.0 & J051243.39-685320.1 & J052059.68-695550.6 & J052925.08-670321.4 & J054323.47-691144.6 \\ 
J045820.57-683251.9 & J051246.61-691129.6 & J052101.66-691417.5 & J052940.53-704246.8 & J054337.27-710040.3 \\ 
J045838.99-685707.5 & J051250.46-685315.1 & J052109.13-700650.9 & J052947.40-701953.7 & J054341.04-710511.1 \\ 
J045847.60-694004.0 & J051308.81-713444.6 & J052111.08-681448.7 & J053017.08-700845.4 & J054346.41-705625.5 \\ 
J045917.75-713040.7 & J051317.88-714841.2 & J052115.64-714805.5 & J053031.73-713933.9 & J054408.06-703932.8 \\ 
J045922.40-715607.0 & J051319.35-694121.3 & J052130.71-712632.6 & J053036.43-701536.1 & J054416.33-690152.6 \\ 
J045926.59-675338.4 & J051329.91-684408.1 & J052140.86-693307.5 & J053046.61-704722.8 & J054418.67-714145.0 \\ 
J045931.60-682023.8 & J051332.89-692142.4 & J052146.65-693642.5 & J053051.67-705528.0 & J054432.41-703622.8 \\ 
J045933.86-684624.2 & J051339.08-703520.2 & J052203.71-723311.5 & J053057.04-702523.9 & J054440.11-691149.0 \\ 
J045940.25-694122.7 & J051343.50-693839.3 & J052206.47-684755.7 & J053059.99-681555.4 & J054455.31-683901.1 \\ 
J045957.29-674813.4 & J051345.11-682731.4 & J052206.92-715017.7 & J053106.69-710934.8 & J054519.91-701353.2 \\ 
J050015.77-681425.8 & J051348.17-675834.5 & J052214.10-691332.8 & J053106.84-700437.5 & J054553.31-700725.2 \\ 
J050040.73-713456.0 & J051349.83-683011.4 & J052217.34-690512.9 & J053110.17-700843.5 & J054558.84-690731.2 \\ 
J050041.33-680604.4 & J051350.28-692947.2 & J052224.81-693201.9 & J053119.11-680412.5 & J054559.01-712456.4 \\ 
J050046.25-690500.5 & J051407.12-693958.2 & J052228.85-712055.1 & J053119.29-675326.7 & J054652.92-705859.6 \\ 
J050116.51-655654.9 & J051426.15-683608.4 & J052234.72-701854.0 & J053127.27-700255.8 & J054658.98-704517.9 \\ 
J050129.50-715649.1 & J051429.40-694002.9 & J052234.74-720743.9 & J053134.58-675522.4 & J054700.48-711003.9 \\ 
J050140.66-702549.3 & J051433.96-684933.7 & J052249.28-683213.9 & J053200.75-695452.4 & J054736.13-704605.5 \\ 
J050141.84-662806.7 & J051438.23-685709.6 & J052251.57-695807.9 & J053204.91-713738.5 & J054739.30-701351.1 \\ 
J050147.53-671009.5 & J051439.86-693820.7 & J052254.97-693651.7 & J053216.47-700827.2 & J054814.65-691856.3 \\ 
J050147.92-714144.7 & J051440.24-700151.0 & J052300.57-703127.6 & J053219.19-701335.5 & J054823.30-710555.9 \\ 
J050227.38-705739.1 & J051446.41-702055.1 & J052302.73-692037.1 & J053219.38-712356.2 & J054930.35-684642.8 \\ 
J050247.52-690203.5 & J051447.35-675743.0 & J052310.12-675006.4 & J053222.15-703655.5 & J055016.28-694249.8 \\ 
J050251.08-671530.9 & J051447.36-684357.7 & J052323.67-695540.1 & J053224.12-710425.0 & J055019.85-700549.0 \\ 
J050304.82-685510.6 & J051452.67-680625.7 & J052325.58-693449.8 & J053236.99-695453.8 & J055131.97-690227.7 \\ 
J050305.60-691742.1 & J051459.63-694025.4 & J052337.54-692939.0 & J053239.93-704641.4 & J055240.26-693530.7 \\ 
J050308.87-711610.6 & J051502.31-701442.2 & J052342.40-702351.9 & J053242.91-702427.2 & J055424.19-694121.0 \\ 
J050313.24-690122.3 & J051516.40-693306.4 & J052346.65-693511.3 & J053250.60-700803.1 & \\ 
J050318.82-702545.2 & J051541.44-695720.8 & J052349.50-702645.0 & J053305.49-715123.9 & \\ 
J050327.66-701517.9 & J051554.95-693049.5 & J052350.38-673234.3 & J053321.08-691637.9 & \\ 
\hline
\label{tab:M}
\end{tabular}}
\end{table*}
\begin{table*}
\scriptsize{
{\renewcommand{\arraystretch}{0.8}
\caption{The new planetary nebula candidates and previously known planetary nebulae in our sample.}
\medskip
\tabcolsep=2pt
\begin{tabular}{clcl}
\hline
Name & Previous Identification & Name & Previous Identification \\ 
\hline
\multicolumn{4}{c}{Uncatalogued planetary nebula candidates}\\ 
\hline
J045108.87-684905.3 & - & J053142.45-683245.3 & YSO/G$^{5}$ \\                         
J045703.49-684513.5 & YSO$^{5}$  & J053208.55-674004.0 & YSO/G$^{5}$ \\                          
J045811.72-662211.3 & YSO$^{5}$,YSO-PE$^{20}$ & J053355.00-672705.5 & G/YSO$^{5}$,QSO$^{9}$ \\               
J050305.59-683337.0 & - & J053609.54-691805.4 & YSO$^{5}$ \\                           
J050432.15-704412.0 & YSO$^{5}$,YSO-hp$^{28}$ & J053615.83-693151.1 & YSO$^{5}$,YSO-PE$^{20}$,YSO-hp$^{28}$ \\ 
J050558.68-693953.6 & YSO$^{5}$  & J053705.13-691406.0 & - \\                                    
J050849.56-684405.1 & - & J053845.34-690251.3 & YSO$^{5}$ \\                         
J050937.01-675508.8 & YSO/G$^{5}$  & J053853.87-690931.2 & YSO$^{5}$,YSO-PE$^{20}$ \\              
J051829.16-691458.2 & YSO$^{5}$  & J053932.39-690005.8 & - \\                                    
J052147.07-675656.6 & YSO$^{5}$,YSO-PE$^{20}$,YSO-hp$^{28}$ & J053938.25-685740.2 & G$^{5}$,YSO$^{28}$ \\                  
J052559.86-671019.4 & YSO$^{5}$,YSO$^{9}$,YSO-hp$^{28}$ & J054000.73-694713.4 & YSO$^{5}$,YSO-PE$^{20}$ \\             
J052609.17-690059.2 & - & J054046.97-700308.7 & - \\                                   
J052633.00-672705.1 & YSO$^{5}$,QSO$^{9}$ & J054142.68-711928.3 & YSO$^{5}$,YSO$^{9}$ \\                 
J052708.12-685230.5 & YSO$^{5}$  & J054309.31-694454.4 & YSO$^{28}$ \\                           
J052852.95-710920.3 & YSO$^{5}$  & J054519.52-711606.9 & Evolved$^{28}$,LPV$^{4}$ \\             
J053132.96-701827.7 & YSO$^{5}$  & J054855.15-700510.7 & YSO$^{5}$,YSO$^{9}$,YSO-hp$^{28}$ \\
\hline
\multicolumn{4}{c}{Previously identified planetary nebulae}\\ 
\hline
J044829.55-690813.0 & PN$^{28}$,QSO$^{9}$ & J052456.62-691530.9 & PN$^{7,28}$ \\                 
J045013.12-693356.9 & PN$^{5,7,28}$,YSO$^{9}$ & J052525.97-685553.8 & PN$^{5,28}$ \\                                                                        
J045201.45-683917.1 & PN$^{7,28}$ & J052743.78-712556.5 & PN$^{5,7,16,28}$ \\                                                                  
J045959.97-702741.2 & PN$^{5,7,16}$ & J052801.54-701331.6 & PN$^{7,16}$ \\                                                                       
J050052.64-701341.0 & PN$^{5,7,16}$ & J052804.80-685947.2 & YSO$^{5}$,YSO-PE$^{20}$,PN$^{28}$ \\                                                 
J050058.17-680748.9 & PN$^{16}$ & J052834.40-703302.7 & PN$^{7,16}$ \\                                                                       
J050341.21-701353.5 & PN$^{5,7,16}$ & J052841.01-673339.2 & PN$^{7,16,28}$ \\                                                                    
J050342.60-700647.4 & PN$^{7,16,28}$ & J052902.79-701924.2 & PN$^{7,16,28}$,QSO$^{9}$ \\                                                          
J050427.64-685811.5 & PN$^{7,16}$ & J052915.66-673247.4 & C$^{8}$,PN$^{5,7,16,28}$ \\                                                          
J050434.20-675221.8 & PN$^{16,28}$,UNK$^{29}$,YSO$^{9}$ & J052918.33-702349.9 & PN$^{7,16}$ \\                                                                       
J050451.97-683909.7 & PN$^{5}$ & J052926.57-723843.1 & PN$^{7}$ \\                                                                          
J050609.37-674527.5 & PN$^{5}$ & J052932.73-701738.7 & PN$^{7,16}$ \\                                                                       
J050623.88-690319.3 & PN$^{5,7,16}$ & J053033.17-704437.9 & PN$^{5}$,PN$^{7,16}$ \\                                                              
J050643.81-691538.1 & PN$^{7,16}$ & J053055.57-672006.2 & YSO$^{5}$,YSO-P$^{20}$,PN$^{16,28}$,yso-L$^{19}$ \\                                  
J050730.70-690808.6 & PN$^{5,16,28}$ & J053059.48-683541.2 & PN$^{5,16,28}$\\                                                                     
J050757.61-685147.5 & PN$^{28}$,YSO$^{9}$ & J053108.97-713640.6 & PN$^{7}$ \\                                                                           
J050803.34-684016.7 & PN$^{5,28}$ & J053115.90-710508.1 & YSO/G$^{5}$,PN$^{10}$,YSO$^{9}$,YSO-hp$^{28}$ \\                                     
J050911.13-673402.6 & PN$^{7,16}$ & J053121.88-704045.3 & PN$^{5,7,16}$,RRLab$^{1}$ \\ 
J050920.18-674725.0 & PN$^{5,7,16,28}$ & J053308.84-711803.3 & PN$^{16}$ \\                                                                          
J050937.20-704908.6 & PN$^{5,7,16,28}$,QSO$^{9}$ &  J053313.06-723647.2 & PN$^{7}$ \\                                                                            
J051009.41-682954.6 & PN$^{5,7,16,28}$ & J053329.77-715228.7 & PN$^{5,7}$\\                                                                          
J051017.11-684822.6 & PN$^{5,7,16,28}$ & J053346.97-683644.2 & C-PN$^{29}$,PN$^{5,7,16,28}$ \\                                                       
J051039.65-683604.8 & PN$^{5,7,16}$ & J053356.10-675308.9 & PN$^{5,7,16,28}$ \\                                                                  
J051102.94-674759.2 & PN$^{5,7,16,28}$ & J053406.23-692618.4 & PN$^{5,7,12,16,28}$ \\                                                                
J051141.99-683459.8 & PN$^{5,7,28}$ & J053421.19-685825.2 & PN$^{5,7,12,16,28}$ \\                                                               
J051327.23-703335.0 & PN$^{7,16}$ & & \\                                                               
J051342.30-681516.4 & PN$^{7,16}$ & J053430.21-702834.8 & PN$^{7,16}$ \\                                                                       
J051546.75-684223.5 & PN$^{7,16,28}$ & J053438.84-701956.7 & PN$^{7,16}$ \\                                                                       
J051702.45-690716.5 & PN$^{10,16,28}$,QSO$^{9}$ &  J053448.02-684835.7 & PN$^{12,16,28}$ \\                                                                     
J051743.59-711533.9 & PN$^{10}$,YSO$^{9}$,YSO-hp$^{28}$ & J053510.23-693939.1 & PN$^{7,12,16,28}$ \\                                                                 
J051929.62-685109.1 & PN$^{7,16,28}$ & J053557.56-695816.9 & PN$^{5,7,12,16,28}$,QSO$^{9}$ \\                                                     
J051954.62-693104.9 & PN$^{5,7,16,28}$ & J053630.83-691817.3 & YSO$^{5}$,PN$^{20}$,YSO-hp$^{28}$ \\                                                  
J052009.35-702538.1 & PN$^{7,16}$ & J053652.96-715338.6 & PN$^{5,7}$ \\                                                                        
J052009.50-695339.0 & PN$^{16,28}$ & J053710.26-712314.3 & O-PN$^{29}$,PN$^{10,16}$,QSO$^{9}$ \\                                                
J052034.66-683518.2 & YSO$^{5}$,PN$^{16,28}$ &  J053737.93-714137.7 & PN$^{7}$ \\                                                                            
J052052.42-700935.5 & C-PN$^{29}$,PN$^{5,7,16,28}$ &  J054008.62-685826.9 & LPV/HII$^{12}$,PN$^{16,28}$,EclBin$^{19}$ \\                                           
J052056.03-700513.1 & PN$^{5,16,28}$ & J054045.04-702806.9 & YSO$^{5}$,HII$^{17}$,YSO-PE$^{20}$,PN$^{16,28}$,YSO$^{9}$ \\                       
J052123.85-683533.8 & PN$^{5,7,16,28}$ & J054224.02-695305.1 & PN$^{12,28}$ \\                                                                       
J052142.85-683924.9 & PN$^{7,16}$ & J054233.17-702924.1 & O-PN$^{29}$,PN$^{7,16,28}$,QSO$^{9}$ \\                                              
J052212.80-694329.1 & PN$^{5,7,16,28}$,YSO$^{9}$ &  J054236.65-700932.0 & C-PN$^{29}$,PN$^{5,28}$ \\                                                             
J052235.20-682425.9 & PN$^{7,16,28}$ & & \\                                                           
J052240.97-711906.6 & PN$^{5,7,16,28}$ & J054434.71-702140.7 & PN$^{5,7,16,28}$,YSO$^{9}$ \\                                                         
J052331.11-690404.6 & PN$^{5,7,10,16,28}$,YSO$^{9}$ &  J054500.14-691819.1 & PN$^{16,28}$,Symb$^{19}$ \\                                                            
J052348.66-691222.0 & PN$^{7,16}$ & J054558.49-711903.9 & Evolved$^{28}$,PN$^{10}$,QSO$^{9}$ \\                                                
J052420.75-700501.3 & PN$^{5,7,10,16,28}$,YSO$^{9}$ &  J054704.54-692733.9 & O-PN$^{29}$,PN$^{5,28}$ \\                                                             
J052427.26-702223.9 & PN$^{7,16}$,QSO$^{9}$ & J054938.74-691000.1 & PN$^{7,16}$ \\                                                                        
J052455.07-713256.3 & PN$^{5,7,16,28}$ &  & \\
\hline
\label{tab:PN}
\end{tabular}}}
\begin{flushleft}
Notes: The column ''Previous identifications'' gives the result of a
positional cross-matching that was done with the catalogues mentioned
in Table~\ref{tab:pagb_param}. Catalogue identifications: YSO $=$young
stellar object, YSO-PE $=$ YSO with PAH emission features and
fine-structure lines, YSO-P $=$ YSO with PAH emission features YSO-hp $=$ high probability YSO candidate, G $=$
background galaxy, QSO $=$ quasi-stellar object, Evolved $=$
evolved object, LPV $=$ long period variable, PN $=$
planetary nebula, C-PN $=$ carbon-rich PN, O-PN $=$ oxygen-rich PN,
LPV/HII $=$ long period variable/ HII region, EclBin $=$ eclipsing binary, C
$=$ carbon star, Symb $=$ Symbiotic, RRLab $=$ RR Lyrae, UNK $=$ object
of unknown type. See catalogued references for full details of the individual object identification.
\end{flushleft}
\end{table*}

\clearpage
 
\section{Objects without $B$ and $I$ magnitudes}
\label{nobv}

In this section, we present the group of 85 objects, from the sample of
581 possible post-AGB/post-RGB and YSO candidates with confirmed LMC
membership, which did not have both $B$ and $I$ magnitudes. These
objects were fed into the STP (see Subsection~\ref{STP}) and their
stellar parameters (\teff,\logg, and [Fe/H]) were determined, but there were not enough data
points to estimate the value of E($B-V$) that minimised the sum of the 
squared differences between the de-reddened observed and the intrinsic 
$B$, $V$, $I$ and $J$ magnitudes and therefore derived the
reddening. We have removed this sample of 85 objects from further
analysis and we carried forth the analysis with a sample of 496 possible post-AGB/post-RGB and YSO
candidates. In Table~\ref{tab:nobv} we present a list of these 85 objects along with
their derived \teff, \logg, [Fe/H], the $V$ band magnitude, the observed
luminosity ($L_{\rm ob}$), and the estimated radial velocity.

\onecolumn
\begin{ThreePartTable}
\small{
\renewcommand{\arraystretch}{1.0}
\medskip
\tabcolsep=1.0pt
\LTcapwidth=\textwidth
\begin{longtable}{lrrrccl>{\scriptsize}l}
\caption{The observational and stellar parameters for the 85 objects
  without $B$ and $V$ magnitudes. \label{tab:nobv}}\\
\hline
Name & $T_{\rm eff}$\,(K) & $\log g$ & [Fe/H] & $V$(mags) & ($L_{\rm ob}$/L$_\odot$) & RV\,(km/s) \\ 
\hline
\endfirsthead
\caption{continued.}\\
\hline
Name & $T_{\rm eff}$\,(K) & $\log g$ & [Fe/H] & $V$(mags) & ($L_{\rm ob}$/L$_\odot$) & RV\,(km/s) \\ 
\hline
\endhead
\hline
\endfoot
\multicolumn{7}{c}{Candidates with [Fe/H] estimates from spectra}\\ 
\hline
J045034.50-701550.1 & 4500 & 0.50 & -0.83 & 15.81 & 1216 & $244 \pm 5$ \\ 
J045129.24-694648.7 & 4929 & 2.07 & -1.02 & 18.10 & 198 & $200 \pm 5$ \\ 
J045304.87-694106.0 & 5617 & 2.93 & -0.91 & 18.16 & 188 & $181 \pm 20$ \\ 
J045835.30-700350.0 & 5250 & 3.50 & 0.50 & 99.99 & 109 & $211 \pm 4$ \\ 
J045906.08-675154.7 & 7632 & 2.00 & -2.50 & 99.99 & 84 & $319 \pm 15$ \\ 
J045931.59-701223.7 & 6179 & 3.21 & -2.50 & 17.23 & 208 & $250 \pm 20$ \\ 
J050131.83-704046.2 & 5546 & 2.23 & -2.50 & 18.16 & 244 & $286 \pm 18$ \\ 
J050247.50-703452.4 & 5386 & 2.17 & -2.50 & 17.82 & 266 & $288 \pm 16$ \\ 
J050504.78-675128.6 & 4500 & 1.42 & -0.70 & 17.50 & 404 & $253 \pm 5$ \\ 
J050906.53-710305.3 & 9520 & 2.00 & 0.09 & 17.96 & 226 & $213 \pm 11$ \\ 
J050931.87-682935.4 & 7626 & 3.50 & 0.50 & 17.27 & 175 & $231 \pm 20$ \\ 
J050939.63-710549.1 & 4759 & 2.46 & -0.71 & 18.02 & 194 & $251 \pm 9$ \\ 
J051109.63-723719.8 & 5611 & 3.38 & -2.50 & 99.99 & 141 & $285 \pm 14$ \\ 
J051200.73-692150.9 & 5747 & 3.00 & -1.24 & 17.47 & 348 & $240 \pm 6$ \\ 
J051255.05-713628.7 & 4633 & 2.17 & -2.07 & 18.08 & 144 & $177 \pm 12$ \\ 
J051305.83-682625.5 & 5119 & 3.50 & -0.59 & 17.54 & 235 & $241 \pm 12$ \\ 
J051427.55-720447.5 & 5053 & 1.50 & -2.50 & 99.99 & 85 & $249 \pm 6$ \\ 
J051529.30-684129.4 & 4500 & 1.87 & -0.94 & 16.86 & 619 & $301 \pm 5$ \\ 
J051551.41-723107.7 & 4500 & 1.00 & -1.29 & 17.78 & 350 & $258 \pm 3$ \\ 
J051616.97-693248.8 & 5733 & 2.98 & -0.45 & 18.11 & 130 & $262 \pm 5$ \\ 
J051712.18-690309.4 & 5938 & 1.00 & -2.50 & 17.52 & 357 & $255 \pm 7$ \\ 
J051825.71-700532.6 & 4500 & 1.71 & -0.14 & 16.97 & 4614 & $255 \pm 2$ \\ 
J051848.35-693334.7 & 5756 & 2.00 & -1.13 & 17.92 & 4652 & $286 \pm 8$ \\ 
J051853.31-730118.0 & 4500 & 1.40 & -0.74 & 16.48 & 2134 & $209 \pm 2$ \\ 
J051926.64-721604.2 & 5393 & 3.00 & -2.50 & 99.99 & 246 & $241 \pm 19$ \\ 
J052002.01-703017.0 & 5546 & 2.12 & -2.50 & 16.25 & 761 & $210 \pm 15$ \\ 
J052108.47-691158.8 & 4850 & 1.82 & -1.71 & 17.66 & 206 & $250 \pm 8$ \\ 
J052214.24-715731.2 & 4500 & 1.93 & -0.46 & 17.01 & 1530 & $250 \pm 2$ \\ 
J052242.48-672330.4 & 7222 & 4.00 & -0.73 & 18.08 & 94 & $260 \pm 19$ \\ 
J052300.09-701831.8 & 4288 & 0.58 & -1.68 & 17.86 & 233 & $264 \pm 9$ \\ 
J052338.56-724600.5 & 4500 & 2.00 & -1.13 & 18.12 & 9149 & $239 \pm 4$ \\ 
J052444.03-695619.5 & 4757 & 1.83 & -0.60 & 18.00 & 312 & $263 \pm 5$ \\ 
J052744.65-710033.8 & 4750 & 2.65 & -1.87 & 17.49 & 148 & $174 \pm 7$ \\ 
J052756.36-700434.0 & 4500 & 1.16 & -1.08 & 17.62 & 824 & $246 \pm 8$ \\ 
J052816.85-693017.0 & 8246 & 1.00 & 0.50 & 17.61 & 9045 & $275 \pm 3$ \\ 
J052856.67-674237.3 & 5620 & 1.88 & -2.13 & 17.98 & 135 & $212 \pm 17$ \\ 
J052906.64-682013.3 & 5075 & 2.29 & -2.50 & 17.37 & 519 & $234 \pm 10$ \\ 
J052918.18-702516.4 & 5439 & 2.28 & -0.94 & 99.99 & 64 & $274 \pm 8$ \\ 
J053024.31-704555.8 & 10809 & 2.00 & 0.50 & 99.99 & 143 & $265 \pm 16$ \\ 
J053108.81-701759.4 & 5865 & 2.59 & -1.53 & 99.99 & 189 & $289 \pm 8$ \\ 
J053136.67-710650.8 & 10500 & 2.00 & 0.50 & 16.74 & 632 & $277 \pm 1$ \\ 
J053140.01-710911.0 & 5774 & 2.50 & -2.50 & 15.29 & 967 & $266 \pm 20$ \\ 
J053246.45-671028.3 & 7628 & 3.50 & 0.50 & 99.99 & 372 & $195 \pm 13$ \\ 
J053255.47-672944.4 & 5606 & 3.00 & -2.50 & 99.99 & 455 & $180 \pm 22$ \\ 
J053327.17-700951.5 & 4867 & 2.00 & -1.17 & 18.00 & 8371 & $276 \pm 6$ \\ 
J053447.57-715908.0 & 4500 & 1.69 & -1.04 & 99.99 & 396 & $262 \pm 2$ \\ 
J053545.26-705720.8 & 6900 & 3.50 & 0.50 & 18.17 & 102 & $230 \pm 7$ \\ 
J053605.89-713053.5 & 4750 & 2.71 & -0.55 & 99.99 & 63 & $287 \pm 8$ \\ 
J053956.85-703214.6 & 5750 & 0.50 & -1.62 & 99.99 & 4019 & $256 \pm 13$ \\ 
J054024.67-702722.7 & 4500 & 1.58 & -0.09 & 17.69 & 598 & $262 \pm 1$ \\ 
J054046.53-704322.2 & 5656 & 0.50 & -1.54 & 15.82 & 1002 & $269 \pm 4$ \\ 
J054236.18-714621.5 & 4500 & 2.66 & -1.50 & 17.88 & 123 & $338 \pm 23$ \\ 
J054237.91-712917.0 & 6036 & 3.00 & -1.21 & 17.89 & 201 & $367 \pm 7$ \\ 
J054311.96-695808.6 & 11500 & 2.50 & 0.50 & 99.99 & 96 & $232 \pm 18$ \\ 
J054354.82-693352.9 & 5248 & 3.00 & -2.09 & 17.21 & 590 & $309 \pm 13$ \\ 
J054506.60-710509.1 & 7626 & 4.00 & -1.18 & 99.99 & 132 & $201 \pm 10$ \\ 
J054643.30-701209.2 & 4628 & 1.00 & -0.86 & 17.28 & 415 & $295 \pm 6$ \\ 
J054738.16-695106.7 & 5799 & 3.00 & -2.50 & 16.30 & 562 & $357 \pm 9$ \\ 
J045451.08-700739.8 & 5518 & 3.50 & -0.99 & 99.99 & 91 & $277 \pm 17$ \\ 
J050823.36-691706.1 & 5222 & 2.64 & -0.66 & 17.09 & 365 & $290 \pm 6$ \\ 
J052738.69-705219.0 & 6009 & 0.97 & -2.14 & 15.54 & 1806 & $278 \pm 7$ \\ 
J052754.75-714036.4 & 4765 & 2.43 & -0.36 & 99.99 & 98 & $264 \pm 15$ \\ 
J053209.95-713117.2 & 6091 & 3.50 & -0.82 & 99.99 & 161 & $261 \pm 15$ \\ 
J054139.18-702208.2 & 5982 & 1.00 & -0.89 & 17.15 & 465 & $253 \pm 11$ \\ 
J054533.77-702126.6 & 5052 & 2.58 & -0.62 & 17.52 & 283 & $273 \pm 3$ \\ 
\hline
\multicolumn{7}{c}{Candidates with [Fe/H] = -0.50}\\ 
\hline
J053241.04-695751.5 & 5005 & 2.16 & -0.5 & 99.99 & 174 & $244 \pm 4$ \\ 
J045239.31-711345.8 & 7500 & 3.0 & -0.5 & 99.99 & 139 & $300 \pm 8$ \\ 
J045338.12-695139.9 & 7253 & 3.5 & -0.5 & 17.07 & 215 & $200 \pm 14$ \\ 
J045537.16-661403.9 & 27947 & 3.5 & -0.5 & 15.89 & 517 & $332 \pm 10$ \\ 
J050151.70-703941.9 & 32178 & 4.63 & -0.5 & 15.06 & 891 & $200 \pm 21$ \\ 
J050414.05-701015.9 & 31000 & 4.24 & -0.5 & 16.39 & 1931 & $284 \pm 14$ \\ 
J050829.25-674232.5 & 7250 & 1.0 & -0.5 & 99.99 & 209 & $289 \pm 21$ \\ 
J051143.88-715105.7 & 12000 & 2.5 & -0.5 & 17.29 & 199 & $323 \pm 14$ \\ 
J052022.20-724604.3 & 6477 & 2.5 & -0.5 & 99.99 & 124 & $285 \pm 18$ \\ 
J052129.82-723148.1 & 7630 & 4.0 & -0.5 & 99.99 & 119 & $236 \pm 22$ \\ 
J052238.63-712206.5 & 10000 & 2.0 & -0.5 & 16.42 & 608 & $316 \pm 22$ \\ 
J052412.04-715449.9 & 7148 & 3.5 & -0.5 & 99.99 & 252 & $282 \pm 19$ \\ 
J052438.81-680653.9 & 10500 & 2.0 & -0.5 & 18.13 & 126 & $287 \pm 19$ \\ 
J053244.04-681131.6 & 7629 & 2.89 & -0.5 & 99.99 & 73 & $348 \pm 19$ \\ 
J054238.36-693324.6 & 6813 & 2.91 & -0.5 & 99.99 & 68 & $276 \pm 32$ \\ 
J054546.43-710225.1 & 10500 & 2.0 & -0.5 & 99.99 & 149 & $265 \pm 15$ \\ 
J054947.47-684636.5 & 7625 & 3.07 & -0.5 & 99.99 & 116 & $269 \pm 15$ \\ 
J055521.66-694235.1 & 8124 & 2.0 & -0.5 & 99.99 & 118 & $302 \pm 10$ \\ 
J053031.62-712257.7 & 7631 & 4.0 & -0.5 & 99.99 & 105 & $295 \pm 28$ \\ 
J054115.32-695809.9 & 8303 & 1.0 & -0.5 & 99.99 & 849 & $282 \pm 18$\\
\hline
\end{longtable}}
\begin{tablenotes}
\item[]Notes: \teff, \logg, and [Fe/H] are estimated using the STP (see
Section~\ref{STP}). '$V$ is the optical $V$-band magnitude and $L_{\rm
  ob}$/L$_\odot$ is the observed luminosity (see
Section~\ref{specobv}). RV is the estimated heliocentric
velocity (see Section~\ref{rv}). \\
\end{tablenotes}
\end{ThreePartTable}
\twocolumn

\clearpage

\section[]{The final sample of high probability post-AGB, post-RGB and YSO candidates}
\label{seds}

Figures~\ref{fig:pagb_sed}\,$-$~\ref{fig:yso_sed} show the SEDs of the objects 
before and after de-reddening (see Section~\ref{sedanalysis}, for full
details).

\begin{figure*}
\centering
\subfloat{\includegraphics[width=15cm]{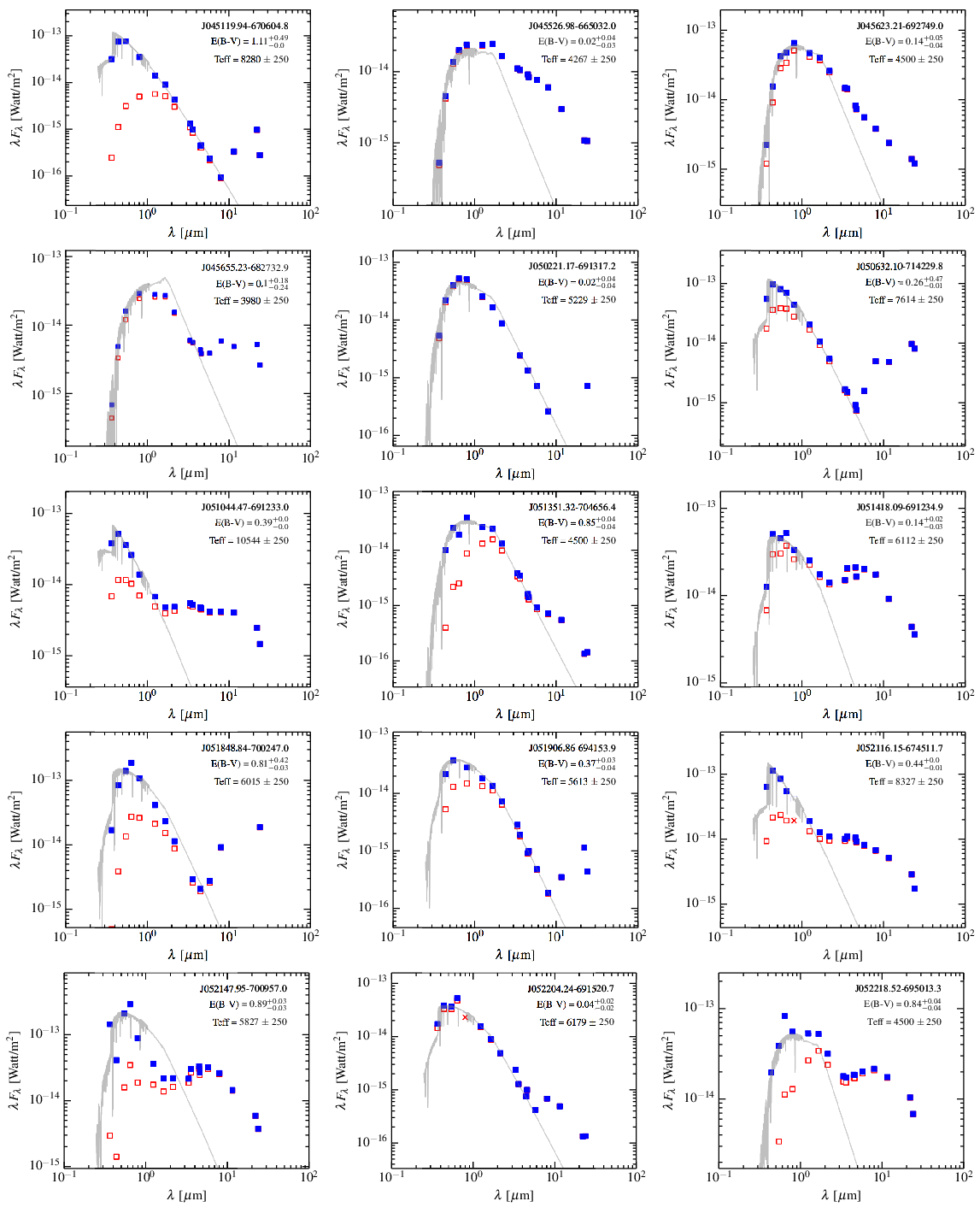}}\,
\caption{SEDs of the post-AGB candidates. The red open squares represent the original broadband 
photometry. The blue filled squares represent the dereddened broadband 
photometry. Up to a wavelength of 10500\AA, we over-plot (grey solid-line) the 
flux-calibrated Munari synthetic spectrum which is estimated to have
the best-fit to the observed spectra (see Section~\ref{STP}). From
10500\AA\, onwards we over-plot the corresponding low-resolution flux distribution
from from the corresponding appropriate ATLAS9 atmospheric model \citep{castelli04}. 
The SED plots also show the name of the individual object, the estimated E(B-V) 
value with error bars (see Section~\ref{reddening}) and the estimated
\teff\, value.}
\label{fig:pagb_sed}
\end{figure*}
\begin{figure*}
\ContinuedFloat
\centering
\subfloat{\includegraphics[width=15cm]{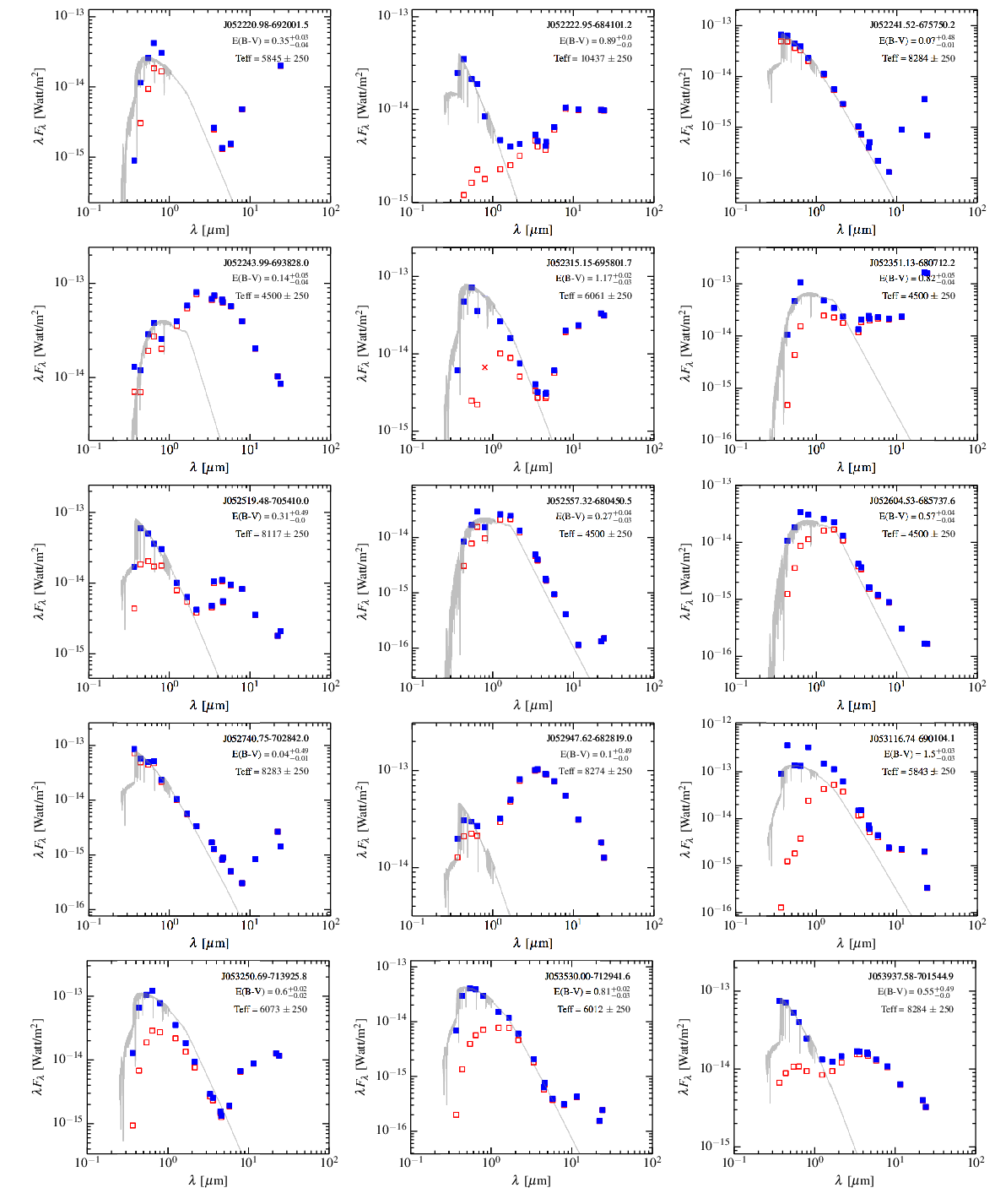}}\,
\caption{Figure~\ref{fig:pagb_sed} continued.}
\end{figure*}
\begin{figure*}
\ContinuedFloat
\centering
\subfloat{\includegraphics[width=15cm]{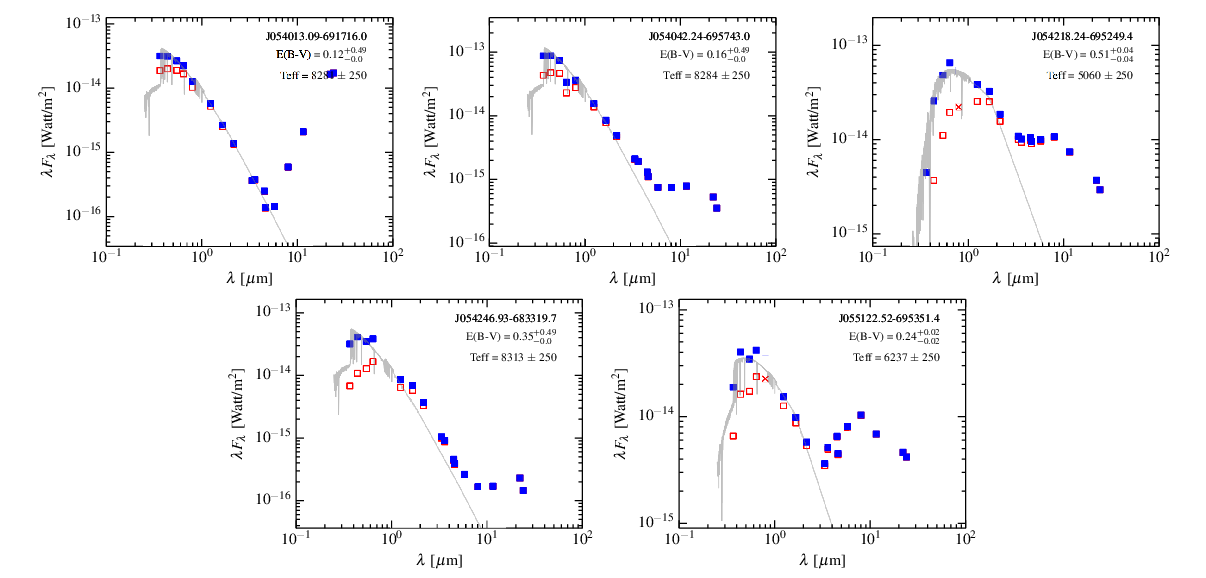}}\,
\caption{Figure~\ref{fig:pagb_sed} continued.}
\end{figure*}

\clearpage
\begin{figure*}
\centering
\subfloat{\includegraphics[width=15cm]{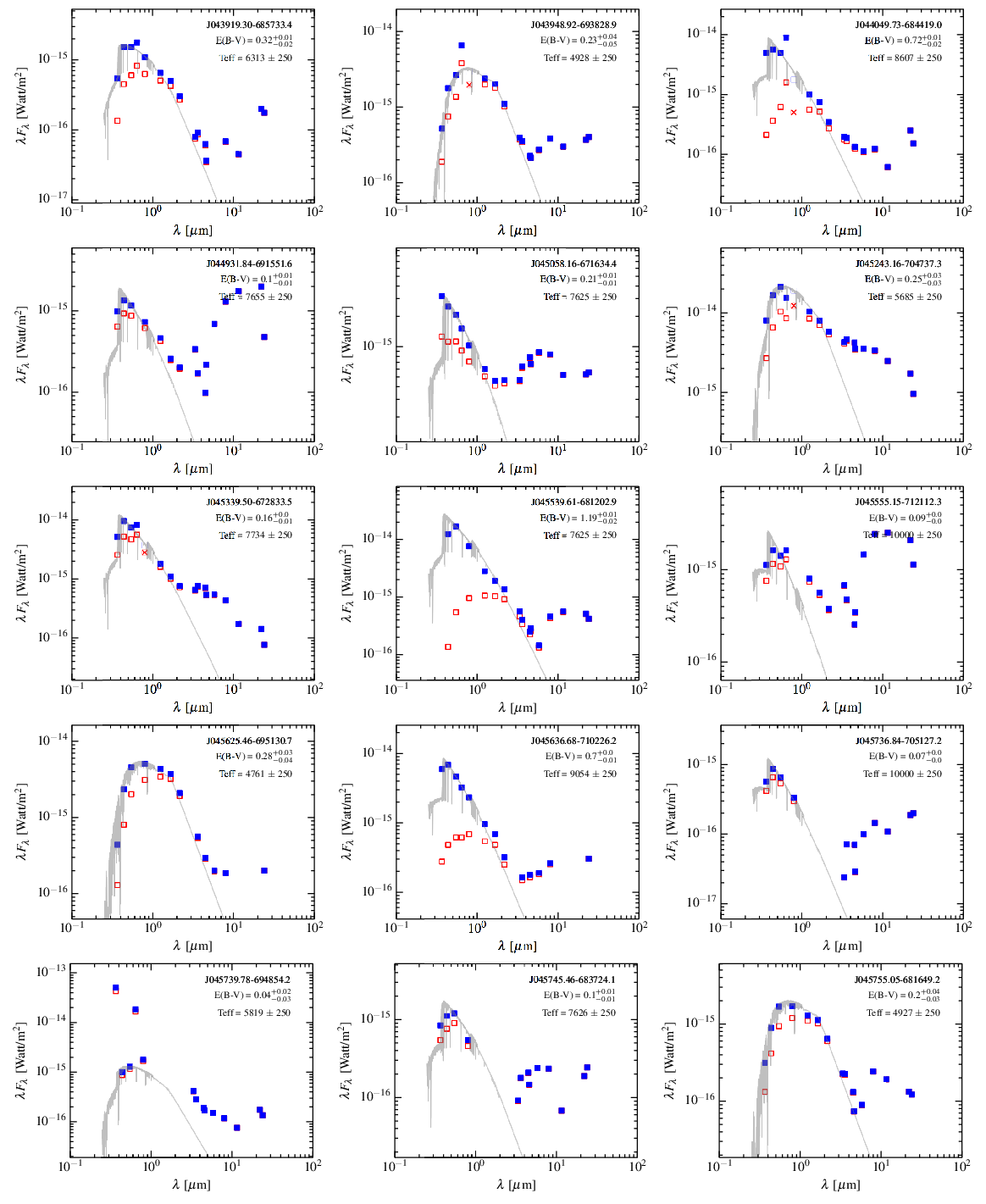}}\,
\caption{Same as Figure ~\ref{fig:pagb_sed}, but for the post-RGB candidates.}
\label{fig:prgb_sed}
\end{figure*}
\begin{figure*}
\ContinuedFloat
\centering
\subfloat{\includegraphics[width=15cm]{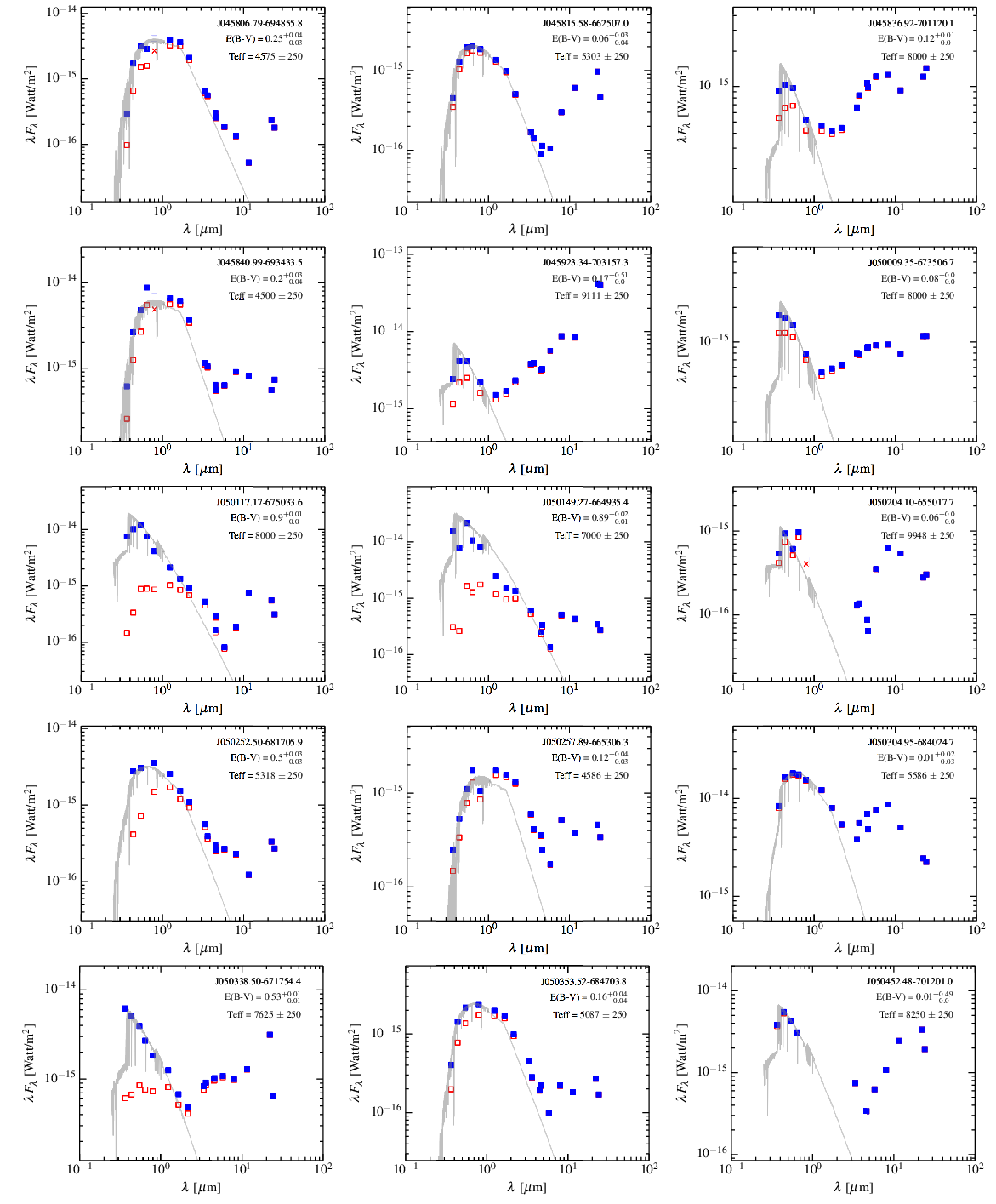}}\,
\caption{Figure ~\ref{fig:prgb_sed} continued.}
\end{figure*}
\begin{figure*}
\ContinuedFloat
\centering
\subfloat{\includegraphics[width=15cm]{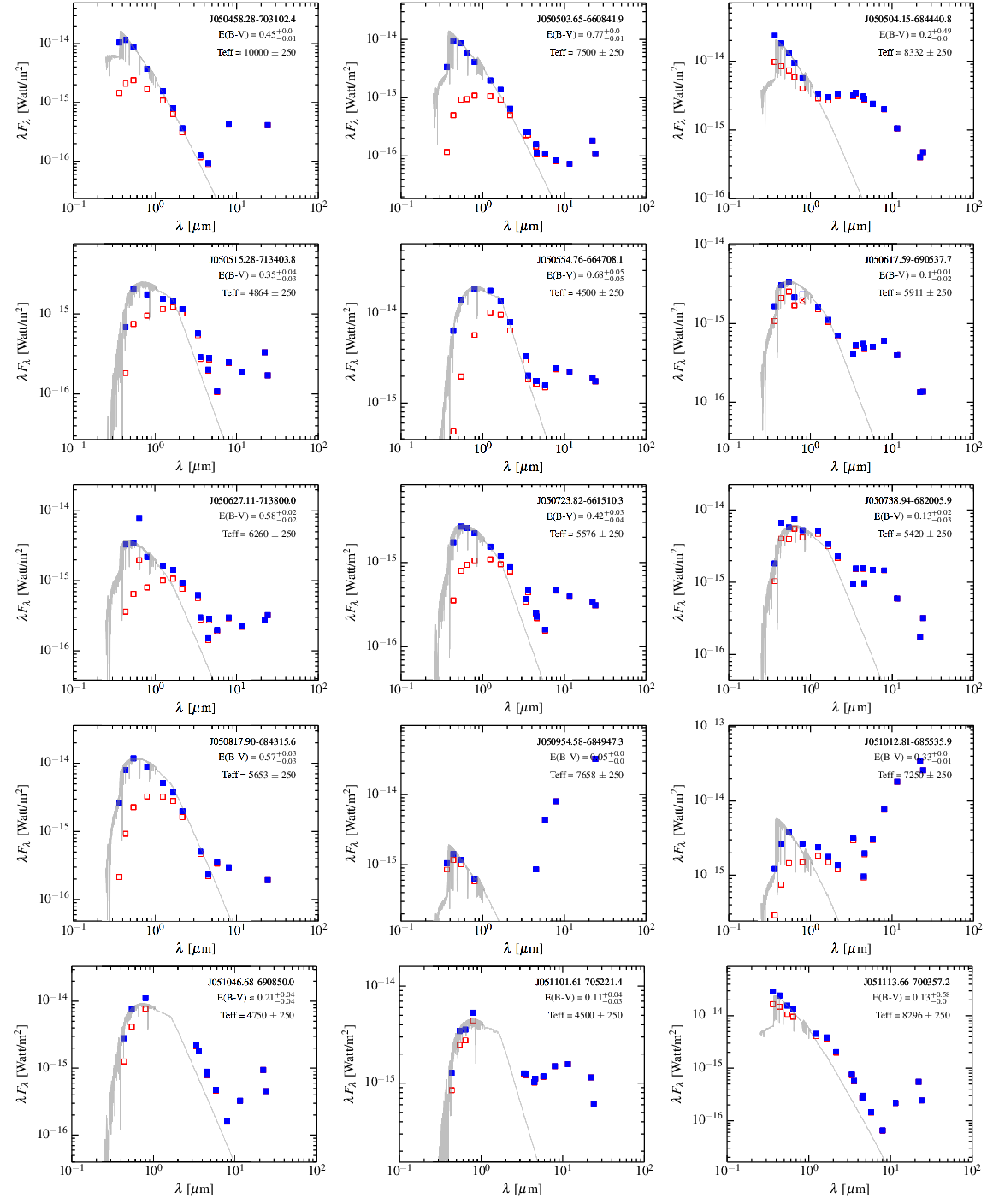}}\,
\caption{Figure ~\ref{fig:prgb_sed} continued.}
\end{figure*}
\begin{figure*}
\ContinuedFloat
\centering
\subfloat{\includegraphics[width=15cm]{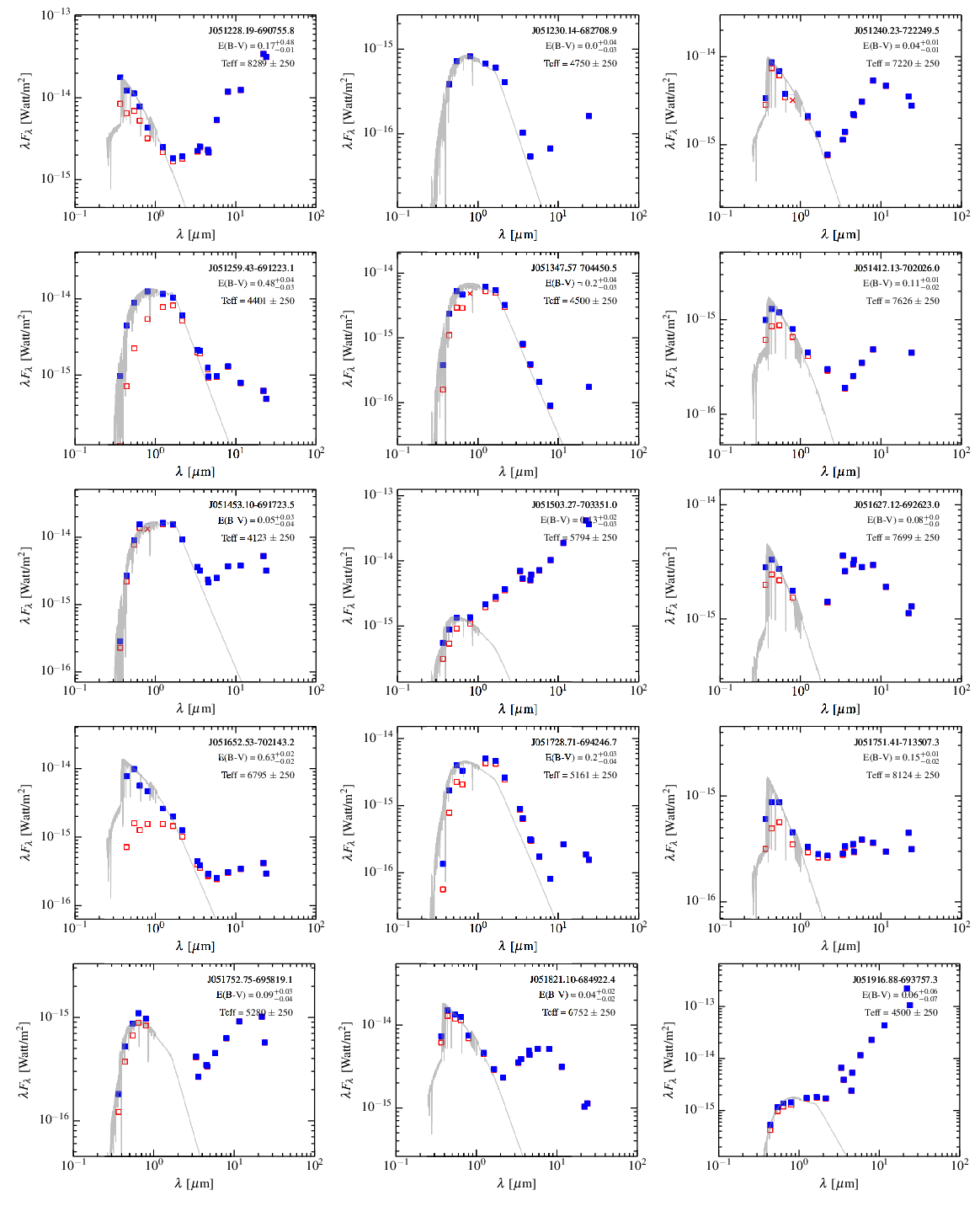}}\,
\caption{Figure ~\ref{fig:prgb_sed} continued.}
\end{figure*}
\begin{figure*}
\ContinuedFloat
\centering
\subfloat{\includegraphics[width=15cm]{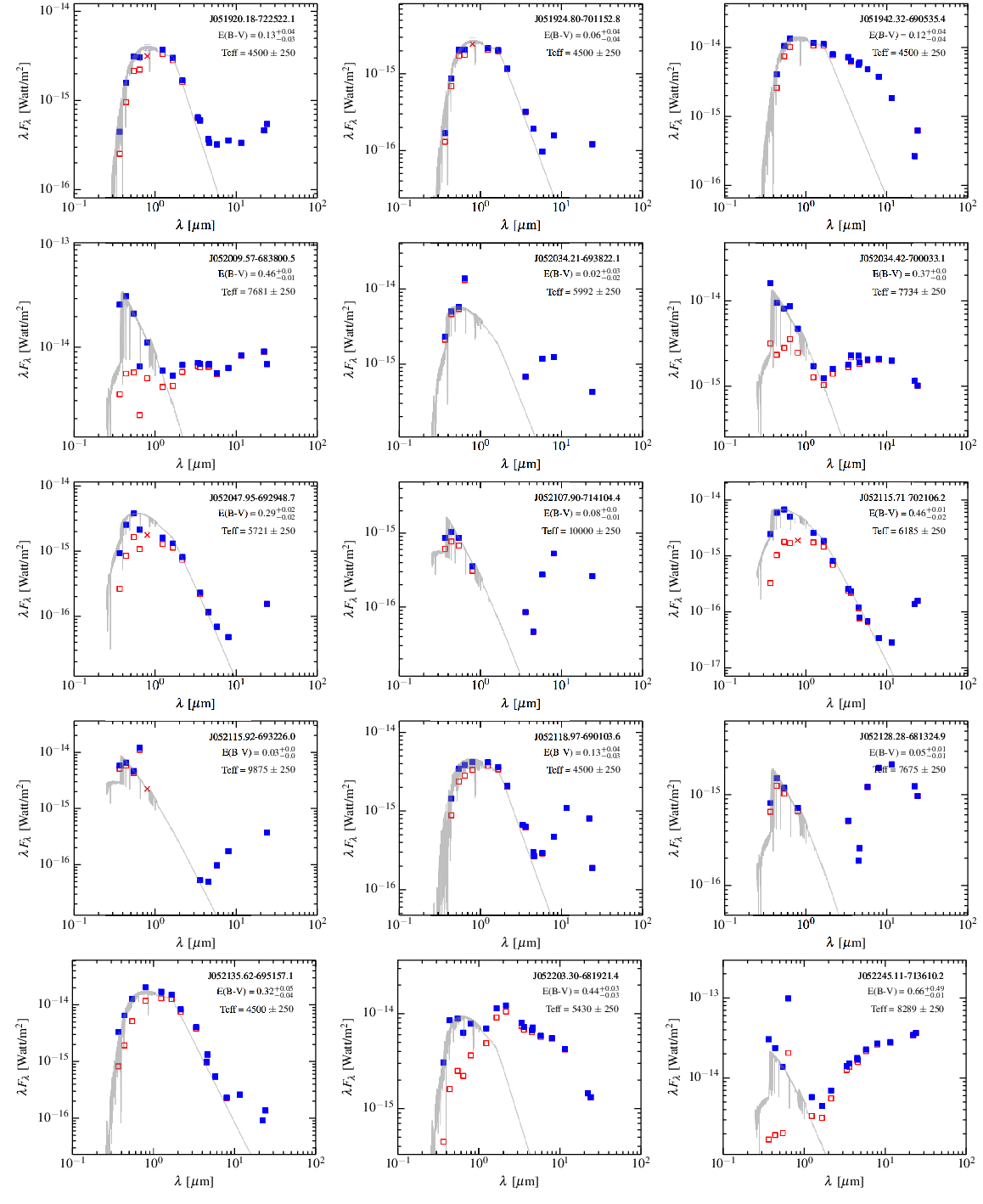}}\,
\caption{Figure ~\ref{fig:prgb_sed} continued.}
\end{figure*}
\begin{figure*}
\ContinuedFloat
\centering
\subfloat{\includegraphics[width=15cm]{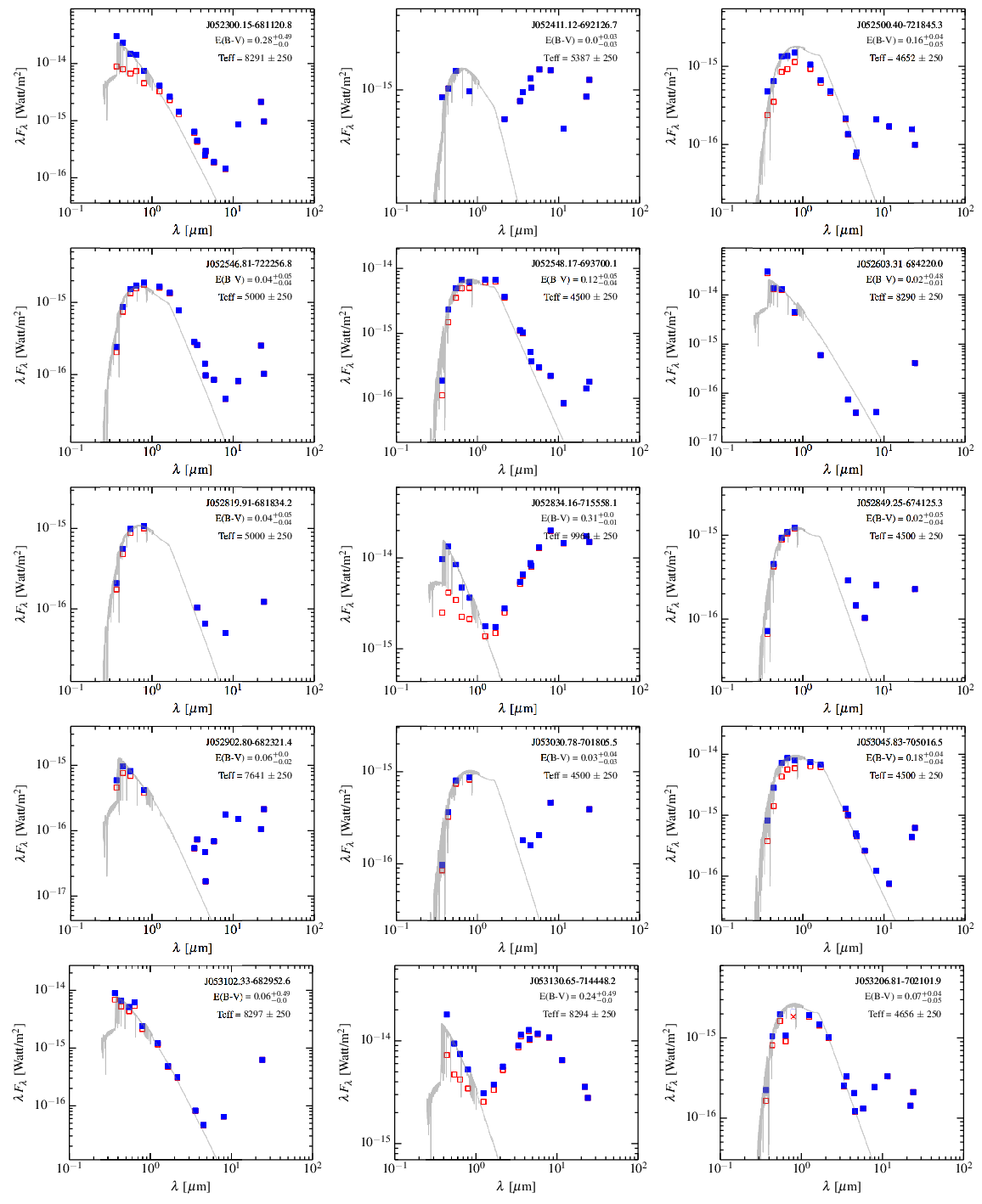}}\,
\caption{Figure ~\ref{fig:prgb_sed} continued.}
\end{figure*}
\begin{figure*}
\ContinuedFloat
\centering
\subfloat{\includegraphics[width=15cm]{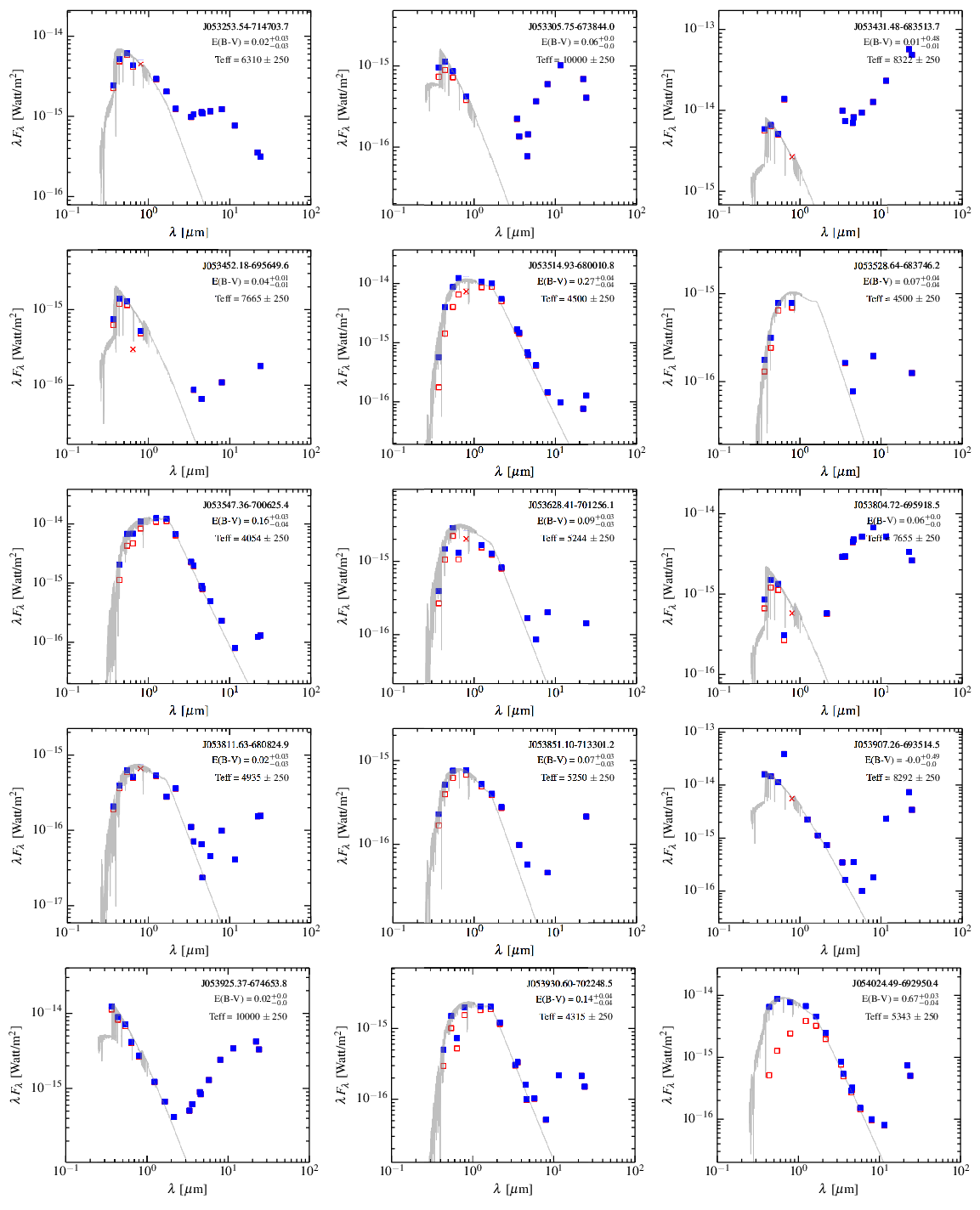}}\,
\caption{Figure ~\ref{fig:prgb_sed} continued.}
\end{figure*}
\begin{figure*}
\ContinuedFloat
\centering
\subfloat{\includegraphics[width=15cm]{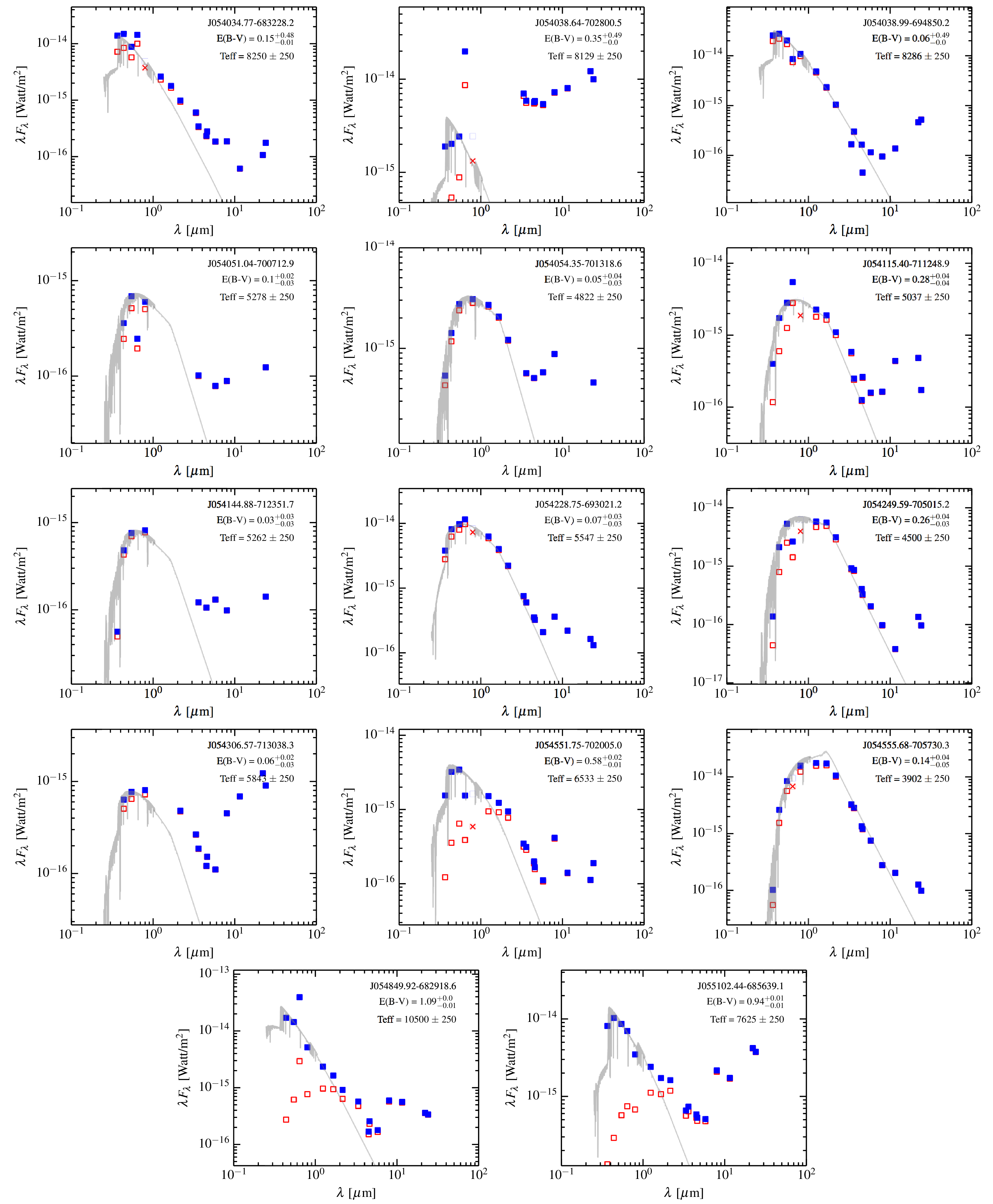}}\,
\caption{Figure ~\ref{fig:prgb_sed} continued.}
\end{figure*}
\clearpage
\begin{figure*}
\centering
\subfloat{\includegraphics[width=15cm]{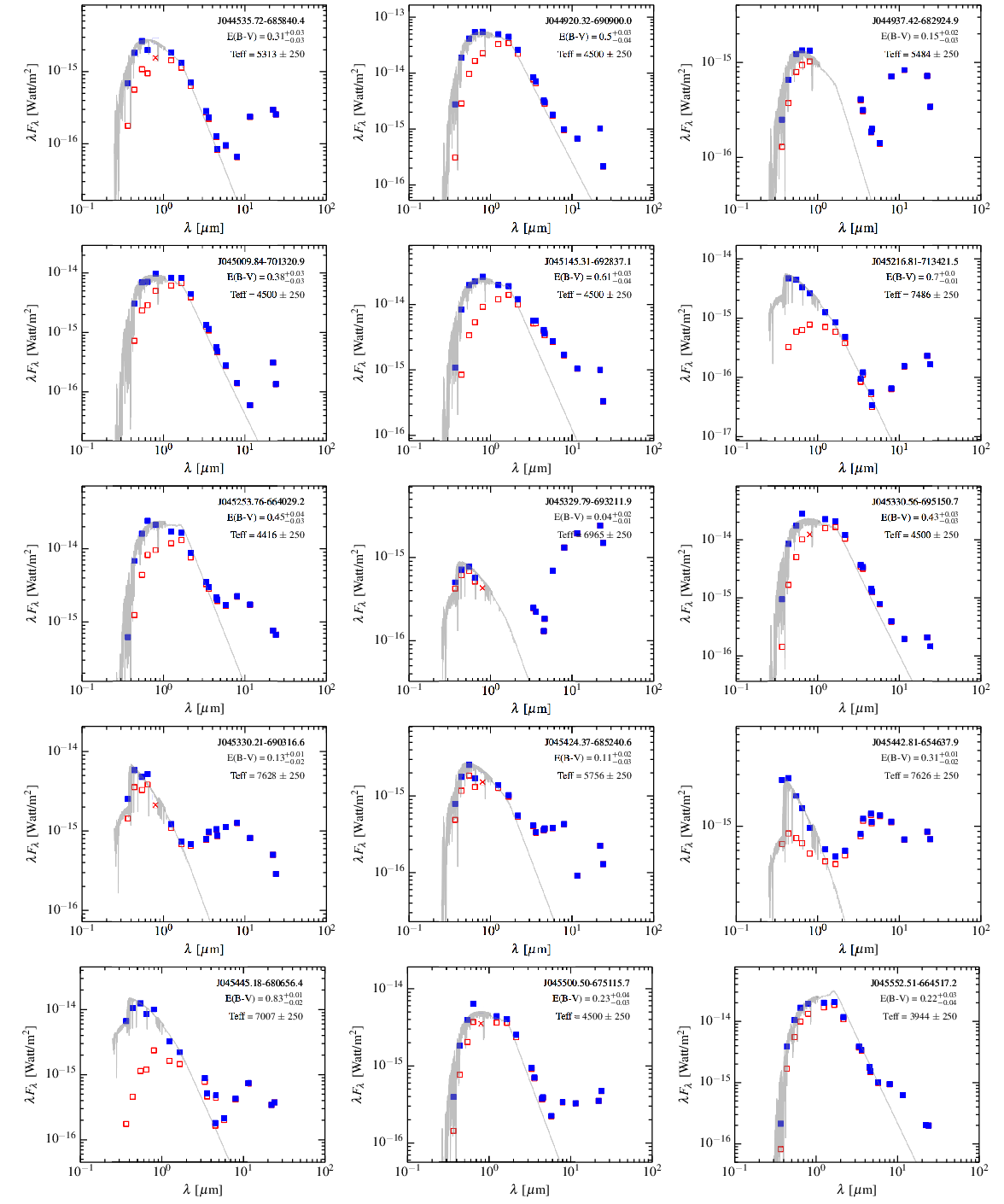}}\,
\caption{Same as Figure ~\ref{fig:pagb_sed}, but for the YSO candidates}
\label{fig:yso_sed}
\end{figure*}
\begin{figure*}
\ContinuedFloat
\centering
\subfloat{\includegraphics[width=15cm]{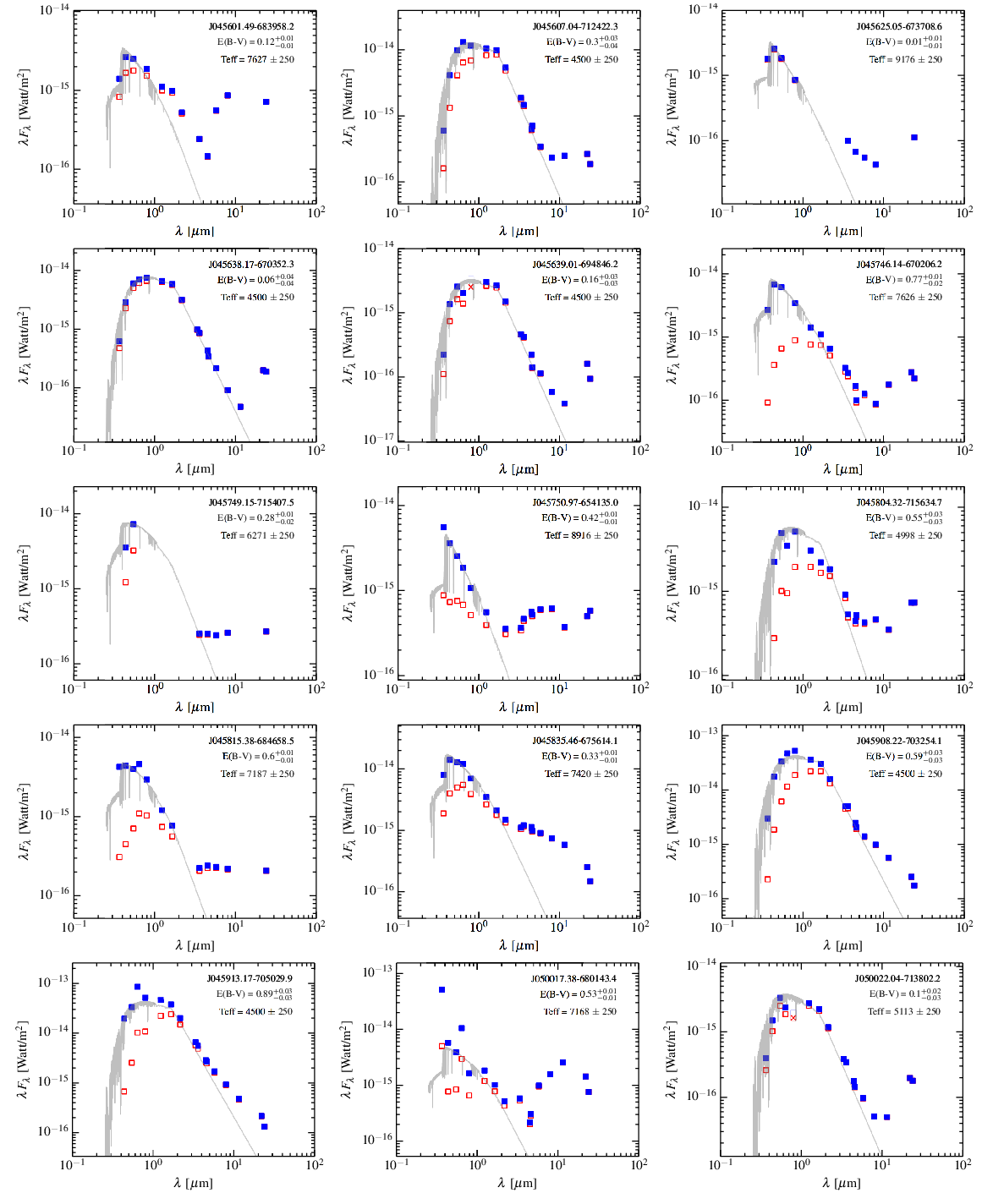}}\,
\caption{Figure~\ref{fig:yso_sed} continued.}
\end{figure*}
\begin{figure*}
\ContinuedFloat
\centering
\subfloat{\includegraphics[width=15cm]{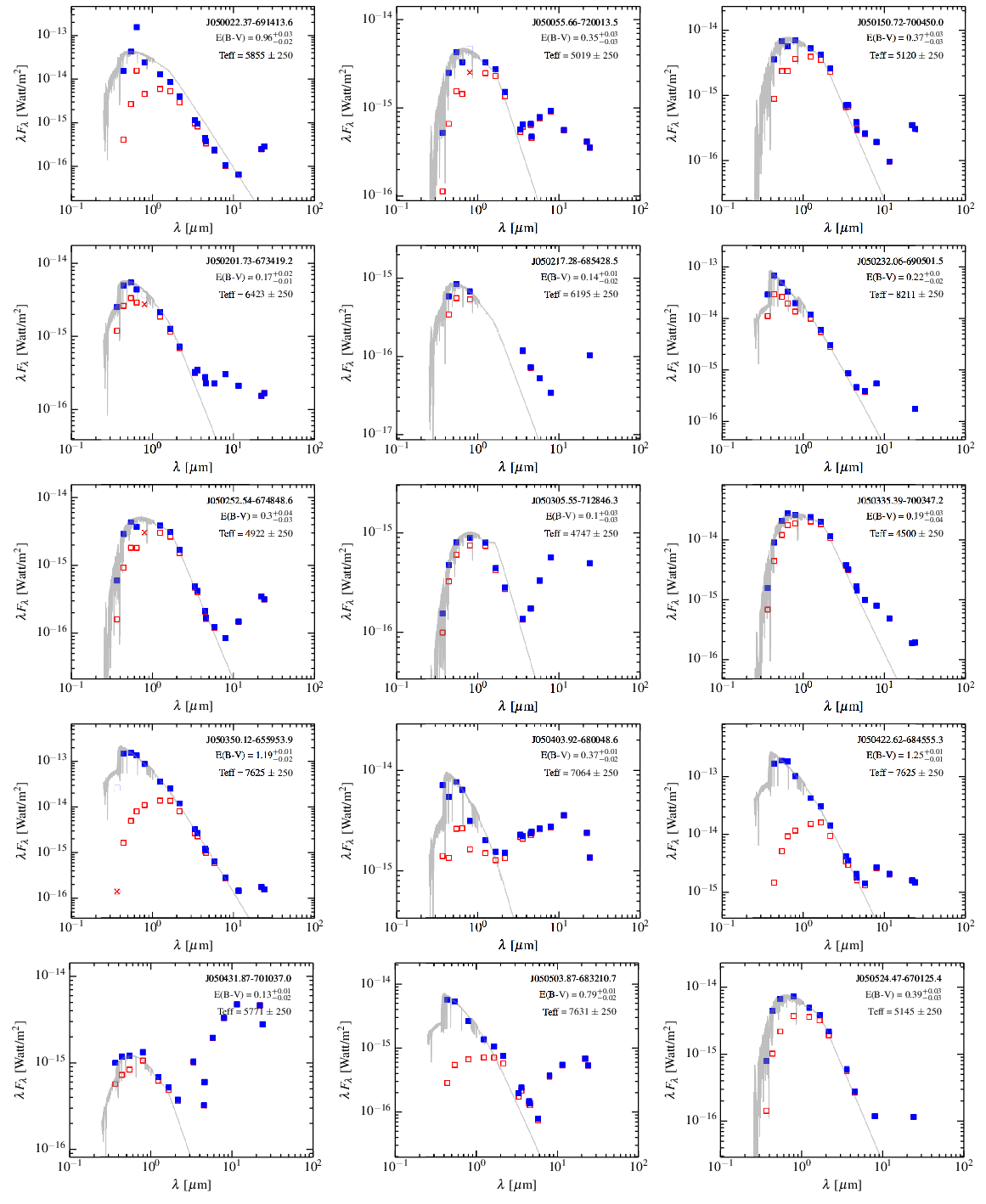}}\,
\caption{Figure~\ref{fig:yso_sed} continued.}
\end{figure*}
\begin{figure*}
\ContinuedFloat
\centering
\subfloat{\includegraphics[width=15cm]{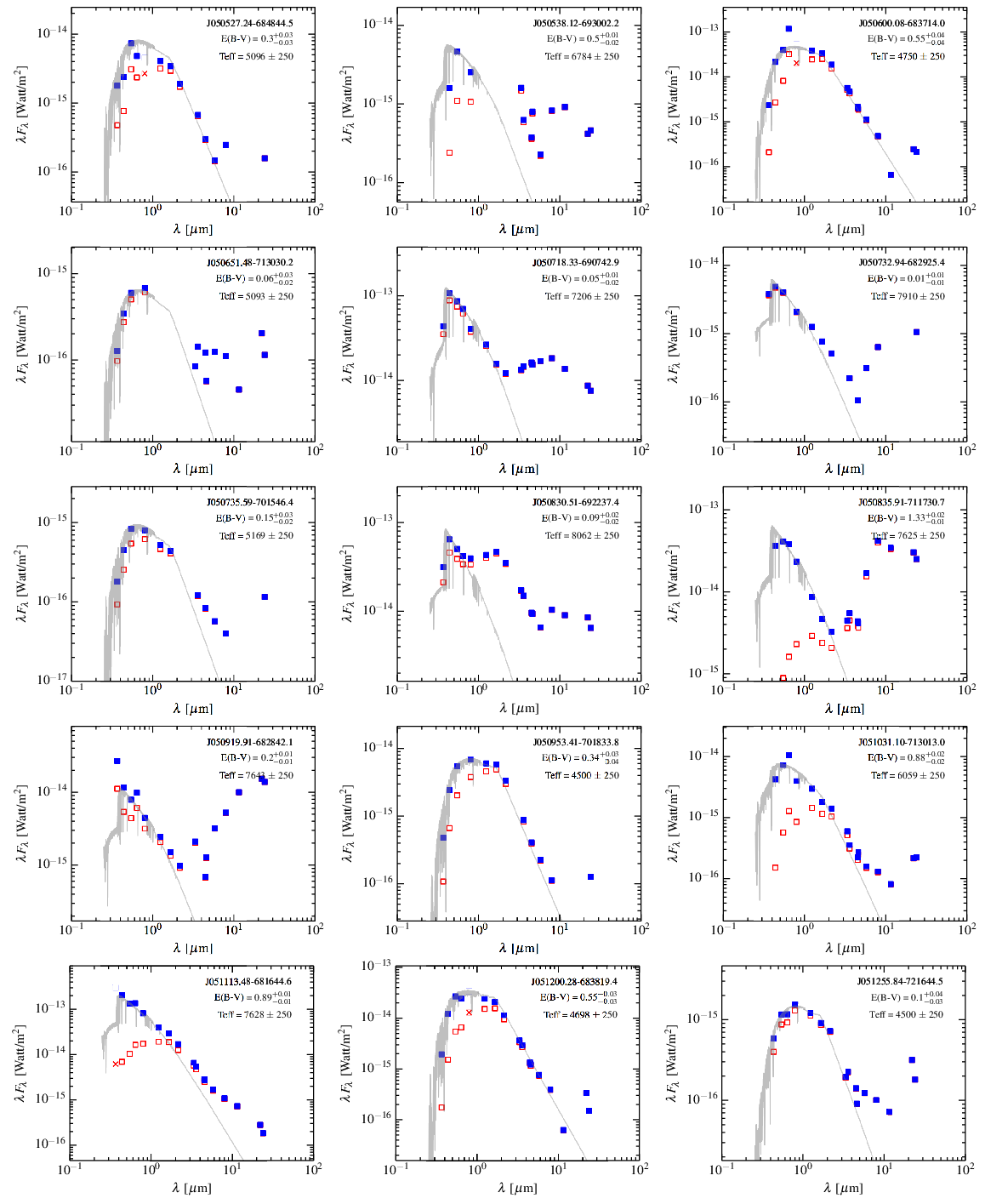}}\,
\caption{Figure~\ref{fig:yso_sed} continued.}
\end{figure*}
\begin{figure*}
\ContinuedFloat
\centering
\subfloat{\includegraphics[width=15cm]{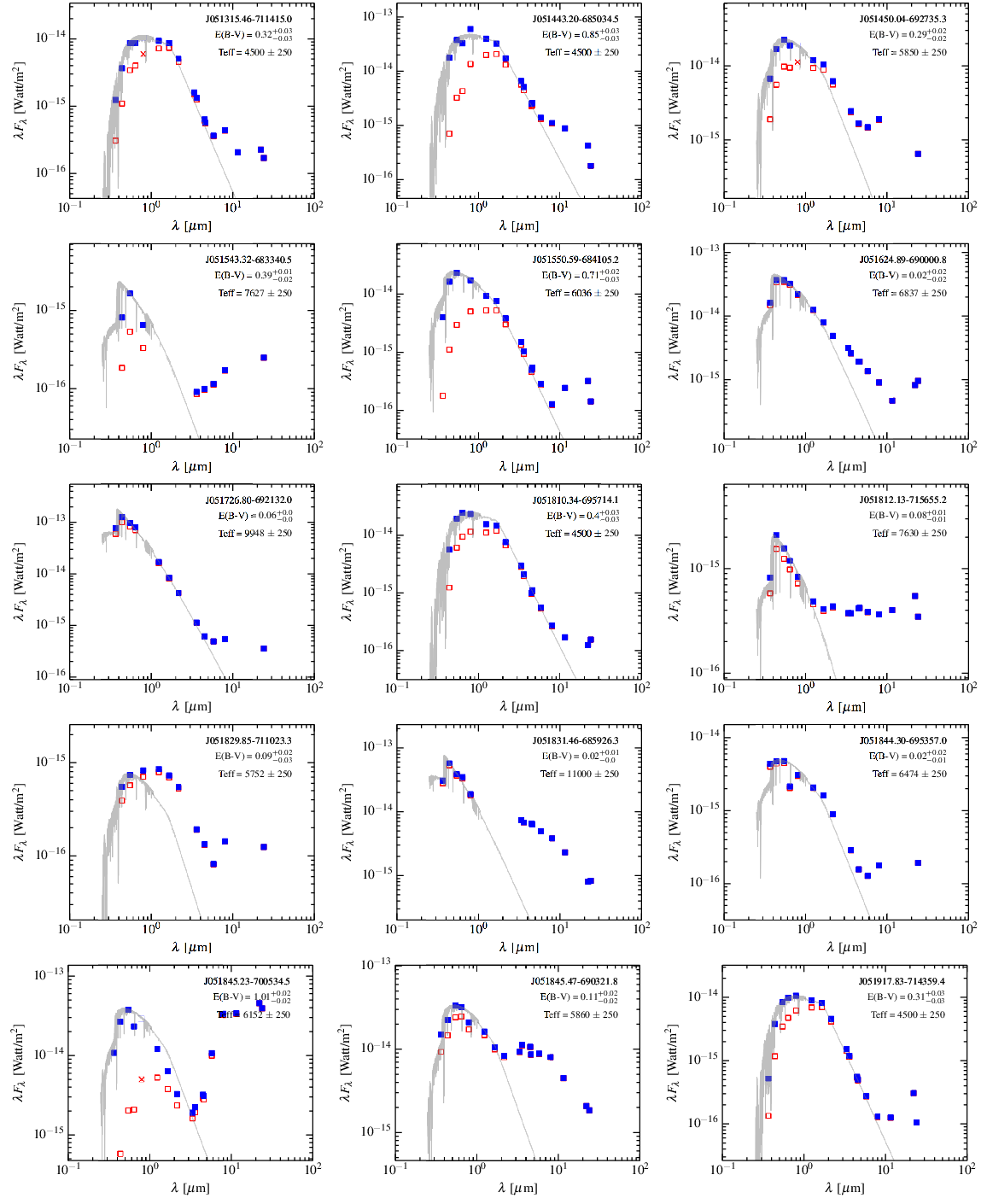}}\,
\caption{Figure~\ref{fig:yso_sed} continued.}
\end{figure*}
\begin{figure*}
\ContinuedFloat
\centering
\subfloat{\includegraphics[width=15cm]{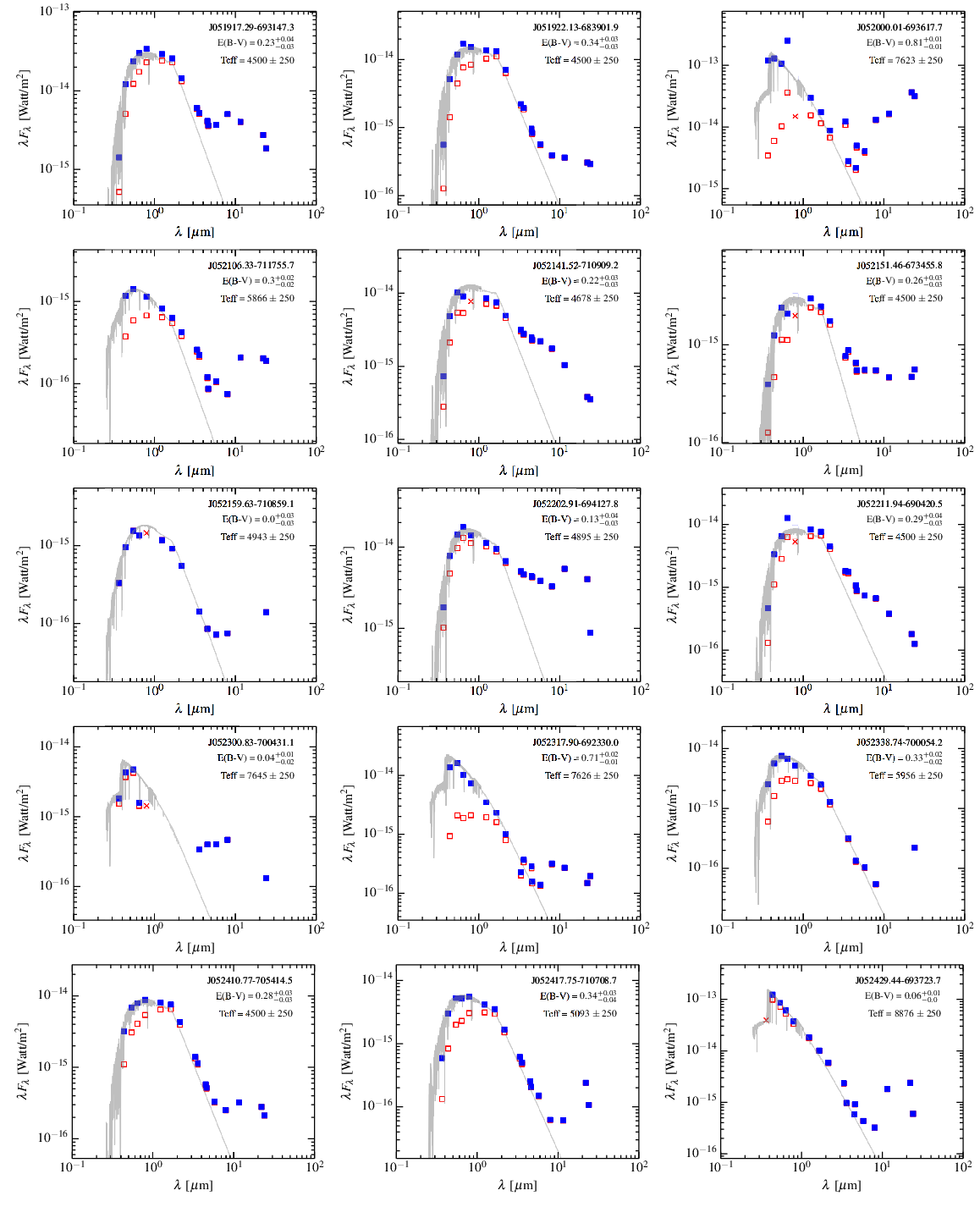}}\,
\caption{Figure~\ref{fig:yso_sed} continued.}
\end{figure*}
\begin{figure*}
\ContinuedFloat
\centering
\subfloat{\includegraphics[width=15cm]{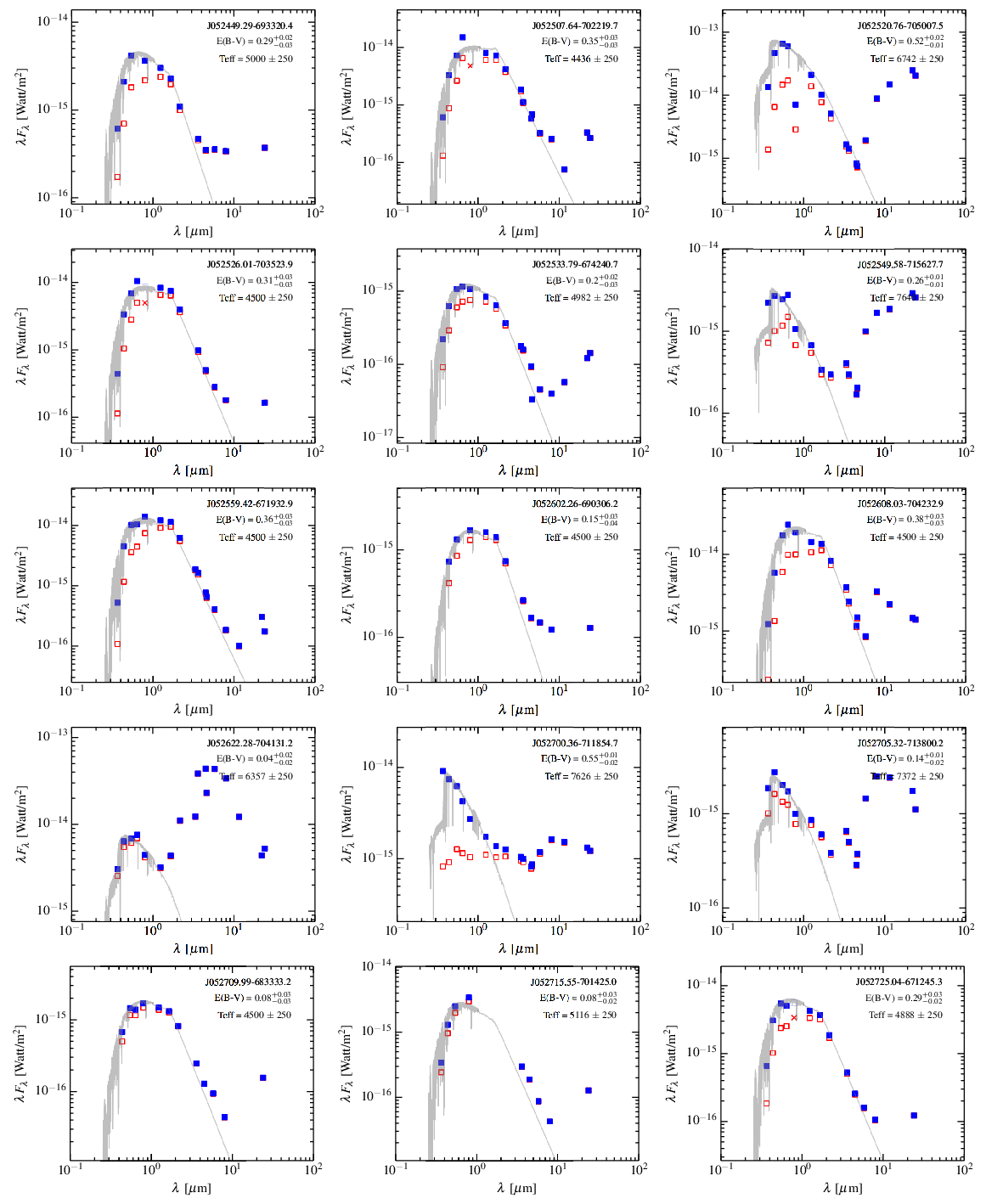}}\,
\caption{Figure~\ref{fig:yso_sed} continued.}
\end{figure*}
\begin{figure*}
\ContinuedFloat
\centering
\subfloat{\includegraphics[width=15cm]{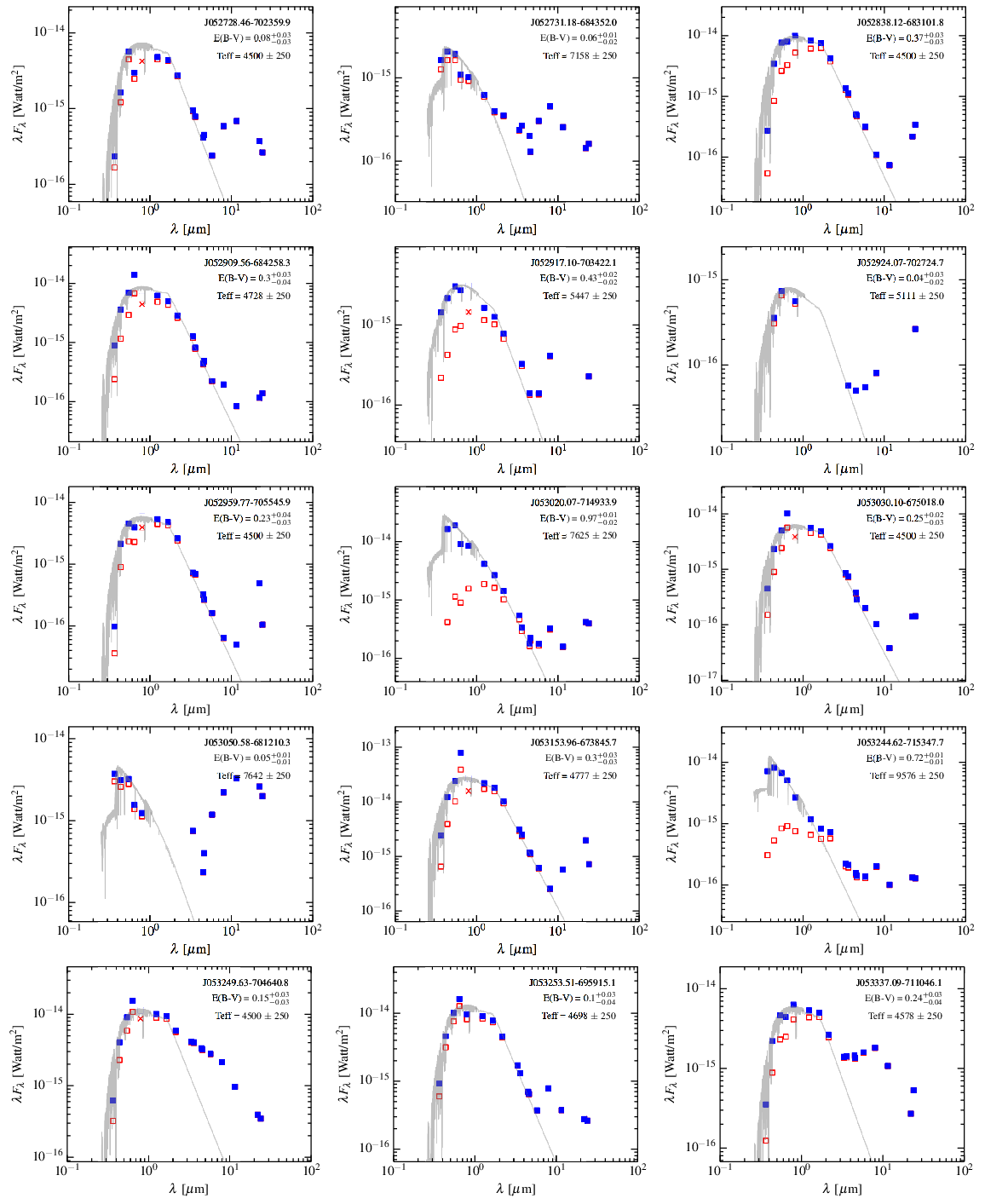}}\,
\caption{Figure~\ref{fig:yso_sed} continued.}
\end{figure*}
\begin{figure*}
\ContinuedFloat
\centering
\subfloat{\includegraphics[width=15cm]{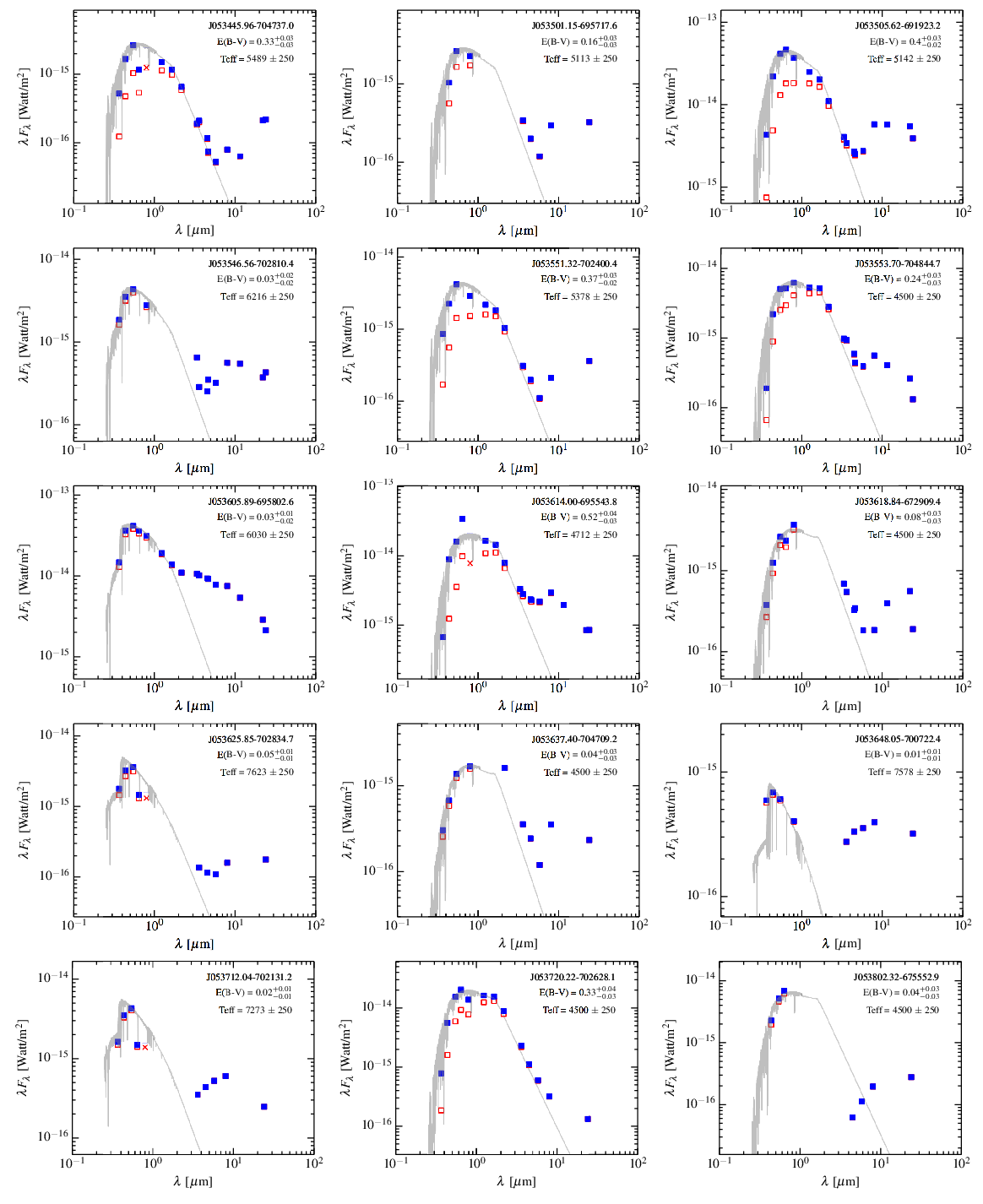}}\,
\caption{Figure~\ref{fig:yso_sed} continued.}
\end{figure*}
\begin{figure*}
\ContinuedFloat
\centering
\subfloat{\includegraphics[width=15cm]{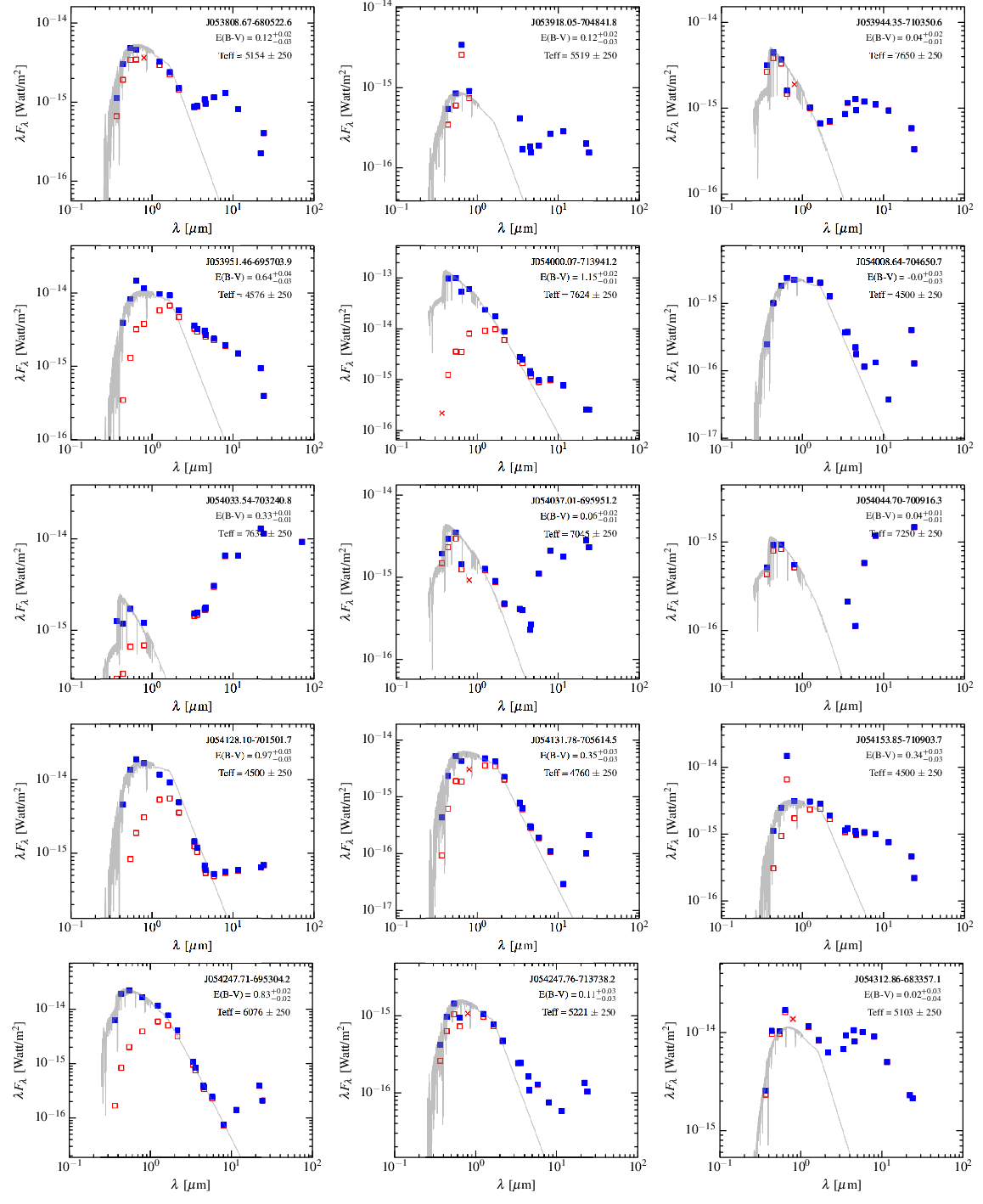}}\,
\caption{Figure~\ref{fig:yso_sed} continued.}
\end{figure*}
\begin{figure*}
\ContinuedFloat
\centering
\subfloat{\includegraphics[width=15cm]{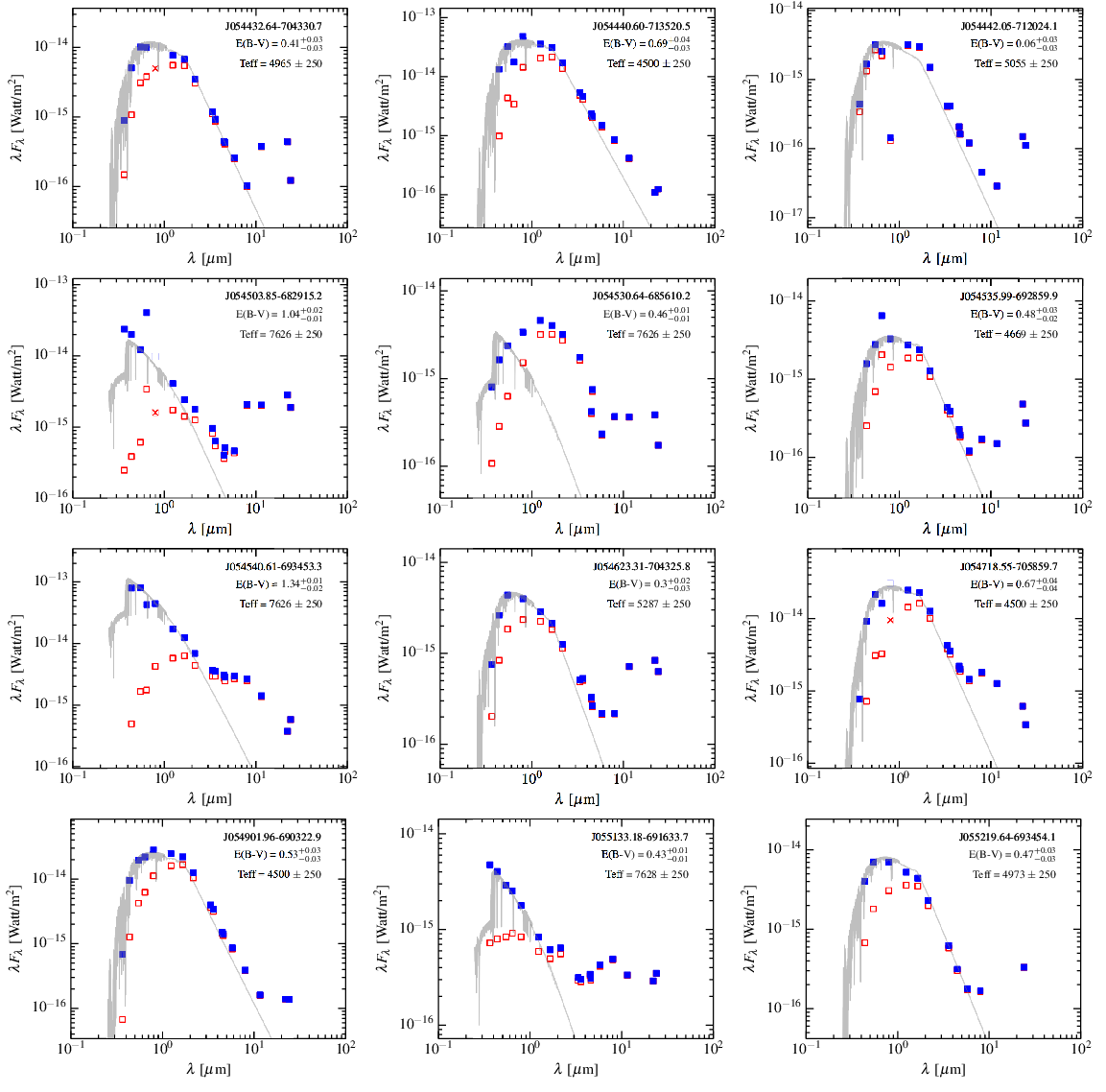}}\,
\caption{Figure~\ref{fig:yso_sed} continued.}
\end{figure*}

\clearpage

\label{lastpage}
\end{document}